\documentclass[12pt]{article}

\global\arraycolsep=1pt

\usepackage{color,amsbsy,amssymb,latexsym,amsfonts, amsmath,cancel}
\usepackage{braket}
\usepackage{mathrsfs}
\usepackage{graphicx}

\numberwithin{equation}{section}
\newcommand{\bel}[1]{\begin{equation}\label{#1}}                     
\newcommand{\bal}[1]{\begin{eqnarray}\label{#1}}   
\newcommand{\be}{\begin{equation}}               
\newcommand{\ba}{\begin{eqnarray}}           
\newcommand{\ee}{\end{equation}}
\newcommand{\ea}{\end{eqnarray}}
\newcommand{\nn}{\nonumber \\}

\newcommand{\ex}{\mathrm{e}}
\newcommand{\de}{\mathrm{d}}

\newcommand{\qq}{\qquad}

\renewcommand{\thefootnote}{\fnsymbol{footnote}}
\newcommand{\pardif}[2]{\frac{\partial #1}{\partial #2}}

\newcommand{\bea}{\begin{equation}}
\newcommand{\eea}{\end{equation}}

\newcommand{\bn}[4]{\boxed{{}^{#1 \; #2}_{#3 \; #4}}}	
\newcommand{\bbn}[4]{\boxed{\boxed{{}^{#1 \; #2}_{#3 \; #4}}}}	
\begin{document}

%
%
\begin{titlepage}
\begin{flushright}
\normalsize
~~~~
OCU-PHYS 433\\
December, 2015
\end{flushright}

\vspace{15pt}

\begin{center}
{\LARGE 
Genus one super-Green function revisited and superstring amplitudes with non-maximal supersymmetry
}
\end{center}

\vspace{23pt}

\begin{center}
{ H. Itoyama$^{a, b} $\footnote{e-mail: itoyama@sci.osaka-cu.ac.jp}
  and 
 Kohei Yano$^a$\footnote{e-mail: kyano@sci.osaka-cu.ac.jp} 
}\\
%
\vspace{18pt}
%

$^a$ \it Department of Mathematics and Physics, Graduate School of Science\\
Osaka City University  and\\
\vspace{5pt}

$^b$ \it Osaka City University Advanced Mathematical Institute (OCAMI) \\

\vspace{5pt}

3-3-138, Sugimoto, Sumiyoshi-ku, Osaka, 558-8585, Japan \\


%

\end{center}
%
\vspace{20pt}
\begin{center}
Abstract\\
\end{center}
We reexamine genus one super-Green functions with general boundary conditions twisted by $(\alpha, \beta)$ for $(\sigma, \tau)$ directions
 in the eigenmode expansion and derive expressions as infinite series of hypergeometric functions.

Using these, we compute one-loop superstring amplitudes with non-maximal supersymmetry, taking an example
of massless vector emissions of open string type I ${\cal Z}_2$ orbifold.


\vfill

\setcounter{footnote}{0}
\renewcommand{\thefootnote}{\arabic{footnote}}

\end{titlepage}

\renewcommand{\thefootnote}{\arabic{footnote}}
\setcounter{footnote}{0}

\section{Introduction}

The study of one-loop superstring amplitudes \cite{GreenSchwarzseries, SchwarzPhysRep} having their bosonic predecessors \cite{KobaNielsen, NeveuScherk, HSV} and that of the attendant genus one Green functions have a long history.
Major results and useful formulas have already been incorporated in standard textbooks \cite{GSWtext, Polchinskitext} and we think that the study of this subject has in that sense been trivialized.
The number of articles devoted to computations and (phenomenological) applications of superstring amplitudes in recent years are,
 however, relatively small and the generalized Green functions that do not satisfy ordinary periodicity or anti-periodicity on the genus one Riemann surfaces in $\sigma$ or $\tau$ directions do not seem to have been systematically studied,
  according to our search \cite{NL}, despite that they are  after all two point functions of QFT free fields.
These Green functions are needed in order to study scattering properties of particles in superstring compactifications \cite{DHVW, KLT, LLS} which carry non-maximal supersymmetry and which are soluble by free fields. 

In the first half of this paper, we study these bosonic and fermionic Green functions as the inverse of Laplacian and Dirac operators respectively,
 exploiting the elementary method of eigen-mode expansion.
In the known special cases, our computation boils down to the formula expressible in terms of the theta functions.
In general, our final expression is given by an infinite series consisting of a hypergeometric function (with its argument successively shifted),
 which is relevant to the genus zero Green function\footnote{Such Green function in fact appears in string theory under constant $B$ field \cite{CIMM}.}.
This is in accord with the picture that the genus one Green functions can be obtained from those of genus zero by putting an infinite number of image charges.
Our result can be represented as super-Green functions with worldsheet supersymmetry broken by the boundary condition.
In the latter half of this paper, we demonstrate the use of these Green functions by a simple yet nontrivial example that has non-maximal supersymmetry.

In the next section, we compute the genus one Green functions with the twist angles  $(\alpha, \beta)$ in $(\sigma, \tau)$ directions respectively, using the eigen-mode expansion.
We make exploit of partial fractions.
In section \ref{sec.Pathint}, we recall superstring one-loop vacuum amplitudes in the worldsheet covariant path integrals and recast them with  those of the light-cone operator formulation
 in order to circumvent the nuisance of the overall normalization.
We review the case of $T^4/{\bf Z}_2$ orbifold .

In section \ref{sec.oneloopamp}, we demonstrate the use of the super-Green functions derived in section \ref{sec.Greenfunc}
 by computing the superannulus contributions to massless vector emission amplitudes,
  taking an example from the case of non-maximal supersymmetry.
Explicit results for $1,2,3$ point amplitudes are compared with the vanishing amplitudes for the case of maximal supersymmetry.

In appendix \ref{notation:app}-\ref{app:propertyofNBCE}, we give some details of computation and background materials quoted in the text.

\section{Genus one Green functions with $(\alpha, \beta)$ boundary condition	\label{sec.Greenfunc}} 

In this section, we compute the genus one Green functions with general boundary condition
 to be designated by $(\alpha, \beta)$, using the eigenmode expansion.
We mainly consider the case of torus here.
The other one-loop geometries, Klein bottle, annulus and M\"obius band, can be constructed by the involution (or, the image method)
 as seen, for example, in \cite{BurgessMorris, IKK}.

Let $z \equiv \sigma^1 + \tau \sigma^2$ and ${\bar z} = \sigma^1 + {\bar \tau} \sigma^2$ ($0 \leq \sigma_1, \sigma_2 \leq 1$) be the complex coordinates
 on the worldsheet torus with modular parameter $\tau \equiv \tau_1 + i \tau_2$.
The Laplacian is defined by $\Delta \equiv 4 \partial_z \partial_{\bar z}$.
We use the plane wave bases
\ba	
	\Phi_{n_1, n_2} \left[{\textstyle {\alpha \atop \beta}} \right] (\sigma^1, \sigma^2)
        &\equiv&	
        \frac{1}{\sqrt{\tau_2}}
        \ex^{2 \pi i (n_1 + \alpha) \sigma^1} \ex^{2 \pi i (n_2 + \beta) \sigma^2}				\nn
        &=&	
        \frac{1}{\sqrt{\tau_2}}
        \ex^{
        	\frac{2 \pi i}{\tau - {\bar{\tau}}}
                \left\{{
                	(n_2 + \beta) - (n_1 + \alpha) {\bar{\tau}}
                }\right\}
                z
            }
        \ex^{
        	- \frac{2 \pi i}{\tau - {\bar{\tau}}}
                \left\{{
                	(n_2 + \beta) - (n_1 + \alpha) \tau
                }\right\}
                {\bar{z}}
            }
		\label{Phi}
\ea
 as our eigenfunctions, where $n_1, n_2 \in {\bf{Z}}$, $0 \leq \alpha, \beta < 1$.
We have imposed the orthonormality on $\Phi_{n_1, n_2} \left[{\textstyle {\alpha \atop \beta}} \right] (\sigma^1, \sigma^2)$
 to determine the normalization factor $\frac{1}{\sqrt{\tau_2}}$. (See appendix \ref{NoramalizationOfPhi}.)
This function possesses the following quasi-periodicities:
\ba	
	\Phi_{n_1, n_2} \left[{\textstyle {\alpha \atop \beta}} \right] (\sigma^1+1, \sigma^2)
        &=&	
        \ex^{2 \pi i \alpha}
        \Phi_{n_1, n_2} \left[{\textstyle {\alpha \atop \beta}} \right] (\sigma^1, \sigma^2)			\nn
        \Phi_{n_1, n_2} \left[{\textstyle {\alpha \atop \beta}} \right] (\sigma^1, \sigma^2+1)
        &=&	
        \ex^{2 \pi i \beta}
        \Phi_{n_1, n_2} \left[{\textstyle {\alpha \atop \beta}} \right] (\sigma^1, \sigma^2)		\:.
        	\label{PeriodicityOfPhi}
\ea

In the subsequent subsections, we will first consider the bosonic and fermionic components
 and then use these components to provide the supertorus Green function.
We will also consider the superannulus Neumann function
 as the involution of the supertorus Green function.

\subsection{bosonic part}

Since the eigenequation is
\be	
	\Delta
         \Phi_{n_1, n_2} \left[{\textstyle {\alpha \atop \beta}} \right] (\sigma^1, \sigma^2)
        =
        \Lambda_{n_1, n_2}^{(\alpha, \beta)}
         \Phi_{n_1, n_2} \left[{\textstyle {\alpha \atop \beta}} \right] (\sigma^1, \sigma^2)		\:,
        	\label{EigenEq}
\ee
 the eigenvalue reads
\ba	
	\Lambda_{n_1, n_2}^{(\alpha, \beta)}
        &\equiv&	
        \frac{4 (2 \pi)^2}{(\tau - {\bar{\tau}})^2}
        \left|{
                (n_2 + \beta) - (n_1 + \alpha) \tau
        }\right|^2						\nn
        &=&	
        - \frac{(2 \pi)^2}{\tau_2^2}
        \left[{
                \{(n_2 + \beta) - (n_1 + \alpha) \tau_1\}^2 + \{(n_1 + \alpha) \tau_2\}^2
        }\right]											\:.
        	\label{Lambda}
\ea
Note that $\Lambda_{0, 0}^{(0, 0)} = 0$.
In the following, we consider the cases of $\alpha \neq 0$ and $(\alpha, \beta) = (0, 0), (0, {1 \over 2})$.

\subsubsection{case of $\alpha \neq 0$}
%

Now we would like to compute the Green function
\ba	
	G \left[{\textstyle {\alpha \atop \beta}} \right] (z, {\bar{z}}|0, 0)
        &\equiv&	
        \sum_{n_1, n_2 = - \infty}^{\infty}
        \frac{1}{\Lambda_{n_1, n_2}^{(\alpha, \beta)}}
        \Phi_{n_1, n_2} \left[{\textstyle {\alpha \atop \beta}} \right] (\sigma^1, \sigma^2)
        \Phi_{n_1, n_2}^* \left[{\textstyle {\alpha \atop \beta}} \right] (0, 0)			\nn
        &=&	
        \frac{1}{\tau_2}
        \sum_{n_1, n_2 = - \infty}^{\infty}
        \frac{1}{\Lambda_{n_1, n_2}^{(\alpha, \beta)}}
        \ex^{2 \pi i (n_1 + \alpha) \sigma^1} \ex^{2 \pi i (n_2 + \beta) \sigma^2}			\:.
        	\label{Green'sfunction}
\ea
By translational invariance, we have chosen $0$ in the second set of arguments.

Exploiting the partial fraction, we decompose $\frac{1}{\Lambda_{n_1, n_2}^{(\alpha, \beta)}}$ into
\be	
	\frac{1}{\Lambda_{n_1, n_2}^{(\alpha, \beta)}}
        =	
        \frac{\tau - {\bar{\tau}}}{4 (2 \pi)^2}
        \frac{1}{n_1 + \alpha}
        \left\{{
        	\frac{1}{(n_2 + \beta) - (n_1 + \alpha) \tau}
                 - \frac{1}{(n_2 + \beta) - (n_1 + \alpha) {\bar{\tau}}}
        }\right\}											\:,
        	\label{partialfractiondecomposition}
\ee
 which is permissible even for $n_1 = 0$, $\alpha \neq 0$.
Using
\be	
	\int_0^1 \de \sigma
        \ex^{
        	- 2 \pi i
                \{{
                	(n_2 + \beta) - (n_1 + \alpha) \tau
                }\}
                 \sigma
            }
        =	
        \frac{
        	1 - 
                \ex^{- 2 \pi i \beta}
                q^{n_1 + \alpha}
             }
             {
             	2 \pi i
                \{{
                	(n_2 + \beta) - (n_1 + \alpha) \tau
                }\}
             }										\:,
        	\label{ToIntegral}
\ee
 we obtain
\ba	
	\frac{1}{\Lambda_{n_1, n_2}^{(\alpha, \beta)}}
        &=&	
        \frac{i(\tau - {\bar{\tau}})}{4 (2 \pi)}
        \frac{1}{n_1 + \alpha}
        \left[{
        	\frac{1}
                     {
                     	1 - 
                	\ex^{- 2 \pi i \beta}
                	q^{n_1 + \alpha}
                     }
                \int_0^1 \de \sigma
        	\ex^{
        		- 2 \pi i
                	\{{
                		(n_2 + \beta) - (n_1 + \alpha) \tau
                	}\}
                	 \sigma
            	    }
        }\right.							\nn
        &\:&	\qq \qq \qq \qq
        -
        \left.{
        	\frac{1}
                     {
                     	1 - 
                	\ex^{- 2 \pi i \beta}
                	{\bar{q}}^{-(n_1 + \alpha)}
                     }
                \int_0^1 \de \sigma
        	\ex^{
        		- 2 \pi i
                	\{{
                		(n_2 + \beta) - (n_1 + \alpha) {\bar{\tau}}
                	}\}
                	 \sigma
            	    }
        }\right]						\;,
        	\label{LambdaInverse}
\ea
  where $q \equiv \ex^{2 \pi i \tau}$.
According to eq. (\ref{ExtendedGSWformula}), we have the following manipulation:
\ba	
	&\:&
	\sum_{n_1, n_2 = - \infty}^\infty
        \frac{1}{n_1 + \alpha}
        \frac{1}
             {
             	1 - 
                \ex^{- 2 \pi i \beta}
                q^{n_1 + \alpha}
             }
        \int_0^1 \de \sigma
        \ex^{
        	- 2 \pi i
               	\{{
         		(n_2 + \beta) - (n_1 + \alpha) \tau
                }\}
                 \sigma
            }
        \ex^{2 \pi i (n_1 + \alpha) \sigma^1} \ex^{2 \pi i (n_2 + \beta) \sigma^2}			\nn
        &\:&	\qq
        =	
        \sum_{n_1 = - \infty}^\infty
        \frac{1}{n_1 + \alpha}
        \frac{1}
             {
             	1 - 
                \ex^{- 2 \pi i \beta}
                q^{n_1 + \alpha}
             }
        \int_0^1 \de \sigma
        \delta (\sigma^2 - \sigma)
        \ex^{
        	+ 2 \pi i \beta (\sigma^2 - \sigma)
                + 2 \pi i (n_1 + \alpha) (\sigma^1 + \tau \sigma)
            }												\nn
        &\:&	\qq
        =	
        \sum_{n_1 = - \infty}^\infty
        \frac{1}{n_1 + \alpha}
        \frac{{\zeta}^{n_1 + \alpha}}
             {
             	1 - 
                \ex^{- 2 \pi i \beta}
                q^{n_1 + \alpha}
             }									\nn
        &\:&	\qq
        {\overset{m \equiv n_1}{=}}	
        \sum_{m = 0}^\infty
        \frac{1}{m + \alpha}
        \frac{{\zeta}^{m + \alpha}}
             {
             	1 - 
                \ex^{- 2 \pi i \beta}
                q^{m + \alpha}
             }
        +
        \sum_{m = 1}^\infty
        \frac{1}{m - \alpha}
        \frac{\ex^{2 \pi i \beta} \left({ \frac{q}{\zeta} }\right)^{m - \alpha}}
             {
             	1 - \ex^{2 \pi i \beta} q^{m - \alpha}
             }									\nn								\nn
        &\:&	\qq
        {\overset{m' \equiv m-1, \: \alpha' \equiv 1 - \alpha}{=}}	
        \sum_{m = 0}^\infty
        \frac{1}{m + \alpha}
        \frac{{\zeta}^{m + \alpha}}
             {
             	1 - 
                \ex^{- 2 \pi i \beta}
                q^{m + \alpha}
             }
        +
        \ex^{2 \pi i \beta}
        \sum_{m' = 0}^\infty
        \frac{1}{m' + \alpha'}
        \frac{
        	\left({
        		\frac{q}{\zeta}
                }\right)^{m' + \alpha'}
             }
             {
             	1 - 
                \ex^{2 \pi i \beta} q^{m' + \alpha'}
             }									\nn
        &\:&	\qq
        {\overset{{\rm eq.} (\ref{ExtendedGSWformula})}{=}}	
        \frac{\Gamma (\alpha)}{\Gamma (1 + \alpha)}
        \sum_{n=0}^\infty
        \left({
        	\ex^{- 2 \pi i \beta}
        }\right)^n
        ({\zeta} q^n)^\alpha
        F(1, \alpha, 1+\alpha; {\zeta} q^n)					\nn
        &\:&	\qq \qq
        +
        \frac{\Gamma (\alpha')}{\Gamma (1 + \alpha')}
        \sum_{n=0}^\infty
        (\ex^{2 \pi i \beta})^{n+1}
        \left({
        	\frac{q^{n+1}}{\zeta}
        }\right)^{\alpha'}
        F \left({1, \alpha', 1+\alpha'; \frac{q^{n+1}}{{\zeta}}}\right)				\:,
        	\label{Preliminary1}
\ea
 where $\zeta \equiv \ex^{2 \pi i z} = \ex^{2 \pi i (\sigma^1 + \tau \sigma^2)}$ and $\alpha' \equiv 1 - \alpha$.
Similarly,
\ba	
	&\:&
        \sum_{n_1, n_2 = - \infty}^\infty
        \frac{1}{n_1 + \alpha}
        \frac{1}
             {
             	1 - 
                \ex^{- 2 \pi i \beta}
                {\bar{q}}^{-(n_1 + \alpha)}
             }
        \int_0^1 \de \sigma
        \ex^{
        	- 2 \pi i
                \{{
                	(n_2 + \beta) - (n_1 + \alpha) {\bar{\tau}}
                }\}
                 \sigma
            }
        \ex^{2 \pi i (n_1 + \alpha) \sigma^1} \ex^{2 \pi i (n_2 + \beta) \sigma^2}			\nn
        &\:&	\qq
        =	
        -
        \frac{\Gamma (\alpha)}{\Gamma (1+\alpha)}
        \sum_{n=0}^\infty
        (\ex^{2 \pi i \beta})^{n+1}
        \left({
                \frac{{\bar q}^{n+1}}{\bar{\zeta}}
        }\right)^{\alpha}
        F \left({1, \alpha, 1+\alpha; \frac{{\bar q}^{n+1}}{\bar{\zeta}}}\right)		\nn
        &\:&	\qq \qq
        -
        \frac{\Gamma (\alpha')}{\Gamma (1+\alpha')}
        \sum_{n=0}^\infty
        (\ex^{- 2 \pi i \beta})^{n}
        \left({
                {\bar{\zeta}} {\bar{q}}^n
        }\right)^{\alpha'}
        F \left({1, \alpha', 1+\alpha'; {\bar \zeta} {\bar q}^n}\right)				\:.
        	\label{Preliminary2}
\ea

Substituting eqs. (\ref{Preliminary1}) and (\ref{Preliminary2}) into eq. (\ref{Green'sfunction}),
 we obtain
\ba	
	&\:&
	G \left[{\textstyle {\alpha \atop \beta}} \right] (z, {\bar{z}}|0, 0)				\nn
        &\:&	\qq
        =	
        - \frac{1}{2 (2 \pi)}
        \left[{
        	\frac{\Gamma (\alpha)}{\Gamma (1 + \alpha)}
        	\sum_{n=0}^\infty
        	\left({
        		\ex^{- 2 \pi i \beta}
        	}\right)^n
        	({\zeta} q^n)^\alpha
        	F(1, \alpha, 1+\alpha; {\zeta} q^n)
        }\right.									\nn
        &\:&	\qq \qq \qq \qq	%
        	+
        	\frac{\Gamma (\alpha')}{\Gamma (1 + \alpha')}
        	\sum_{n=0}^\infty
        	(\ex^{2 \pi i \beta})^{n+1}
        	\left({
        		\frac{q^{n+1}}{\zeta}
        	}\right)^{\alpha'}
        	F \left({1, \alpha', 1+\alpha'; \frac{q^{n+1}}{{\zeta}}}\right)		\nn
        &\:&	\qq \qq \qq \qq	%
        	+
        	\frac{\Gamma (\alpha)}{\Gamma (1+\alpha)}
        	\sum_{n=0}^\infty
        	(\ex^{2 \pi i \beta})^{n+1}
        	\left({
                	\frac{{\bar q}^{n+1}}{\bar{\zeta}}
        	}\right)^{\alpha}
        	F \left({1, \alpha, 1+\alpha; \frac{{\bar{q}}^{n+1}}{{\bar{\zeta}}}}\right)			\nn
        &\:&	\qq \qq \qq \qq	%
        \left.{
        	+
        	\frac{\Gamma (\alpha')}{\Gamma (1+\alpha')}
        	\sum_{n=0}^\infty
        	(\ex^{- 2 \pi i \beta})^n
        	\left({
                	{\bar{\zeta}} {\bar{q}}^n
        	}\right)^{\alpha'}
        	F (1, \alpha', 1+\alpha'; {\bar{\zeta}} {\bar{q}}^n)
        }\right]										\:.
\ea

\subsubsection{case of $(\alpha, \beta) = (0, 0)$}
%

In this case, it is necessary to exclude $(n_1, n_2) = (0, 0)$ at the sum in eq. (\ref{Green'sfunction}):
\be	
        G_{++} (z, {\bar{z}}|0, 0)
        \equiv	
	G \left[{\textstyle {0 \atop 0}} \right] (z, {\bar{z}}|0, 0)
        =	
        \frac{1}{\tau_2}
        \sum_{\begin{subarray}{c}n_1, n_2 = - \infty \\ (n_1, n_2) \neq (0,0) \end{subarray}}^{\infty}
        \frac{1}{\frac{4 (2 \pi)^2}{(\tau - {\bar \tau})^2} |n_2 - n_1 \tau|^2}
        \ex^{2 \pi i n_1 \sigma^1} \ex^{2 \pi i n_2 \sigma^2}				\:.
        	\label{G_++:sec2}
\ee

As the result of the calculation in appendix \ref{app:G++}, we obtain
\ba	
	&\:&
        G_{++} (z, {\bar{z}}|0, 0)				\nn
        &\:&
        {\overset{{\rm eq.} (\ref{summary:G_++calc.App})}{=}}	
        \frac{1}{2 \pi}
        \ln \left|{
        		\frac{\vartheta \left[{\textstyle {{1 \over 2} \atop {1 \over 2}}} \right] (z)}
             		     {\vartheta' \left[{\textstyle {{1 \over 2} \atop {1 \over 2}}} \right] (0)}
            	  }\right|
        -
        {1 \over 2}
        \frac{\left({ {\rm Im} \: z }\right)^2}{\tau_2}
        +
        \left[{
        	\frac{1}{2 \pi}
                2
        	\sum_{n=1}^\infty
        	\ln |1 - q^n|
                -
                \frac{1}{2}
                ({\rm Im} \: z)
                +
                \frac{1}{2 \pi}
                \ln (2 \pi)
                +
                2 \tau_2 \cdot \frac{\pi^2}{6}
        }\right]						\:.		\nn
        	\label{def:G_++}
\ea
The terms in the bracket [...] vanish when acting on $\Delta = 4 \partial_{z} \partial_{\bar z}$.

\subsubsection{case of $(\alpha, \beta) = (0, {1 \over 2})$}
%

Here we consider
\be
        G_{+-} (z, {\bar{z}}|0, 0)
        \equiv	
	G \left[{\textstyle {0 \atop {1 \over 2}}} \right] (z, {\bar{z}}|0, 0)
        =	
        \frac{1}{\tau_2}
        \sum_{n_1, n_2 = - \infty}^{\infty}
        \frac{1}{\frac{4 (2 \pi)^2}{(\tau - {\bar \tau})^2} \left|{ \left({ n_2 + {1 \over 2} }\right) - n_1 \tau }\right|^2}
        \ex^{2 \pi i n_1 \sigma^1} \ex^{2 \pi i \left({ n_2 + {1 \over 2} }\right) \sigma^2}			\:.
        	\label{G_{+-}:def:sec2}
\ee
Now we divide this sum into $n_1 \neq 0$ part and $n_1 = 0$ part
 to use the partial fraction decomposition in eq. (\ref{partialfractiondecomposition}).

As the result of the calculation in appendix \ref{app:G+-}, we obtain
\be	
        G_{+-} (z, {\bar{z}}|0, 0)
        {\overset{{\rm eq.} (\ref{summary:G_+-})}{=}}	
        \frac{1}{2 \pi}
        \left[{
        	\ln |1 - \zeta|
                +
        	\sum_{m=1}^\infty
        	(-1)^{m} \ln |1 - \zeta q^m| \left|{ 1 - \frac{q^{m}}{\zeta} }\right|
        }\right]
        -
        \pi^2 \tau_2					\:.
        	\label{def:G_+-}
\ee

\subsection{fermionic part}

The eigen-equations are
\ba	
	(-i) \partial_{\bar{z}}
         \Phi_{n_1, n_2} \left[{\textstyle {\alpha \atop \beta}} \right] (\sigma^1, \sigma^2)
        &=&	
        \kappa_{n_1, n_2}^{(\alpha, \beta)}
         \Phi_{n_1, n_2} \left[{\textstyle {\alpha \atop \beta}} \right] (\sigma^1, \sigma^2)			\nn
        (-i) \partial_z
         \Phi_{n_1, n_2} \left[{\textstyle {\alpha \atop \beta}} \right] (\sigma^1, \sigma^2)
        &=&	
	{\bar{\kappa}}_{n_1, n_2}^{(\alpha, \beta)}
         \Phi_{n_1, n_2} \left[{\textstyle {\alpha \atop \beta}} \right] (\sigma^1, \sigma^2)				\:.
        	\label{Eigeneq:S_barS}
\ea
The eigenvalues can be written as
\ba	
	\kappa_{n_1, n_2}^{(\alpha, \beta)}
        &=&	
        - \frac{2 \pi}{\tau - {\bar{\tau}}}
          \{ (n_2 + \beta) - (n_1 + \alpha) \tau \}
        =	%
        + \frac{\pi i}{\tau_2}
          \{ (n_2 + \beta) - (n_1 + \alpha) \tau \}			\nn
        {\bar{\kappa}}_{n_1, n_2}^{(\alpha, \beta)}
        &=&	
        + \frac{2 \pi}{\tau - {\bar{\tau}}}
          \{ (n_2 + \beta) - (n_1 + \alpha) {\bar{\tau}} \}
        =	%
        - \frac{\pi i}{\tau_2}
          \{ (n_2 + \beta) - (n_1 + \alpha) {\bar{\tau}} \}		\;.
		\label{kappa_barakappa}
\ea
Note that $\kappa_{0, 0}^{(0, 0)} = {\bar{\kappa}}_{0, 0}^{(0, 0)} = 0$.

\subsubsection{case of $(\alpha, \beta) \neq (0, 0)$}
%

Here we calculate the Green function
\ba	
	{\cal{S}} \left[{\textstyle {\alpha \atop \beta}} \right] (z, {\bar{z}}|0, 0)
        &=&	
        \frac{1}{\tau_2}
        \sum_{n_1, n_2 = - \infty}^{\infty}
        \frac{1}{\kappa_{n_1, n_2}^{(\alpha, \beta)}}
        \ex^{2 \pi i (n_1 + \alpha) \sigma^1} \ex^{2 \pi i (n_2 + \beta) \sigma^2}			\nn
        {\bar{\cal{S}}} \left[{\textstyle {\alpha \atop \beta}} \right] (z, {\bar{z}}|0, 0)
        &=&	
        \frac{1}{\tau_2}
        \sum_{n_1, n_2 = - \infty}^{\infty}
        \frac{1}{{\bar{\kappa}}_{n_1, n_2}^{(\alpha, \beta)}}
        \ex^{- 2 \pi i (n_1 + \alpha) \sigma^1} \ex^{- 2 \pi i (n_2 + \beta) \sigma^2}			\:.
        	\label{GreenFunc:fermion}
\ea

We obtain
\ba	
	&\:&
	{\cal{S}} \left[{\textstyle {\alpha \atop \beta}} \right] (z, {\bar{z}}|0, 0)			\nn
        &\:&	\qq
        {\overset{{\rm eqs. \:} (\ref{ToIntegral}), \: (\ref{kappa_barakappa}), \: (\ref{GreenFunc:fermion})}{=}}	
        - \frac{1}{\tau_2} (\tau - {\bar{\tau}}) i
        \sum_{n_1, n_2 = - \infty}^{\infty}
        \frac{1}{1 - \ex^{- 2 \pi i \beta} q^{n_1 + \alpha}}			\nn
        &\:&	\qq \qq \qq \qq \qq \qq \qq \qq \qq	
        \times
        \int_0^1 \de \sigma
        \ex^{2 \pi i (\sigma^2 - \sigma) n_2} \ex^{2 \pi i \beta (\sigma^2 - \sigma)} \ex^{2 \pi i (n_1 + \alpha) (\sigma^1 + \tau \sigma^2)}			\nn
        &\:&	\qq
        =	
        - \frac{(\tau - {\bar{\tau}}) i}{\tau_2}
        \sum_{n_1 = - \infty}^{\infty}
        \frac{{\zeta}^{n_1 + \alpha}}{1 - \ex^{- 2 \pi i \beta} q^{n_1 + \alpha}}
        {\overset{{\rm eq. \:} (\ref{b=aq,/1-aToTheta})}{=}}	
        \frac{i}{\pi}
        \frac{
        	\vartheta \left[{\textstyle {\alpha - {1 \over 2}  \atop  {1 \over 2}- \beta}} \right] (z|\tau)
             }
             {
             	\vartheta \left[{\textstyle {\alpha - {1 \over 2}  \atop  {1 \over 2}- \beta}} \right] (0|\tau)
             }
        \frac{
        	\vartheta' \left[{\textstyle {{1 \over 2} \atop  {1 \over 2}}} \right] (0|\tau)
             }
             {
             	\vartheta \left[{\textstyle {{1 \over 2}  \atop  {1 \over 2}}} \right] (z|\tau)
             }						\:.
		\label{calc.:FermionPart1}
\ea
Similarly, using
\be	
	\overline{
        	\vartheta' \left[{\textstyle {{1 \over 2} \atop {1 \over 2}}} \right] (0|\tau)
        }
        =
        \vartheta' \left[{\textstyle {{1 \over 2} \atop {1 \over 2}}} \right] (0| -{\bar \tau})		\;,
        	\label{barTheta'[1/2,1/2](0|tau)}
\ee
\ba
	{\bar{\cal{S}}} \left[{\textstyle {\alpha \atop \beta}} \right] (z, {\bar{z}}|0, 0)
        &=&	
	-
        \frac{i}{\pi}
        \frac{
        	\vartheta \left[{\textstyle {\alpha - {1 \over 2}  \atop  {1 \over 2}+ \beta}} \right] (-{\bar z}| -{\bar \tau})
             }
             {
             	\vartheta \left[{\textstyle {\alpha - {1 \over 2}  \atop  {1 \over 2}+ \beta}} \right] (0| -{\bar \tau})
             }
        \frac{
        	\vartheta' \left[{\textstyle {{1 \over 2} \atop  {1 \over 2}}} \right] (0| -{\bar \tau})
             }
             {
             	\vartheta \left[{\textstyle {{1 \over 2}  \atop  {1 \over 2}}} \right] (-{\bar z}| -{\bar \tau})
             }							\nn
        &{\overset{ {\rm eqs. \:} (\ref{alpha+1_beta+1}), \: (\ref{bar_theta})}{=}}&	
        -
        \frac{i}{\pi}
        \frac{
        	\overline{
                	\vartheta \left[{\textstyle {\alpha - {1 \over 2}  \atop {1 \over 2} - \beta}} \right] (z| \tau)
                }
             }
             {
             	\overline{
                	\vartheta \left[{\textstyle {\alpha - {1 \over 2} \atop {1 \over 2} - \beta}} \right] (0| \tau)
                }
             }
        \frac{
        	\overline{
                	\vartheta' \left[{\textstyle {{1 \over 2} \atop {1 \over 2}}} \right] (0| \tau)
                }
             }
             {
             	\overline{
                	\vartheta \left[{\textstyle {{1 \over 2} \atop {1 \over 2}}} \right] (z|  \tau)
                }
             }						\:.
\ea
This time, we have used
\be
	\int_0^1 \de \sigma
        \ex^{
        	2 \pi i
                \{{
                	(n_2 + \beta) - (n_1 + \alpha) {\bar{\tau}}
                }\}
                 \sigma
            }
        =	
        -
        \frac{
        	1 - 
                \ex^{2 \pi i \beta}
                {\bar{q}}^{n_1 + \alpha}
             }
             {
             	2 \pi i
                \{{
                	(n_2 + \beta) - (n_1 + \alpha) {\bar{\tau}}
                }\}
             }
        	\label{ToIntegral2}
\ee
 instead of eq. (\ref{ToIntegral}),
  avoiding getting $\delta (\sigma^2 + \sigma)$ which vanishes in the original domain.

In particular, when $(\alpha, \beta) = ({1 \over 2}, {1 \over 2}), ({1 \over 2}, 0), (0, {1 \over 2})$, we obtain
\ba	
	{\cal{S}}_{--} (z, {\bar{z}}|0, 0)
        &\equiv&	
	{\cal{S}} \left[{\textstyle {{1 \over 2} \atop {1 \over 2}}} \right] (z, {\bar{z}}|0, 0)
        =	
        \frac{i}{\pi}
        \frac{
        	\vartheta \left[{\textstyle {0 \atop 0}} \right] (z|\tau)
             }
             {
             	\vartheta \left[{\textstyle {0 \atop 0}} \right] (0|\tau)
             }
        \frac{
        	\vartheta' \left[{\textstyle {{1 \over 2} \atop {1 \over 2}}} \right] (0|\tau)
             }
             {
             	\vartheta \left[{\textstyle {{1 \over 2} \atop {1 \over 2}}} \right] (z|\tau)
             }										\nn
	{\cal{S}}_{-+} (z, {\bar{z}}|0, 0)
        &\equiv&	
	{\cal{S}} \left[{\textstyle {{1 \over 2} \atop 0}} \right] (z, {\bar{z}}|0, 0)
        =	
        \frac{i}{\pi}
        \frac{
        	\vartheta \left[{\textstyle {0 \atop {1 \over 2}}} \right] (z|\tau)
             }
             {
             	\vartheta \left[{\textstyle {0 \atop {1 \over 2}}} \right] (0|\tau)
             }
        \frac{
        	\vartheta' \left[{\textstyle {{1 \over 2} \atop {1 \over 2}}} \right] (0|\tau)
             }
             {
             	\vartheta \left[{\textstyle {{1 \over 2} \atop {1 \over 2}}} \right] (z|\tau)
             }										\nn
	{\cal{S}}_{+-} (z, {\bar{z}}|0, 0)
        &\equiv&	
	{\cal{S}} \left[{\textstyle {0 \atop {1 \over 2}}} \right] (z, {\bar{z}}|0, 0)
        =	
        \frac{i}{\pi}
        \frac{
        	\vartheta \left[{\textstyle {{1 \over 2} \atop 0}} \right] (z|\tau)
             }
             {
             	\vartheta \left[{\textstyle {{1 \over 2} \atop 0}} \right] (0|\tau)
             }
        \frac{
        	\vartheta' \left[{\textstyle {{1 \over 2} \atop {1 \over 2}}} \right] (0|\tau)
             }
             {
             	\vartheta \left[{\textstyle {{1 \over 2} \atop {1 \over 2}}} \right] (z|\tau)
             }
        	\label{S_{--},S_{-+},S_{+-}}
\ea
 and the complex conjugates of these.

\subsubsection{case of $(\alpha, \beta) = (0, 0)$}
%

In this case, we need to exclude the zero mode $(n_1, n_2) = (0, 0)$ in the sum:
\ba	
	{\cal{S}}_{++} (z, {\bar{z}}|0, 0)
        &\equiv&	
	{\cal{S}} \left[{\textstyle {0 \atop 0}} \right] (z, {\bar{z}}|0, 0)
        =	
        \frac{1}{\tau_2}
        \sum_{\begin{subarray}{c}n_1, n_2 = - \infty \\ (n_1, n_2) \neq (0,0) \end{subarray}}^{\infty}
        \frac{1}
             {
             	\left({
        	- \frac{2 \pi}{\tau - {\bar{\tau}}}
        	}\right)
                (n_2 - n_1 \tau)
             }
        \ex^{2 \pi i n_1 \sigma^1} \ex^{2 \pi i n_2 \sigma^2}			\nn
        {\bar{\cal{S}}}_{++} (z, {\bar{z}}|0, 0)
        &\equiv&	
	{\bar{\cal{S}}} \left[{\textstyle {0 \atop 0}} \right] (z, {\bar{z}}|0, 0)
        =	
        \frac{1}{\tau_2}
        \sum_{\begin{subarray}{c}n_1, n_2 = - \infty \\ (n_1, n_2) \neq (0,0) \end{subarray}}^{\infty}
        \frac{1}{
        	+ \frac{2 \pi}{\tau - {\bar{\tau}}}
          		(n_2 - n_1 {\bar{\tau}})
        	}
        \ex^{- 2 \pi i n_1 \sigma^1} \ex^{- 2 \pi i n_2 \sigma^2}			\:.
        	\label{def.:S_{++}and{bar S}_{++}}
\ea
Here we use the relation
\be	
	{\cal{S}}_{++} (z, {\bar{z}}|0, 0)
        {\overset{{\rm eqs.} (\ref{G_++:sec2}), (\ref{def.:S_{++}and{bar S}_{++}})}{=}}	
        4 i
        \pardif{}{z}
        G_{++} (z, {\bar{z}}|0, 0)
        	\label{S_{++}=4ipartial_z G_{++}}
\ee
 to calculate eq. (\ref{def.:S_{++}and{bar S}_{++}}),
  because eq. (\ref{b=aq,/1-aToTheta}) appears not to work well when $(\alpha, \beta) = (0, 0)$.
Eq. (\ref{S_{++}=4ipartial_z G_{++}}) can be easily understood by using the last line in eq. (\ref{Phi}).
From eqs. (\ref{S_{++}=4ipartial_z G_{++}}), (\ref{def:G_++}), we obtain
\be	
	{\cal{S}}_{++} (z, {\bar{z}}|0, 0)
        =	
        \frac{i}{\pi}
        \frac{\vartheta' \left[{\textstyle {{1 \over 2} \atop {1 \over 2}}} \right] (z)}
             {\vartheta \left[{\textstyle {{1 \over 2} \atop {1 \over 2}}} \right] (z)}
        -
        2
        \frac{({\rm Im} \: z)}{\tau_2}
        -
        1
        	\label{S_++}					\:.
\ee
In addition,
\be	
	{\bar{\cal{S}}}_{++} (z, {\bar{z}}|0, 0)
        =	
        {\overline{{\cal{S}}_{++} (z, {\bar{z}}|0, 0)}}
        {\overset{(\ref{S_++})}{=}}	
        -
        \frac{i}{\pi}
        \frac{\overline{\vartheta' \left[{\textstyle {{1 \over 2} \atop {1 \over 2}}} \right] (z)}}
             {\overline{\vartheta \left[{\textstyle {{1 \over 2} \atop {1 \over 2}}} \right] (z)}}
        -
        2
        \frac{({\rm Im} \: z)}{\tau_2}
        -
        1							\:.
        	\label{barS_{++}}
\ee
The last term, namely, $-1$ in eqs. (\ref{S_++}) and (\ref{barS_{++}}) vanishes when acting with $(-i) \partial_{\bar z}$ or $(-i) \partial_{z}$.

\subsection{supertorus Green function and superannulus Neumann function}

\subsubsection{supertorus Green function}
%

We define the supertorus Green function ($\nu_{\rm f} = (-,-) {\: {\rm or} \:} (-,+) {\: {\rm or} \:} (+,-)$) by
\ba	
	{\bf G}_{\shortstack{$+ \pm$ \\ $\nu_{\rm f}$}}^{\rm{supertorus}} (z_I, {\bar{z}}_I|z_J, {\bar{z}}_J)
        &\equiv&	
	G_{+ \pm} (z_I, {\bar{z}}_I|z_J, {\bar{z}}_J)
        +
        \frac{\theta_I \theta_J}{4}
        {\cal{S}}_{\nu_{\rm f}} (z_I, {\bar{z}}_I|z_J, {\bar{z}}_J)
        -
        \frac{{\bar \theta}_I {\bar \theta}_J}{4}
        {\overline{\cal{S}}_{\nu_{\rm f}}} (z_I, {\bar{z}}_I|z_J, {\bar{z}}_J)			\:,		\nn
\ea
 where $\theta$, ${\bar \theta}$ are Grassmann coordinates
  and $G_{++}$, $G_{+-}$ and ${\cal{S}}_{\nu_{\rm f}}$ are given in eqs. (\ref{def:G_++}), (\ref{def:G_+-}) and (\ref{S_{--},S_{-+},S_{+-}}), respectively.
According to appendix \ref{G_STandGSS}, we can see that ${\bf G}_{\shortstack{$+ \pm$ \\ $\nu_{\rm f}$}}^{\rm{supertorus}} \sim {\bf G}^{\rm supersphere}$
 when $z_I \sim z_J$, where ${\bf G}^{\rm supersphere}$ is the supersphere Green function.
The worldsheet supersymmetry is broken in general by the boundary condition,
 but it is still useful to consider this object, which we demonstrate in section \ref{sec.oneloopamp}.

\subsubsection{superannulus Neumann function}
%

Using the image method as in \cite{ItoMox} (appendix \ref{App:imagemethod:SA}),
 the superannulus Neumann function can be written as
\ba	
	&\:&
	{\bf N}_{\shortstack{$+ \pm$ \\ $\nu_{\rm f}$}}^{\rm{superannulus}} \left({ z, {\bar{z}}'; z, {\bar{z}}' \left|{ \frac{i \tau_2}{2} }\right.}\right)			\nn
        &\:&	\qq
        =	
        {1 \over 2}
        \left\{{
        	{\bf G}_{\shortstack{$+ \pm$ \\ $\nu_{\rm f}$}}^{\rm{supertorus}}
                 \left({\left.{
                 	\frac{z}{2}, \frac{z'}{2}; \frac{\theta}{\sqrt{2}}, \frac{\theta'}{\sqrt{2}}
                 }\right| \frac{i \tau_2}{2} }\right)
                + {\bf G}_{\shortstack{$+ \pm$ \\ $\nu_{\rm f}$}}^{\rm{supertorus}}
                  \left({\left.{
                 	\frac{{\tilde{z}}}{2}, \frac{z'}{2}; \frac{{\tilde{\theta}}}{\sqrt{2}}, \frac{\theta'}{\sqrt{2}}
                  }\right| \frac{i \tau_2}{2} }\right)
        }\right.		\nn
        &\:&	\qq
        \left.{
                 + {\bf G}_{\shortstack{$+ \pm$ \\ $\nu_{\rm f}$}}^{\rm{supertorus}}
                  \left({\left.{
                 	\frac{z}{2}, \frac{{\tilde{z}}'}{2}; \frac{\theta}{\sqrt{2}}, \frac{{\tilde{\theta}}'}{\sqrt{2}}
                  }\right| \frac{i \tau_2}{2} }\right)
                 + {\bf G}_{\shortstack{$+ \pm$ \\ $\nu_{\rm f}$}}^{\rm{supertorus}}
                  \left({\left.{
                 	\frac{{\tilde{z}}}{2}, \frac{{\tilde{z}}'}{2}; \frac{{\tilde{\theta}}}{\sqrt{2}}, \frac{{\tilde{\theta}}'}{\sqrt{2}}
                  }\right| \frac{i \tau_2}{2} }\right)
        }\right\}		\nn
        &\:&	\qq
        =	
        {1 \over 2}
        \left\{{
        	{\bf G}_{\shortstack{$+ \pm$ \\ $\nu_{\rm f}$}}^{\rm{supertorus}}
                 \left({\left.{
                 	\frac{z}{2}, \frac{z'}{2}; \frac{\theta}{\sqrt{2}}, \frac{\theta'}{\sqrt{2}}
                 }\right| \frac{i \tau_2}{2} }\right)
                + {\bf G}_{\shortstack{$+ \pm$ \\ $\nu_{\rm f}$}}^{\rm{supertorus}}
                  \left({\left.{
                 	\frac{-{\bar{z}}}{2}, \frac{z'}{2}; \frac{\pm i{\bar{\theta}}}{\sqrt{2}}, \frac{\theta'}{\sqrt{2}}
                  }\right| \frac{i \tau_2}{2} }\right)
        }\right.		\nn
        &\:&	\qq
        \left.{
                 + {\bf G}_{\shortstack{$+ \pm$ \\ $\nu_{\rm f}$}}^{\rm{supertorus}}
                  \left({\left.{
                 	\frac{z}{2}, \frac{-{\bar{z}}'}{2}; \frac{\theta}{\sqrt{2}}, \frac{\pm i{\bar{\theta}}'}{\sqrt{2}}
                  }\right| \frac{i \tau_2}{2} }\right)
                 + {\bf G}_{\shortstack{$+ \pm$ \\ $\nu_{\rm f}$}}^{\rm{supertorus}}
                  \left({\left.{
                 	\frac{-{\bar{z}}}{2}, \frac{-{\bar{z}}'}{2}; \frac{\pm i{\bar{\theta}}}{\sqrt{2}}, \frac{\pm i{\bar{\theta}}'}{\sqrt{2}}
                  }\right| \frac{i \tau_2}{2} }\right)
        }\right\}					\:,		\nn
        	\label{N:SA_GeneralForm}
\ea
 where ${\tilde z}$, ${\tilde z}'$, ${\tilde \theta}$ and ${\tilde \theta}$ denote respectively
  the conjugate points of $z$, $z'$, $\theta$ and $\theta'$.

\section{Path integral of an NSR fermionic string and genus one vacuum amplitudes	\label{sec.Pathint}} 

In this section, quoting the formula (\ref{pathintformula}) in appendix \ref{App:pathint:review} for the path integral
 of an NSR fermionic string valid for any genus, we briefly review genus one vacuum amplitudes.
Introducing notation to represent the contributions of the path integrals from a worldsheet chiral boson
 and a fermion obeying the general boundary condition specified by
  $
  	\left(
		\begin{array}{cc}
		 \alpha_{\rm b} & \beta_{\rm b} \\
		 \alpha_{\rm f} & \beta_{\rm f}
		\end{array}
	\right)
  $,
   we formulate our discussion to cover a large class of cases with toroidal compactification and its orbifolding.
Here in this notation, we have labelled the bosonic case by b and the fermionic case by f.

\subsection{path integral formula for an NSR fermionic string}

Let the bosonic coordinates, the fermionic ones, zweibein and the Rarita - Schwinger field
 be $X^M$, $\psi_{\rm Maj \:}{}_\alpha^M$, $e_\alpha{}^m$ and $\chi_\alpha{}^m$ respectively.
We denoted by $M, N, ...$, $\alpha, \beta, ...$, $m, n, ...$ and $a, b, ...$, ten dimensional vector indices, two dimensional spinor indices, two dimensional worldsheet indices,
 and two dimensional local Lorentz indices respectively.
According to Appendix \ref{App:pathint:review},
 the path integral formula can be written as
\ba	
	\langle \prod_I O_I \rangle
        &=&
        \sum_{\rm{top.}} \sum_{\rm{s.s.}}
         \int \frac{{\cal{D}} e_m{}^a}{\Omega (\rm{D}) \Omega (\rm{W}) \Omega (\rm{L})}
         \int \frac{{\cal{D}} \chi_m{}^a}{\Omega (\rm{S}) \Omega (\rm{SW})}
        \int {\cal{D}} X^M \int {\cal{D}}\psi_{\rm Maj}{}^M
         e^{-S} \prod_I O_I											\nn
        &=&
        \sum_{\rm{top.}} \sum_{\rm{s.s.}}
        \int \prod_i \de \tau_i \frac{1}{\Omega({\rm{CKV}})}
        {\rm{det}}' (P_1^\dag P_1)^{1 \over 2}
        {\rm{det}} \langle \psi_i | \psi_j \rangle^{-{1 \over 2}}
        {\rm{det}} \left\langle{ \psi_i \left|{ \pardif{e_m{}^a}{\tau_i} }\right. }\right\rangle	\nn
        &\;&	\qq
        \times
        \int \prod_i \de a_i \frac{1}{\Omega({\rm{CKS}})}
        {\rm{det}}' (P_{1/2}^\dag P_{1/2})^{-{1 \over 2}}
        {\rm{det}} \langle \Psi_i | \Psi_j \rangle^{1 \over 2}
        {\rm{det}} \langle \Psi_i | \Phi_j \rangle^{-1}							\nn
        &\:&	\qq\qq
        \times
        \int {\cal{D}} X^M \int {\cal{D}}\psi_{\rm Maj}{}^M
        \ex^{-S}
        \prod_I O_I
        	\label{pathintformula:sec.3}
\ea
 where the action is
\ba	
	S
        &=&
        \frac{1}{2 \pi \alpha'} \int d^2\sigma \sqrt{g}
        \left\{{
        	{1 \over 2} g^{m n} \partial_m X^M \partial_n X_M
                  - {i \over 2} \psi_{\rm Maj}{}^M \gamma^a \nabla_a \psi_{\rm Maj \:}{}_M
        }\right.				\nn
        &\:&	\qq\qq\qq\qq
        \left.{
                - {1 \over 2}
                 \left({
                 	\psi_{\rm Maj}{}^M \gamma^a \gamma^b \chi_a
                 }\right)
                 \left({
                 	\partial_b X_M - {1 \over 4} \chi_b \psi_{\rm Maj \:}{}_M
                 }\right)
        }\right\},
        	\label{action:sec.3}
\ea
  and
 \ba	
 	\chi_a &=& e_a{}^m \chi_m, \;\; \partial_b = e_b{}^m \partial_m		\nn
        \nabla_a &=& e_a{}^m \left({ \partial_m - \omega_m {1 \over 2} \gamma^5 }\right)		\nn
        \omega_m &=& e_m{}^a \varepsilon^{pq} \partial_p e_q{}^b \delta_{ab}			\:.
        	\label{componentsinaction:sec.3}
 \ea
For the notation in eqs. (\ref{pathintformula:sec.3}), (\ref{action:sec.3}), (\ref{componentsinaction:sec.3}),
 please look at appendix \ref{App:pathint:review}.
Below, at one-loop, we will recast the expressions eq. (\ref{pathintformula:sec.3}) into
 that from the light-cone operator formalism.

\subsection{superstring genus one vacuum amplitudes in flat ten dimensions}

\subsubsection{torus}

	In the case of torus, the boundary conditions for fermions in flat ten dimensions
 are specified by $+$ (periodic) or $-$ (antiperiodic)
  for both $\sigma^1$ and $\sigma^2$ directions:
\ba
	\psi(\sigma^1 + 1, \sigma^2) &=& r \psi(\sigma^1, \sigma^2), \:\: r = \pm 1	\label{psiperiod1} \\
        \psi(\sigma^1, \sigma^2 + 1) &=& s \psi(\sigma^1, \sigma^2), \:\: s = \pm 1	\:.
\ea
Similar expressions hold for ${\bar \psi}$.
In the notation of section \ref{sec.Greenfunc}, $r = \ex^{2 \pi i \alpha}$, $s = \ex^{2 \pi i \beta}$, so that
\ba
	r = +1	&\Leftrightarrow&	\alpha = 0			\:,	\qq
        r = -1   \Leftrightarrow	\alpha = {1 \over 2}			\nn
        s = +1	&\Leftrightarrow&	\beta = 0			\:,	\qq
        s = -1   \Leftrightarrow	\beta = {1 \over 2}			\qq
        {\rm modulo} \: 1		\:.
\ea
Except for the $(+,+)$ spin structure, there is neither conformal killing spinor, nor supermoduli.
For the $(+,+)$ spin structure, its contribution to the vacuum amplitude vanishes
 due to the integrations of $\psi$, ${\bar \psi}$ fermionic zero modes.
The torus vacuum amplitude for IIB/IIA in flat ten dimensions is, therefore, simply written as
\ba
	Z_{\rm{flat}}^{\rm{IIB/IIA}}
        &=&
        V_{\rm{E}}
        \sum_{(r,s)} \sum_{(r',s')} C_{rs} {\bar{C}}_{r's'}
        {1 \over 2} \int_{\cal{F}} \prod_{i=1,2} \de \tau_i \det{\left({ P_1^\dag P_1}\right)^{1 \over 2}}
        \left[{ \det{\Delta_{\hat{g}}} }\right]^{-5}					\nn
        &\:&
        \det{\left\langle{ \psi_i {\left|{\pardif{e_m{}^a}{\tau_j}}\right.} }\right\rangle}
        \det{\left\langle{ \psi_i | \psi_j }\right\rangle}^{-{1 \over 2}}
        \left({ \int \de^2 \sigma \sqrt{{\hat{g}}} }\right)^5				\nn
        &\:&
        \left[{
        	{\rm det}'\left({ P_{1/2}^\dag P_{1/2} }\right)^{-{1 \over 2}}
                {\rm det} \left({ \gamma \cdot \partial }\right)^{5}
        }\right]_{(r,s), (r',s')}			\:,
\ea
 where ${\hat{g}}_{ab} = \left[
				\begin{array}{cc}
				 1 & \tau_1 \\
				 \tau_1 & \tau_1^2 + \tau_2^2
				\end{array}
			\right]$,
  and we have chosen $C_{--} = - C_{-+} = - C_{+-} = {1 \over 2}$, ${\bar{C}}_{--} = - {\bar{C}}_{-+} = - {\bar{C}}_{+-} = {1 \over 2}$, $C_{++} = {\bar{C}}_{++} = 0$
   in accordance with the GSO projection \cite{GSO} of the IIB superstring that implements the modular invariance.
We have denoted by $\cal{F}$ the fundamental region of the torus and the factor ${1 \over 2}$ is accounted for by $SL (2, {\bf Z})/PSL (2, {\bf Z}) = \{\{ \pm {\bf 1}_{2} \}\}$.
The Euclidean volume is denoted by $V_{\rm{E}}$.
Omitting the calculations of the Weil Petersen measure factor and those of the functional determinants,
 we obtain
\be
	Z_{\rm{flat}}^{\rm{IIB/IIA}}
        =
        K V_{\rm{E}} {1 \over 2}
        \int_{\cal{F}} \frac{\de^2 \tau}{(\tau_2)^2} \frac{1}{\tau_2^4}
        \frac{1}{\left|{ \eta(\tau) }\right|^{16}} |T(\tau)|^2,
 \label{ZIIBtorus*}
\ee
 where
\be
	T(\tau)
        =
        \frac{1}{\eta(\tau)^4}
        \left({
        	C_{--} \vartheta \left[{\textstyle {0 \atop 0}} \right]^4
                + C_{-+} \vartheta \left[{\textstyle {0 \atop {1 \over 2}}} \right]^4
                + C_{+-} \vartheta \left[{\textstyle {{1 \over 2} \atop 0}} \right]^4
        }\right)
\ee
 and $\eta(\tau) = \ex^{\frac{i \pi \tau}{12}} \displaystyle\prod_{n=1}^\infty \left({ 1 - \ex^{2 \pi i n \tau} }\right)$ is the Dedekind eta function.

In this paper, we take a short cut to proceed further and to determine the normalization factor $K$ by comparing the last expression eq. (\ref{ZIIBtorus*})
 with the vacuum amplitude evaluated in the light cone gauge operator formalism,
  written in terms of the $so(8)$ characters.
(The overall normalization can also be seen by the one-loop free energy in local field theory):
\be
	\Gamma_{\rm{flat}}^{\rm{IIB/IIA}}
        =
        - \frac{V_{\rm{E}}}{2 (4 \pi^2 \alpha')^5}
        \int_{\cal{F}} \frac{\de^2 \tau}{\tau_2^2}
        ({\bar{\chi}} X \chi)_{\rm{IIB/IIA, \: flat}},
 \label{GammaIIBtorus**}
\ee
 where
\be
	({\bar{\chi}} X \chi)
        \equiv
        \sum_{i,j = {\rm NS, R}} {\bar{\chi}}_i X_{ij} \chi_j
        =
        |V_8 - S_8/C_8|^2 \frac{1}{\tau_2^4 |\eta|^{16}}
\ee
\be
	V_8 = \frac{\vartheta \left[{\textstyle {0 \atop 0}} \right]^4 - \vartheta \left[{\textstyle {0 \atop {1 \over 2}}} \right]^4}{2 \eta^4},
        \:\:
        S_8/C_8 = \frac{\vartheta \left[{\textstyle {{1 \over 2} \atop 0}} \right]^4 \pm \vartheta \left[{\textstyle {{1 \over 2} \atop {1 \over 2}}} \right]^4}{2 \eta^4}
            = \frac{\vartheta \left[{\textstyle {{1 \over 2} \atop 0}} \right]^4}{2 \eta^4}			\:.
\ee
Identifying eq. (\ref{ZIIBtorus*}) with eq. (\ref{GammaIIBtorus**}), we obtain
\be
	K = - \frac{1}{(4 \pi^2 \alpha')^5}			\:.
\ee
Note that, from the point of view of one-loop free energy in local field theory,
 $\frac{1}{(4 \pi^2 \alpha' \tau_2)^{1 \over 2}}$ comes from a gaussian integration over one momentum,
  and $- {1 \over 2} \int \frac{\de^2 \tau}{\tau_2} \cdots$ comes from a proper time representation of $\log {\rm Det}$.

\subsubsection{Klein bottle, annulus, m\"obius strip}

	Let us write eq. (\ref{ZIIBtorus*}) as
\be
	Z_{\rm{torus}}^{\rm{IIB}}
        =
        {1 \over 2} K V_{\rm{E}} {\cal{T}}
\ee
\be
	{\cal{T}}
        =
        \int_{\cal{F}} \frac{\de^2 \tau}{\tau_2^2} |\chi(\tau)|_{\rm{torus}}^2 \: , \:\:
        |\chi(\tau)|_{\rm{torus}}^2 = {\rm{Tr}} q^{L_0^{\rm{(cyl)}}} {\bar{q}}^{{\bar{L}}_0^{\rm{(cyl)}}} ,
\ee
 where $L_0^{\rm{(cyl)}}$ and ${\bar{L}}_0^{\rm{(cyl)}}$ are the right and left Hamiltonian of the closed fermionic string
  on the cylinder in the light-cone gauge operator formalism.

	To construct an unoriented string, namely the type I superstring , one first makes the closed string sector by the $\Omega$ (twist) projection:
 ${\rm{Tr}} q^{L_0^{\rm{(cyl)}}} {\bar{q}}^{{\bar{L}}_0^{\rm{(cyl)}}} \rightarrow {\rm{Tr}} {\left({ \frac{1 + \Omega}{2} }\right)} q^{L_0^{\rm{(cyl)}}} {\bar{q}}^{{\bar{L}}_0^{\rm{(cyl)}}}$.
We obtain
\be
	Z_{\rm{closed, \: one-loop}}^{\rm{I}}
        =
        {1 \over 2} K V_{\rm{E}}
        \frac{{\cal{T}} + {\cal{K}}}{2}
\ee
\be
	{\cal{K}}
        =
        \int_0^\infty \frac{\de \tau_2}{\tau_2^2}
        \sum_{i = {\rm NS, R}} \chi_i (2 i \tau_2)					\:.
\ee
 By the similar procedure, the vacuum amplitude of the open string sector reads
\be
	Z_{\rm{open, \: one-loop}}^{\rm{I}}
        =
        {1 \over 2} K V_{\rm{E}} \frac{{\cal{A}} + {\cal{M}}}{2}
\ee
\ba
	{\cal{A}}
        &=&
        \int_0^\infty \frac{\de \tau_2}{\tau_2^2}
        \sum_{i = {\rm NS, R}} \chi_i {\left({ {i \over 2} \tau_2 }\right)} (\rm{cpf})^2		\\
        {\cal{M}}
        &=&
        \int_0^\infty \frac{\de \tau_2}{\tau_2^2}
        \sum_{i = {\rm NS, R}} {\tilde{\chi}}_i {\left({ {i \over 2} \tau_2 + {1 \over 2} }\right)} (\rm{cpf}) \epsilon		\label{partitionfunc:MSpartSec3}		\:.
\ea
Here, cpf denotes the Chan Paton factor, $\epsilon = \pm 1$ and ${\tilde{\chi}} {\left({ {i \over 2} \tau_2 + {1 \over 2} }\right)}$ indicates that the replacement
 by $\tau \rightarrow {i \over 2} \tau_2 + {1 \over 2}$ in the argument is to be made only for the oscillator part.
These replacements ${\cal{K}}: \tau \rightarrow 2 i \tau_2$, ${\cal{A}}: \tau \rightarrow {i \over 2} \tau_2$, ${\cal{M}}: \tau \rightarrow {i \over 2} \tau_2 + {1 \over 2}$
 are understood both by the twist projection in the operator formalism and by the worldsheet involutions
  of the worldsheet path integrals with torus as the double of the respective open Riemann surfaces.

	Finally, the infrared stability seen as the cancellation of the massless poles in $Z_{\rm{closed, \: one-loop}}^{\rm{I}} + Z_{\rm{open, \: one-loop}}^{\rm{I}}$
 in the transverse channel (or equivalently the cancellation of dilaton tadpoles \cite{ItoMox, DouglasGrinstein, PolCai, AIKT, SagAngreview, GPol} or infinity cancellation \cite{GSInfty, ItoMox})
  selects ${\rm{cpf}} = 2^5 = 32$, $\epsilon = -1$ and the gauge group $SO(32)$ \cite{GSanomalycancel}.

\subsection{generalization to cases of toroidal compactification and their orbifolding}

In order to proceed even further and to prepare for calculation of string scattering amplitudes in section \ref{sec.oneloopamp},
 we will introduce notation for the integrand of the string one-loop partition function.
Let us, in particular, write $({\bar \chi} X \chi)_{\rm IIB/IIA \: flat}$ as
\ba	
	&\:&
	({\bar \chi} X \chi)_{\rm IIB/IIA \: flat}			\nn
        &\:&
        =	
        {1 \over 2}
        \left({
        	\bn{+}{+}{-}{-}^8
                -
                \bn{+}{+}{-}{+}^8
                -
                \bn{+}{+}{+}{-}^8
                \mp
                \bn{+}{+}{+}{+}^8
        }\right)
        {1 \over 2}
        \overline{\left({
             	\bn{+}{+}{-}{-}^8
                -
                \bn{+}{+}{-}{+}^8
                -
                \bn{+}{+}{+}{-}^8
                \mp
                \bn{+}{+}{+}{+}^8
        }\right)}							\nn
        &\:&
        \equiv	
        \sum_{\nu, {\bar \nu}} {\cal J}_{\nu, {\bar \nu}, {\rm IIB/IIA, flat}}			\:.
		\label{{bar chi}Xchi:IIB/IIA,flat}
\ea
Here we have introduced
\be	
	\bn{r_{\rm b}}{s_{\rm b}}{r_{\rm f}}{s_{\rm f}}
        \left({
        	\equiv		
        	\bbn{\alpha_{\rm b}}{\beta_{\rm b}}{\alpha_{\rm f}}{\beta_{\rm f}}
        }\right)
        	\label{box:rb,sb,rf,sf}
\ee
 in order to represent the contribution from a single chiral boson and fermion
  obeying the boundary conditions $(\alpha_{\rm b}, \beta_{\rm b})$ and $(\alpha_{\rm f}, \beta_{\rm f})$ respectively:
\be	
	\bn{+}{+}{r_{\rm f}}{s_{\rm f}}
        =	
        \bbn{0}{\:\:\: 0}{\alpha_{\rm f}}{\beta_{\rm f}}
        =	
        {\sqrt{
        	\frac{\vartheta \left[{\textstyle {{1 \over 2}+\alpha_{\rm f} \atop {1 \over 2}+\beta_{\rm f}}} \right] (0)}{\eta^3}
        }}			\:.
        	\label{box:+,+,rf,sf:concrete}
\ee
The power $8 = 10 - 2$ seen in eq. (\ref{{bar chi}Xchi:IIB/IIA,flat}) permits covariant interpretation as
 the $2$d metric and $2$d gravitino fields obey the same boundary condition as the worldsheet bosons and fermions do respectively.
As a simple prototypical example, let us consider IIB string on $T^4 (= (S^1)^4)/{\bf Z}_2$ with radii of $S^1$ being $R_I$, $I = 5,6,7,8$.
\ba
	&\:&
        ({\bar \chi} X \chi)_{{\rm IIB}, T^4/{\bf Z}_2}					\nn
        &\:&
        =	
        {1 \over 2}
        \left({
        	\prod_{I=5,6,7,8} F_2 (a_I, \tau)
        }\right)
        ({\bar \chi} X \chi)_{{\rm IIB, \: flat}}				\nn
        &\:&	%
        +
        {1 \over 2}
        \left({
        	\prod_{I=5,6,7,8} a_I \sqrt{\tau_2}
        }\right)
        {1 \over 2}
        \left({
        	\bn{+}{+}{-}{-}^4
                \bn{+}{-}{-}{+}^4
                -
                \bn{+}{+}{-}{+}^4
                \bn{+}{-}{-}{-}^4
                -
                \bn{+}{+}{+}{-}^4
                \bn{+}{-}{+}{+}^4
                -
                \bn{+}{+}{+}{+}^4
                \bn{+}{-}{+}{-}^4
        }\right)								\nn
        &\:&	\qq\qq\qq\qq
        \times
        {1 \over 2}
        \left({\overline{
        	\bn{+}{+}{-}{-}^4
                \bn{+}{-}{-}{+}^4
                -
                \bn{+}{+}{-}{+}^4
                \bn{+}{-}{-}{-}^4
                -
                \bn{+}{+}{+}{-}^4
                \bn{+}{-}{+}{+}^4
                -
                \bn{+}{+}{+}{+}^4
                \bn{+}{-}{+}{-}^4
        }}\right)								\nn
        &\:&	%
        +
        {1 \over 2}
        \left({
        	\prod_{I=5,6,7,8} a_I \sqrt{\tau_2}
        }\right)
        {1 \over 2}
        \left({
        	\bn{+}{+}{-}{-}^4
                \bn{-}{+}{+}{-}^4
                -
                \bn{+}{+}{-}{+}^4
                \bn{-}{+}{+}{+}^4
                -
                \bn{+}{+}{+}{-}^4
                \bn{-}{+}{-}{-}^4
                -
                \bn{+}{+}{+}{+}^4
                \bn{-}{+}{-}{+}^4
        }\right)								\nn
        &\:&	\qq\qq\qq\qq
        \times
        {1 \over 2}
        \left({\overline{
        	\bn{+}{+}{-}{-}^4
                \bn{-}{+}{+}{-}^4
                -
                \bn{+}{+}{-}{+}^4
                \bn{-}{+}{+}{+}^4
                -
                \bn{+}{+}{+}{-}^4
                \bn{-}{+}{-}{-}^4
                -
                \bn{+}{+}{+}{+}^4
                \bn{-}{+}{-}{+}^4
        }}\right)								\nn
        &\:&	%
        +
        {1 \over 2}
        \left({
        	\prod_{I=5,6,7,8} a_I \sqrt{\tau_2}
        }\right)
        {1 \over 2}
        \left({
        	\bn{+}{+}{-}{-}^4
                \bn{-}{-}{+}{+}^4
                -
                \bn{+}{+}{-}{+}^4
                \bn{-}{-}{+}{-}^4
                -
                \bn{+}{+}{+}{-}^4
                \bn{-}{-}{-}{+}^4
                -
                \bn{+}{+}{+}{+}^4
                \bn{-}{-}{-}{-}^4
        }\right)								\nn
        &\:&	\qq\qq\qq\qq
        \times
        {1 \over 2}
        \left({\overline{
        	\bn{+}{+}{-}{-}^4
                \bn{-}{-}{+}{+}^4
                -
                \bn{+}{+}{-}{+}^4
                \bn{-}{-}{+}{-}^4
                -
                \bn{+}{+}{+}{-}^4
                \bn{-}{-}{-}{+}^4
                -
                \bn{+}{+}{+}{+}^4
                \bn{-}{-}{-}{-}^4
        }}\right)								\nn
        &\:&
        \equiv	
        \sum_{\nu, {\bar \nu}} {\cal J}_{\nu, {\bar \nu}, {\rm IIB}, T^4/{\bf Z}_2}
\ea
 where
 $
 	F_2 (a_I, \tau)
        \equiv	
        a_I {\sqrt{\tau_2}}
        \displaystyle\sum_{\ell, m \in {\bf Z}}
        q^{{1 \over 4} (m a_I + {\ell \over {a_I}})^2}
        {\bar q}^{{1 \over 4} (m a_I - {\ell \over {a_I}})^2}
 $
  and $a_I \equiv \frac{\sqrt{\alpha'}}{R_I}$.
The first line represents the contribution from the $T^4$ compactification without ${\bf Z}_2$ insertion,
 the second, the third and the fourth lines represent the contributions from the untwisted sector with ${\bf Z}_2$ insertion,
  the ${\bf Z}_2$ twisted sector and the ${\bf Z}_2$ twisted sector with the ${\bf Z}_2$ insertion respectively.
In each term inside the bracket, the first bin represents the spacetime part and the second bin the internal part.
Referring to the character of $c=1$, ${\bf Z}_2$ orbifold, we are able to see
\ba	
	\bn{r_{\rm b}}{s_{\rm b}}{r_{\rm f}}{s_{\rm f}}
        &=&	
        \bbn{\alpha_{\rm b}}{\beta_{\rm b}}{\alpha_{\rm f}}{\beta_{\rm f}}
        =	
        {\sqrt{
        	\frac{2 \eta}{\vartheta \left[{\textstyle {{1 \over 2}+\alpha_{\rm b} \atop {1 \over 2}+\beta_{\rm b}}} \right] (0)}
        }}
        {\sqrt{
        	\frac{\vartheta \left[{\textstyle {{1 \over 2}+\alpha_{\rm f} \atop {1 \over 2}+\beta_{\rm f}}} \right] (0)}{\eta}
        }}
        =	
        {\sqrt{
        	\frac{2 \vartheta \left[{\textstyle {{1 \over 2}+\alpha_{\rm f} \atop {1 \over 2}+\beta_{\rm f}}} \right] (0)}
                     {\vartheta \left[{\textstyle {{1 \over 2}+\alpha_{\rm b} \atop {1 \over 2}+\beta_{\rm b}}} \right] (0)}
        }}						\nn
        &\:&	\qq\qq\qq\qq\qq\qq
        {\rm for} \: (\alpha_{\rm b}, \beta_{\rm b}) = \left({ 0, {1 \over 2} }\right), \left({ {1 \over 2}, 0 }\right), \left({ {1 \over 2}, {1 \over 2} }\right)		\:.
        	\label{box:rb,sb,rf,sf:concrete}
\ea
Here, the arguments of the theta constants are modulo $1$ and the non-integer parts are understood to be taken.
Note that, in this notation, we have included the contribution from the $2^4 = 16$ fixed points in the twisted sector in eq. (\ref{box:rb,sb,rf,sf:concrete}).

\subsection{Open superstring on $T^4/{\bf Z}_2$	\label{sec.opensstrT4/Z2}}

Another prototypical example which we will consider in the next section is
 the open string sector in the type I superstring on $T^4(=(S^1)^4)/{\bf Z}_2$.
The partition function is
\be	
	Z_{{\rm I}, T^4/{\bf Z}_2}
        =	
        - \frac{V_{\rm E}}{2} \frac{1}{(4 \pi^2 \alpha')^5}
        \frac{{\cal A}_{{\rm I}, T^4/{\bf Z}_2} + {\cal M}_{{\rm I}, T^4/{\bf Z}_2}}{2}
\ee
\be	
	{\cal A}_{{\rm I}, T^4/{\bf Z}_2}
        =	
        \int_0^\infty \frac{\de \tau_2}{\tau_2}
        \frac{1}{\tau_2^5}
        \left.{
        	{\cal J}_{{\rm I}, T^4/{\bf Z}_2}
        }\right|_{\tau = {i \over 2} \tau_2}
\ee
\be	
	{\cal M}_{{\rm I}, T^4/{\bf Z}_2}
        =	
        \int_0^\infty \frac{\de \tau_2}{\tau_2}
        \frac{1}{\tau_2^5}
        \left.{
        	{\tilde {\cal J}}_{{\rm I}, T^4/{\bf Z}_2}
        }\right|_{\tau = {i \over 2} \tau_2 + \frac{1}{2}}			\:.
        	\label{calM_{I,T4/Z2}:sec.3}
\ee
Among the many possibilities discussed in \cite{PraSag, GPol, SagAngreview}, where the dilaton tadpoles cancel,
 we will consider the simplest case where the gauge group is $U(n=16)_{(9)} \times U(d=16)_{(5)}$
  with all of the D$5$ branes at the same fixed point and the first and the second subscripts indicate D$9$ and D$5$ brane respectively\footnote{Other aspects of this series of model are discussed in \cite{Sagrev[26]}-\cite{Sagrev[141]}.}.
\ba	
	&\:&
	{\cal J}_{{\rm I}, T^4/{\bf Z}_2}					\nn
        &\:&
        =	
        {1 \over 2} {\cal J}_{{\rm I}, T^4}
        +
        {1 \over 2} {\cal J}_{{\rm I}, T^4}^{({\bf Z}_2)}			\nn
        &\:&
        =	
        (2n)^2
        {1 \over 2}
        \left({ \prod_{I=5,6,7,8} F_1 (a_I, \tau_2) }\right)
        {1 \over 2}
        \left({
        	\bn{+}{+}{-}{-}^8 - \bn{+}{+}{-}{+}^8 - \bn{+}{+}{+}{-}^8 - \bn{+}{+}{+}{+}^8
        }\right)								\nn
        &\:&	%
        -
        (2 d)^2
        {1 \over 2}
        \left({ \prod_{I=5,6,7,8} (a_I \sqrt{\tau_2}) }\right)
        {1 \over 2}
        \left({
        	\bn{+}{+}{-}{-}^4 \bn{+}{-}{-}{+}^4
                 - \bn{+}{+}{-}{+}^4 \bn{+}{-}{-}{-}^4
                 - \bn{+}{+}{+}{-}^4 \bn{+}{-}{+}{+}^4
                 - \bn{+}{+}{+}{+}^4 \bn{+}{-}{+}{-}^4
        }\right)								\nn
        &\:&
        \equiv	
        {1 \over 2}
        \sum_{\nu} {\cal J}_{\nu}
        +
        {1 \over 2}
        \sum_{\nu} {\cal J}_{\nu}^{({\bf Z}_2)}				\:,
        	\label{calJ_{I,T4/Z2}}
\ea
 where
 $
 	F_1 (a_I, \tau_2)
        =	
        a_I {\sqrt{\tau_2}}
        \displaystyle\sum_{p_I}
        \ex^{- t p_I p^I}
 $, $\pi \tau_2 \equiv \frac{t}{\alpha'}$.
See also \cite{AFIV, IRU, Poppitz, Kiritsis, IYano1}.

\section{One-loop superstring amplitudes with non-maximal supersymmetry		\label{sec.oneloopamp}} 

In this section, we apply the genus one super Green function constructed
 under the general twists in the $(\sigma, \tau)$ directions to superstring amplitudes.
For simplicity, we illustrate this by the annulus contribution
 to the open superstring amplitudes of the compactification in section \ref{sec.opensstrT4/Z2},
 but our procedure is applicable to a large class of toroidal models and their orbifolding
  of closed and open superstrings including heterotic string \cite{GHMR} compactifications.

\subsection{Neumann functions with arguments on the boundary}

In order to proceed to the computation, we need the Neumann function for the superannulus
 under a variety of boundary conditions for a worldsheet boson and a worldsheet fermion specified
  by $\left({\textstyle {\nu_{\rm b} \atop \nu_{\rm f}}} \right)$ and with the arguments set on the same boundary.
The Neumann function for the M\"obius strip case can be read off from the annulus case by the change of the arguments
 in the theta functions and will not be discussed explicitly here.
For the case of $\left({\textstyle {++ \atop \nu_{\rm f}}} \right)$, which is always needed,
\ba	
	&\:&
	{\bf N}_{\shortstack{$++$ \\ $\nu_{\rm f}$}}^{\rm{superannulus}} \left({ z_I, {\bar{z}}_I; z_J, {\bar{z}}_J \left|{ \frac{i \tau_2}{2} }\right.}\right)			\nn
        &\:&	\qq
        \overset{{\rm{on \: }} z={\tilde{z}}, \theta={\tilde{\theta}}}{=}	
        {1 \over 2}
        \left.{
        4 {\bf G}_{\shortstack{$++$ \\ $\nu_{\rm f}$}}^{\rm{supertorus}}
                 \left({\left.{
                 	\frac{z_I}{2}, \frac{z_J}{2}; \frac{\theta_I}{\sqrt{2}}, \frac{\theta_J}{\sqrt{2}}
                 }\right| \frac{i \tau_2}{2} }\right)
        }\right|_{(z, \theta) = ({\tilde{z}}, {\tilde{\theta}}) \equiv (-{\bar{z}}, \pm i{\bar{\theta}})}		\nn
        &\:&	\qq
        =	
        \frac{4}{2}
        \left[{
        	G_{++} \left({\left.{ \frac{z_I}{2}; \frac{z_J}{2} }\right| \frac{i \tau_2}{2} }\right)
        	+
        	\frac{\frac{\theta_I}{\sqrt{2}} \frac{\theta_J}{\sqrt{2}}}{4}
        	{\cal{S}}_{\nu_{\rm f}} \left({\left.{ \frac{z_I}{2}; \frac{z_J}{2} }\right| \frac{i \tau_2}{2} }\right)
        }\right.				\nn
        &\:&	\qq\qq\qq\qq	%
        \left.{\left.{
        	-
        	\frac{ \frac{{\bar \theta}_I}{\sqrt{2}} \frac{{\bar \theta}_J}{\sqrt{2}} }{4}
                {\overline{\cal{S}}_{\nu_{\rm f}}} \left({\left.{ \frac{z_I}{2}; \frac{z_J}{2} }\right| \frac{i \tau_2}{2} }\right)
        }\right]}\right|_{(z, \theta) = ({\tilde{z}}, {\tilde{\theta}}) \equiv (-{\bar{z}}, \pm i{\bar{\theta}})}		\nn
        &\:&	\qq
        =	
        2
        \left[{
        	\left.{G_{++} \left({\left.{ \frac{z_I}{2}; \frac{z_J}{2} }\right| \frac{i \tau_2}{2} }\right)}\right|_{z = {\tilde z} \equiv - {\bar z}}
        	+
        	\frac{\frac{\theta_I}{\sqrt{2}} \frac{\theta_J}{\sqrt{2}}}{4}
        	\left.{{\cal{S}}_{\nu_{\rm f}} \left({\left.{ \frac{z_I}{2}; \frac{z_J}{2} }\right| \frac{i \tau_2}{2} }\right)}\right|_{z = {\tilde z} \equiv - {\bar z}}
        }\right.				\nn
        &\:&	\qq\qq\qq\qq	%
        \left.{
        	-
        	\frac{ \frac{(\mp i \theta_I)}{\sqrt{2}} \frac{(\mp i \theta_J)}{\sqrt{2}} }{4}
                \left.{{\overline{\cal{S}}_{\nu_{\rm f}}} \left({\left.{ \frac{z_I}{2}; \frac{z_J}{2} }\right| \frac{i \tau_2}{2} }\right)}\right|_{z = {\tilde z} \equiv - {\bar z}}
        }\right]				\nn
        &\:&	\qq
        {\overset{{\rm eqs.} (\ref{alpha+1_beta+1}), (\ref{bar_theta}), (\ref{barTheta'[1/2,1/2](0|tau)})}{=}}	
        2
        \left[{
        	\frac{1}{2 \pi}
        	\ln \left|{
        			\frac{\vartheta \left[{\textstyle {{1 \over 2} \atop {1 \over 2}}} \right] \left({ \frac{z_I}{2} - \frac{z_J}{2} \left|{\frac{i\tau_2}{2}}\right. }\right)}
        	     		     {\vartheta' \left[{\textstyle {{1 \over 2} \atop {1 \over 2}}} \right] \left({ 0 \left|{\frac{i\tau_2}{2}}\right. }\right)}
        	    	  }\right|
        	+
        	\frac{(z_I - z_J)^2}{4 \tau_2}
        }\right.				\nn
        &\:&	\qq\qq	%
        +
        \left\{{
        	\frac{1}{2 \pi}
                2
        	\sum_{n=1}^\infty
        	\ln |1 - q^n|
                -
                \frac{z_I - z_J}{2 \cdot 2 i}
                +
                \frac{1}{2 \pi}
                \ln (2 \pi)
                +
                2 \frac{\tau_2}{2} \cdot \frac{\pi^2}{6}
        }\right\}				\nn
        &\:&	\qq\qq	%
        +
        \left.{
        	\frac{\frac{\theta_I}{\sqrt{2}} \frac{\theta_J}{\sqrt{2}}}{4}
        	{\cal{S}}_{\nu_{\rm f}} \left({\left.{ \frac{z_I}{2}; \frac{z_J}{2} }\right| \frac{i \tau_2}{2} }\right)
        	+
        	\frac{\frac{\theta_I}{\sqrt{2}} \frac{\theta_J}{\sqrt{2}}}{4}
        	{\cal{S}}_{\nu_{\rm f}} \left({\left.{ \frac{z_I}{2}; \frac{z_J}{2} }\right| \frac{i \tau_2}{2} }\right)
        }\right]								\nn
        &\:&	\qq
        =	
        \frac{2}{2 \pi}
        \ln \left|{
        		\frac{\vartheta \left[{\textstyle {{1 \over 2} \atop {1 \over 2}}} \right] \left({ \frac{z_I}{2} - \frac{z_J}{2} \left|{\frac{i\tau_2}{2}}\right. }\right)}
             		     {\vartheta' \left[{\textstyle {{1 \over 2} \atop {1 \over 2}}} \right] \left({ 0 \left|{\frac{i\tau_2}{2}}\right. }\right)}
            	  }\right|
        +
        \frac{2 (z_I - z_J)^2}{4 \tau_2}
        +
        \frac{2 \cdot 2}{4}
        \frac{\theta_I}{\sqrt{2}} \frac{\theta_J}{\sqrt{2}}
        {\cal{S}}_{\nu_{\rm f}} \left({\left.{ \frac{z_I}{2}; \frac{z_J}{2} }\right| \frac{i \tau_2}{2} }\right)			\nn
        &\:&	\qq\qq	%
        +
        2
        \left\{{
        	\frac{1}{2 \pi}
                2
        	\sum_{n=1}^\infty
        	\ln |1 - q^n|
                -
                \frac{z_I - z_J}{2 \cdot 2 i}
                +
                \frac{1}{2 \pi}
                \ln (2 \pi)
                +
                2 \frac{\tau_2}{2} \cdot \frac{\pi^2}{6}
        }\right\}								\nn
	&\:&	\qq
        =	
        \frac{1}{\pi}
        \ln \left|{
        		\frac{\vartheta \left[{\textstyle {{1 \over 2} \atop {1 \over 2}}} \right] \left({ \frac{z_I}{2} - \frac{z_J}{2} \left|{\frac{i\tau_2}{2}}\right. }\right)}
             		     {\vartheta' \left[{\textstyle {{1 \over 2} \atop {1 \over 2}}} \right] \left({ 0 \left|{\frac{i\tau_2}{2}}\right. }\right)}
            	  }\right|
        +
        \frac{(z_I - z_J)^2}{2 \tau_2}
        +
        \frac{\theta_I}{\sqrt{2}} \frac{\theta_J}{\sqrt{2}}
        {\cal{S}}_{\nu_{\rm f}} \left({\left.{ \frac{z_I}{2}; \frac{z_J}{2} }\right| \frac{i \tau_2}{2} }\right)			\nn
        &\:&	\qq\qq	%
        +
        2
        \left\{{
        	\frac{1}{2 \pi}
                2
        	\sum_{n=1}^\infty
        	\ln |1 - q^n|
                -
                \frac{z_I - z_J}{2 \cdot 2 i}
                +
                \frac{1}{2 \pi}
                \ln (2 \pi)
                +
                2 \frac{\tau_2}{2} \cdot \frac{\pi^2}{6}
        }\right\}					\:.
        	\label{NSuperannulus}
\ea
The last line of eq. (\ref{NSuperannulus}) can be dropped in the calculation of amplitudes
 as the source ${\bf J}$ satisfies $\int \de^2 z \de \theta \de {\bar \theta} {\bf J} = 0$.
We need the case $\left({\textstyle {+- \atop \nu_{\rm f}}} \right)$ as well:
\ba	
	&\:&
	{\bf N}_{\shortstack{$+ -$ \\ $\nu_{\rm f}$}}^{\rm{superannulus}} \left({ z_J, {\bar{z}}_J; z_K, {\bar{z}}_K \left|{ \frac{i \tau_2}{2} }\right.}\right)			\nn
        &\:&	\qq
        \overset{{\rm{on \: }} z={\tilde{z}}, \theta={\tilde{\theta}}}{=}	
        2
        \left[{
        	\left.{G_{+-} \left({\left.{ \frac{z_J}{2}; \frac{z_K}{2} }\right| \frac{i \tau_2}{2} }\right)}\right|_{z = {\tilde z} \equiv - {\bar z}}
        	+
        	\frac{\frac{\theta_J}{\sqrt{2}} \frac{\theta_K}{\sqrt{2}}}{4}
        	\left.{{\cal{S}}_{\nu_{\rm f}} \left({\left.{ \frac{z_J}{2}; \frac{z_K}{2} }\right| \frac{i \tau_2}{2} }\right)}\right|_{z = {\tilde z} \equiv - {\bar z}}
        }\right.				\nn
        &\:&	\qq\qq\qq\qq	%
        \left.{
        	-
        	\frac{ \frac{(\mp i \theta_J)}{\sqrt{2}} \frac{(\mp i \theta_K)}{\sqrt{2}} }{4}
                \left.{{\overline{\cal{S}}_{\nu_{\rm f}}} \left({\left.{ \frac{z_J}{2}; \frac{z_K}{2} }\right| \frac{i \tau_2}{2} }\right)}\right|_{z = {\tilde z} \equiv - {\bar z}}
        }\right]				\nn
        &\:&	\qq
        =	
        2 \left.{G_{+-} \left({\left.{ \frac{z_J}{2}; \frac{z_K}{2} }\right| \frac{i \tau_2}{2} }\right)}\right|_{z = {\tilde z} \equiv - {\bar z}}
        +
        \frac{\theta_J}{\sqrt{2}} \frac{\theta_K}{\sqrt{2}}
        {\cal{S}}_{\nu_{\rm f}} \left({\left.{ \frac{z_J}{2}; \frac{z_K}{2} }\right| \frac{i \tau_2}{2} }\right)			\:.
        	\label{N_+-nu:onz=-barz}
\ea
Note that, for closed string models with ${\bf Z}_2$ insertion,
 ${\bf G}_{\shortstack{$-+$ \\ $\nu_{\rm f}$}}^{\rm{supertorus}}$ and 
  ${\bf G}_{\shortstack{$--$ \\ $\nu_{\rm f}$}}^{\rm{supertorus}}$ are needed
   in order to evaluate the contributions from the twisted sectors.
Likewise,
 ${\bf N}_{\shortstack{$-+$ \\ $\nu_{\rm f}$}}^{\rm{superannulus}}$, ${\bf N}_{\shortstack{$--$ \\ $\nu_{\rm f}$}}^{\rm{superannulus}}$
  are needed in the case of a $5$-$9$ string.

\subsection{Koba-Nielsen type formula for genus one superstring amplitudes}

Let $\displaystyle\prod_{I=1}^N O_I$ be the product of $N$ vertex operators
 $\zeta_I^{({\rm P})} \cdot \int \de z_I \int \de \theta_I D_I {\bf X} (z_I, \theta_I) \ex^{i k_I \cdot {\bf X} (z_I, \theta_I)}$, $I = 1, ..., N$,
  for massless vector emission of an open superstring.
It can be written as
\be	
	\prod_{I=1}^N O_I
        =	
        \prod_{J=1}^N
        \left({
        	\zeta_J^{({\rm P})} \cdot \int \de \eta_J \int \de z_J \int \de \theta_J
        }\right)
        \exp \left[{
             	i \int \de^2 z \de \theta \de {\bar \theta}
                {\bf J} (z, {\bar z}, \theta, {\bar \theta})
                \cdot
                {\bf X} (z, {\bar z}, \theta, {\bar \theta})
             }\right]				\:,
\ee
\be	
	{\bf J}^M (z, {\bar{z}}, \theta, {\bar{\theta}})
        =	
        \sum_{I=1}^N (k_I^M - i \eta_I^M D_I)
         \delta^{(2)}(z - z_I) (\theta - \theta_I) ({\bar{\theta}} - {\bar{\theta}}_I)		\:.
\ee
Here we have introduced the grassmann source $\eta_J$, $J = 1, 2, 3, ..., N$, for this representation.
Following appendix \ref{App:pathint:review},
 we carry out the gaussian integration\footnote{See \cite{ItoMox}. See also \cite{AtickSen, KLLSW, MetTseytlin, AndTseytlin, JonesTye} for different approaches of computation.}
  and the sum ${\cal S}$ over the boundary conditions.

Let ${\cal S} = {\cal S}' \oplus {\cal S}_{(++)}$, where ${\cal S}_{(++)}$ is the part of the sum which contains $(++)$ to some power in $\nu_{\rm f}$.
For these parity-violating cases \cite{Clavelli}, it is well-known that the amplitudes for lower $N$ vanish.
Ignoring these cases in this paper, let us denote the remaining part of the $N$ point amplitude for the case labelled by $\bullet$ by
\be	
	A'{}_N^\bullet
        =	
        - {1 \over 2}
        \frac{\left({ V_{\rm E} \delta }\right) g^N}{(4 \pi^2 \alpha')^5}
        \frac{{\cal A}'{}_N^\bullet + {\cal M}'{}_N^\bullet}{2}			\:.
\ee
Here, we have denoted by $\left({ V_{\rm E} \delta }\right)$ a product of the momentum conserving delta functions $(2 \pi)^{d_{\rm dim}} \delta^{(d_{\rm dim})} \left({ \displaystyle\sum_I k_I }\right)$
 (that appear from the integrations of the zero modes of the bosonic coordinates) and the volume of the compactification $V_{\rm c}$.
The annulus and the M\"obius strip contributions are denoted by ${\cal A}'{}_N^\bullet$
 and ${\cal M}'{}_N^\bullet$ respectively, and
\ba	
	{\cal A}'{}_N^\bullet
        &=&	
        \int_0^\infty \frac{\de \tau_2}{\tau_2^6}
        \sum_{\nu \in {\cal S}'} {\cal J}_{\nu}^\bullet
        \prod_{J=1}^N
        \left({
        	\zeta_J^{({\rm P})} \cdot \int \de \eta_J \int \de z_J \int \de \theta_J
        }\right)					\nn
        &\:&	\qq\qq	%
        \exp \left.{\left[{
        	\pi \alpha'
                \sum_{I,J=1}^N
                (k_I - i \eta_I D_I)^M (k_J - i \eta_J D_J)^L
                {\bf N}'{}_{ML, \nu}
             }\right]}\right|_{\tau = {i \over 2} \tau_2}
        	\label{calA':N,bullet}
\ea
\ba	
	{\cal M}'{}_N^\bullet
        &=&	
        \int_0^\infty \frac{\de \tau_2}{\tau_2^6}
        \sum_{\nu \in {\cal S}'} {\tilde {\cal J}}_{\nu}^\bullet
        \prod_{J=1}^N
        \left({
        	\zeta_J^{({\rm P})} \cdot \int \de \eta_J \int \de z_J \int \de \theta_J
        }\right)					\nn
        &\:&	\qq\qq	%
        \exp \left.{\left[{
        	\pi \alpha'
                \sum_{I,J=1}^N
                (k_I - i \eta_I D_I)^M (k_J - i \eta_J D_J)^L
                {\bf N}'{}_{ML, \nu}
             }\right]}\right|_{\tau = {i \over 2} \tau_2 + \frac{1}{2}}			\:,
        	\label{calM':N,bullet}
\ea
 in accordance with eqs. (\ref{partitionfunc:MSpartSec3}) and (\ref{calM_{I,T4/Z2}:sec.3}).
We will restrict our attention to the annulus case from now on.

We have denoted by $\displaystyle\sum_{\nu \in {\cal S}'} {\cal J}_{\nu}^\bullet$
 the part of the integrand which has appeared in the vacuum amplitude, (for instance, eq. (\ref{calJ_{I,T4/Z2}}))
  and ${\bf N}'{}_{ML, \nu}$ indicates the superannulus Neumann function specified by the boundary condition
   $\nu = \left({\textstyle {\nu_{\rm b} \atop \nu_{\rm f}}} \right)$
    which is determined by the spacetime indices $M, L$.
The prime ${}'$ indicates the omission of the bosonic and fermionic zero modes.

Let us analyze the exponential part of the integrand in eq. (\ref{calA':N,bullet})
\ba	
	\exp [\cdots]
        &=&	
        \exp \left[{
        	2 \alpha'
                \sum_{1 \leq I < J \leq N}
                k_I^M k_J^L
                \pi N'{}_{ML, \nu_{\rm b}}^{IJ}
             }\right]						\nn
        &\:&	%
        \exp \left[{
        	2 \alpha'
                \sum_{1 \leq I < J \leq N}
                \left\{{
                	i k_I^M \theta_I k_J^L \theta_J B_{ML, \nu_{\rm f}}^{IJ}
                }\right.
             }\right.						\nn
        &\:&	\qq	%
        		+ (- \eta_I^M k_J^L \theta_J + \eta_J^L k_I^M \theta_I) B_{ML, \nu_{\rm f}}^{IJ}
                	+ (\eta_I^M \theta_I k_J^L - \eta_J^L \theta_J k_I^M) C_{ML, \nu_{\rm b}}^{IJ}			\nn
        &\:&	\qq	%
        		- i \eta_I^M \eta_J^L B_{ML, \nu_{\rm f}}^{IJ}				\nn
        &\:&	\qq	%
             \left.{
             	\left.{
                	+ \eta_I^M \eta_J^L \theta_I \theta_J E_{ML, {\nu_{\rm b}}}^{IJ}
                }\right\}
             }\right]					\:,
        	\label{exp[...]}
\ea
 where the $I=J$ part vanishes by the on-shell condition.
Following \cite{ItoMox}, let us label the first, the second, the third and the fourth line
 of the exponent by the number of $\eta$'s and by the number of $\theta$'s,
  namely, $[0,2]$, $[1,1]$, $[2,0]$, $[2,2]$ respectively.
Also upon compactification, namely the division of the pair of indices $(M, L)$ into the spacetime part $(\mu, \lambda)$
 and the internal part $(\ell, \ell')$, we set the internal part of the momenta $k_I^\ell = 0$ for simplicity.
Index structure of $\pi N_{ML, ++}^{IJ}$, $\pi N_{ML, +-}^{IJ}$, $B_{ML, \nu_{\rm f}}^{IJ}$,
 $C_{ML, ++}^{IJ}$, $C_{ML, +-}^{IJ}$, $E_{ML, ++}^{IJ}$, $E_{ML, +-}^{IJ}$
  are of the form $\displaystyle\sum_{\bullet \bullet} g_{\bullet \bullet} X_{\nu (\bullet \bullet)}^{IJ}$
   and the expressions are read off from
\ba	
	\pi N_{++}^{IJ}
        &\equiv&	%
        \ln \left|{
        	\frac{\vartheta \left[{\textstyle {{1 \over 2} \atop {1 \over 2}}} \right] \left({ \frac{z_I}{2} - \frac{z_J}{2} \left|{\frac{i\tau_2}{2}}\right. }\right)}
        	     		     {\vartheta' \left[{\textstyle {{1 \over 2} \atop {1 \over 2}}} \right] \left({ 0 \left|{\frac{i\tau_2}{2}}\right. }\right)}
            }\right|
         + {\pi \over 2} \frac{(z_I - z_J)^2}{\tau_2}			\nn
        B_{\nu_{\rm f}}^{IJ}
        &\equiv&	%
        \frac{1}{2} \frac{\pi}{i} {\cal{S}}_{\nu_{\rm f}} (\frac{z_I}{2} - \frac{z_J}{2} | \frac{i \tau_2}{2})
        =	
        \frac{1}{2}
        \frac{
        	\vartheta_{\nu_{\rm f}} \left({ \frac{z_I}{2} - \frac{z_J}{2} \left|{ \frac{i \tau}{2} }\right.}\right)
             }
             {
             	\vartheta_{\nu_{\rm f}} \left({ 0 \left|{ \frac{i \tau}{2} }\right.}\right)
             }
        \frac{
        	\vartheta' \left[{\textstyle {{1 \over 2} \atop {1 \over 2}}} \right] \left({ 0 \left|{ \frac{i \tau}{2} }\right.}\right)
             }
             {
             	\vartheta \left[{\textstyle {{1 \over 2} \atop {1 \over 2}}} \right] \left({ \frac{z_I}{2} - \frac{z_J}{2} \left|{ \frac{i \tau}{2} }\right.}\right)
             }								\nn
        C_{++}^{IJ}
        &\equiv&	%
        \pardif{}{z_I}
        \left.{\left[{ \pi N_{++}^{IJ} }\right]}\right|_{z = {\tilde z} \equiv - {\bar z}}
        =	%
        \frac{1}{2}
        \frac{\vartheta' \left[{\textstyle {{1 \over 2} \atop {1 \over 2}}} \right] (\frac{z_I}{2} - \frac{z_J}{2} | \frac{i \tau_2}{2})}{\vartheta \left[{\textstyle {{1 \over 2} \atop {1 \over 2}}} \right] (\frac{z_I}{2} - \frac{z_J}{2} | \frac{i \tau_2}{2})}
	 + \pi \frac{z_I - z_J}{\tau_2}					\nn
        E_{++}^{IJ}
        &\equiv&	%
        \pardif{}{z_I} \pardif{}{z_J}
        \left.{\left[{ \pi N_{++}^{IJ} }\right]}\right|_{z = {\tilde z} \equiv - {\bar z}}			\nn
        &=&	%
        \frac{1}{4}
        \left\{{
          \frac{\vartheta'' \left[{\textstyle {{1 \over 2} \atop {1 \over 2}}} \right] \left({ \frac{z_I}{2} - \frac{z_J}{2} \left|{\frac{i\tau_2}{2}}\right. }\right)}
               {\vartheta \left[{\textstyle {{1 \over 2} \atop {1 \over 2}}} \right] \left({ \frac{z_I}{2} - \frac{z_J}{2} \left|{\frac{i\tau_2}{2}}\right. }\right)}
          - \left({
        	\frac{\vartheta' \left[{\textstyle {{1 \over 2} \atop {1 \over 2}}} \right] \left({ \frac{z_I}{2} - \frac{z_J}{2} \left|{\frac{i\tau_2}{2}}\right. }\right)}
        	     {\vartheta \left[{\textstyle {{1 \over 2} \atop {1 \over 2}}} \right] \left({ \frac{z_I}{2} - \frac{z_J}{2} \left|{\frac{i\tau_2}{2}}\right. }\right)}
            }\right)^2
        }\right\}
         + \frac{\pi}{\tau_2}
		\label{DefABCE}
\ea
\ba
	\pi N_{+-}^{IJ}
        &\equiv&	%
        2 \pi
        \left.{G_{+-} \left({\left.{ \frac{z_I}{2}; \frac{z_J}{2} }\right| \frac{i \tau_2}{2} }\right)}\right|_{z = {\tilde z} \equiv - {\bar z}}
        -
        \left[{
        	2 \pi
        	{\rm Im} \frac{z_J}{2}
                -
        	(2 \pi)
        	\pi^2 \tau_2
        }\right]			\nn
        &=&	
        \ln \left|{ \sqrt{\zeta_I} - \sqrt{\zeta_J} }\right|
        +
        \sum_{m=1}^\infty
        (-1)^{m} \ln \left|{ 1 - \frac{\sqrt{\zeta_I} (\sqrt{|q|})^m}{\sqrt{\zeta_J}} }\right| \left|{ 1 - \frac{\sqrt{\zeta_J} (\sqrt{|q|})^{m}}{\sqrt{\zeta_I}} }\right|			\nn
	C_{+-}^{IJ}
        &\equiv&	%
        {1 \over 2}
        \left({
        	\pardif{}{z_I} - \pardif{}{z_J}
        }\right)
        \left.{\left[{
        	\pi N_{+-}^{IJ}
        }\right]}\right|_{z = {\tilde z} \equiv - {\bar z}}				\nn
        &=&	
        \frac{\pi i}{2}
        \frac{{\sqrt{\zeta_I}} + {\sqrt{\zeta_J}}}{{\sqrt{\zeta_I}} - {\sqrt{\zeta_J}}}
        +
        \pi i
        \sum_{m=1}^\infty
        (-1)^{m}
        \left\{{
               	\frac{1}{1 - \frac{\sqrt{\zeta_J}}{\sqrt{\zeta_I} (\sqrt{|q|})^m}}
                -
                \frac{1}{1 - \frac{\sqrt{\zeta_I}}{\sqrt{\zeta_J} (\sqrt{|q|})^{m}}}
        }\right\}											\nn
        E_{+-}^{IJ}
        &\equiv&	%
        (-1)
        {1 \over 2}
        \left({
        	\pardif{}{z_I} - \pardif{}{z_J}
        }\right)
        {1 \over 2}
        \left({
        	\pardif{}{z_J} - \pardif{}{z_I}
        }\right)
        \left.{\left[{
        	\pi N_{+-}^{IJ}
        }\right]}\right|_{z = {\tilde z} \equiv - {\bar z}}			\nn
        &=&	
        -
        (\pi i)^2
        \frac{\sqrt{\zeta_I} \sqrt{\zeta_J}}
        {\left({ {\sqrt{\zeta_I}} - {\sqrt{\zeta_J}} }\right)^2}
        -
        (\pi i)^2
        \sum_{m=1}^\infty
        (-1)^{m}
        \left\{{
        	\frac{\frac{\sqrt{\zeta_J}}{\sqrt{\zeta_I} (\sqrt{|q|})^m}}
                     {\left({ 1 - \frac{\sqrt{\zeta_J}}{\sqrt{\zeta_I} (\sqrt{|q|})^m} }\right)^2}
                +
                \frac{\frac{\sqrt{\zeta_I}}{\sqrt{\zeta_J} (\sqrt{|q|})^{m}}}
                     {\left({ 1 - \frac{\sqrt{\zeta_I}}{\sqrt{\zeta_J} (\sqrt{|q|})^{m}} }\right)^2}
        }\right\}							\:.		\nn
        	\label{Def:N,C,E:+-}
\ea
See appendix \ref{app:propertyofNBCE} for these properties.

\subsection{Analysis and evaluation of $N=1,2,3$ cases}

We will now analyze a few simplest cases.
Let us first obtain a few generic features of the amplitudes from the integral representation.
First, in order to obtain a non-vanishing amplitude, all grassmann integrations must be saturated.
Also, under the assumption made in the last subsection, $\displaystyle\sum_I k_I^M = 0$ for $M=0,1,...,9$.
Note that, the zero mode is absent in the expansion of $X^\ell$, $\ell = 5, 6, 7, 8$,
 and that momentum conservation is not ensured.

\begin{enumerate}
\renewcommand{\labelenumi}{\Roman{enumi})}
 \item	$N=1$; the amplitude vanishes generally and trivially as such case is absent in the integrand.
 \item	$N=2$; there is no contribution from $[0,2][2,0]$ or from $[1,1]^2$ by $k_I \cdot k_J = k_I \cdot \zeta_J^{({\rm P})} = 0$ for $I,J=1,2$.
 		Neither is there any contribution from $[2,2]$ which does not involve $\nu_{\rm f}$ by the same reason
                 that the vacuum amplitude vanishes by the Jacobi identity or supersymmetry.
 \item	$N=3$; this case poses the general question of the presence or absence of the vertex correction.
 		There is no contribution form the parts in which $[0,2]$ is involved as $k_I \cdot k_J = 0$ for $I,J = 1,2,3$.
 		The remaining possibilities for a non-vanishing amplitude are $[1,1]^3$ and $[1,1][2,2]$.
                Among them, the parts which do not involve $B_{\bullet \bullet, \nu_{\rm f}}^{IJ}$ vanish
                 after the summation over $\nu_{\rm f}$ by the Jacobi identity or supersymmetry.
                We conclude that the possibilities are contained in $[1,1]$ of
                 $B_{ML, \nu_{\rm f}}^{12} B_{ML, \nu_{\rm f}}^{23} B_{ML, \nu_{\rm f}}^{31}$ type
                  and of $(B_{ML, \nu_{\rm f}}^{IJ})^2 C_{ML, \nu_{\rm b}}^{IJ}$ type.
\end{enumerate}

\subsubsection{case of maximal supersymmetry}

In this case, namely, in the case of flat $10$d and its toroidal compactifications,
 it is well-known that the vanishing of these two types after the summation over $\nu_{\rm f}$
  is established, (see, for example, \cite{DP}) by the Riemann identity eq. (\ref{Riemannid}).
In fact
\be
	{\cal J}_{\nu}
        =	
        (2n)^2
        \left({
        	\prod_I F_1 (a_I, \tau_2)
        }\right)
        C_{\nu}
        \frac{\vartheta_{\nu} (0)^4}{\eta^{12}}
\ee
 according to eqs. (\ref{box:+,+,rf,sf:concrete}) and (\ref{calJ_{I,T4/Z2}}) and
\ba	
	&\:&
        \frac{\eta^{12}}{(2n)^2 \left({ \displaystyle\prod_I F_I (a_I, \tau_2) }\right)}
        \sum_{\nu_{\rm f}} {\cal J}_{\nu_{\rm f}}
        (B_{\nu_{\rm f}}^{IJ})^2				\nn
        &\:&
        =	
        \sum_{\nu_{\rm f}}
        C_{\nu_{\rm f}}
        \vartheta_{\nu_{\rm f}} (0)^4
        (B_{\nu_{\rm f}}^{IJ})^2				\nn
        &\:&
        =	
        \sum_{\nu_{\rm f}}
        C_{\nu_{\rm f}}
        \vartheta_{\nu_{\rm f}} (0)^4
        \left({
        	\frac{1}{2}
        	\frac{\vartheta_{\nu_{\rm f}} (\frac{z_I}{2} - \frac{z_J}{2})}{\vartheta_{\nu_{\rm f}} (0)} \frac{\vartheta' \left[{\textstyle {{1 \over 2} \atop {1 \over 2}}} \right] (0)}{\vartheta \left[{\textstyle {{1 \over 2} \atop {1 \over 2}}} \right] (\frac{z_1}{2} - \frac{z_2}{2})}
        }\right)^2								\nn
        &\:&
        {\overset{{\rm eq.} (\ref{Riemannid})}{=}}	
        \frac{1}{2}
        \left({
        	\frac{1}{2}
        	\frac{\vartheta' \left[{\textstyle {{1 \over 2} \atop {1 \over 2}}} \right] (0)}{\vartheta \left[{\textstyle {{1 \over 2} \atop {1 \over 2}}} \right] (\frac{z_1}{2} - \frac{z_2}{2})}
        }\right)^2
        2 \vartheta \left[{\textstyle {{1 \over 2} \atop {1 \over 2}}} \right] \left({ \frac{z_1}{2} - \frac{z_2}{2} }\right)
           \vartheta \left[{\textstyle {{1 \over 2} \atop {1 \over 2}}} \right] \left({ - \left({ \frac{z_1}{2} - \frac{z_2}{2} }\right)}\right)
            \vartheta \left[{\textstyle {{1 \over 2} \atop {1 \over 2}}} \right] (0)
             \vartheta \left[{\textstyle {{1 \over 2} \atop {1 \over 2}}} \right] (0)			\nn
	&\:&
        =	
        0						\:,
        	\label{theta(0)^4(B^IJ)^2}
\ea
 where
\ba
	x_1 &=& {1 \over 2} \left\{{ 0 + 0 + \left({ \frac{z_1}{2} - \frac{z_2}{2} }\right) + \left({ \frac{z_1}{2} - \frac{z_2}{2} }\right) }\right\} = \frac{z_1}{2} - \frac{z_2}{2}			\nn
        y_1 &=& {1 \over 2} \left\{{ 0 + 0 - \left({ \frac{z_1}{2} - \frac{z_2}{2} }\right) - \left({ \frac{z_1}{2} - \frac{z_2}{2} }\right) }\right\} = -\left({ \frac{z_1}{2} - \frac{z_2}{2} }\right)			\nn
        u_1 &=& {1 \over 2} \left\{{ 0 - 0 + \left({ \frac{z_1}{2} - \frac{z_2}{2} }\right) - \left({ \frac{z_1}{2} - \frac{z_2}{2} }\right) }\right\} = 0			\nn
        v_1 &=& {1 \over 2} \left\{{ 0 - 0 - \left({ \frac{z_1}{2} - \frac{z_2}{2} }\right) + \left({ \frac{z_1}{2} - \frac{z_2}{2} }\right) }\right\} = 0				\:.
\ea
 and we have used $\vartheta \left[{\textstyle {{1 \over 2} \atop {1 \over 2}}} \right] (0) = 0$.
Similarly, 
\ba	
	&\:&
        \frac{\eta^{12}}{(2n)^2 \left({ \displaystyle\prod_I F_I (a_I, \tau_2) }\right)}
        \sum_{\nu_{\rm f}} {\cal J}_{\nu_{\rm f}}
        B_{\nu_{\rm f}}^{12} B_{\nu_{\rm f}}^{23} B_{\nu_{\rm f}}^{13}				\nn
        &\:&
        =	
	\sum_{\nu_{\rm f}}
        C_{\nu_{\rm f}}
        \vartheta_{\nu_{\nu_{\rm f}}} (0)^4
        B_{\nu_{\rm f}}^{12} B_{\nu_{\rm f}}^{23} B_{\nu_{\rm f}}^{13}				\nn
        &\:&
        =	
        \sum_{\nu_{\rm f}}
        C_{\nu_{\rm f}}
        \vartheta_{\nu_{\rm f}} (0)^4
        \left({
        	\frac{1}{2}
        }\right)^3
        \left({
        	\frac{\vartheta_{\nu_{\rm f}} (\frac{z_1}{2} - \frac{z_2}{2})}{\vartheta_{\nu_{\rm f}} (0)} \frac{\vartheta' \left[{\textstyle {{1 \over 2} \atop {1 \over 2}}} \right] (0)}{\vartheta \left[{\textstyle {{1 \over 2} \atop {1 \over 2}}} \right] (\frac{z_1}{2} - \frac{z_2}{2})}
        }\right)							\nn
        &\:&	\qq\qq	%
        \times
        \left({
        	\frac{\vartheta_{\nu_{\rm f}} (\frac{z_2}{2} - \frac{z_3}{2})}{\vartheta_{\nu_{\rm f}} (0)} \frac{\vartheta' \left[{\textstyle {{1 \over 2} \atop {1 \over 2}}} \right] (0)}{\vartheta \left[{\textstyle {{1 \over 2} \atop {1 \over 2}}} \right] (\frac{z_2}{2} - \frac{z_3}{2})}
        }\right)
        \left({
        	\frac{\vartheta_{\nu_{\rm f}} (\frac{z_1}{2} - \frac{z_3}{2})}{\vartheta_{\nu_{\rm f}} (0)} \frac{\vartheta' \left[{\textstyle {{1 \over 2} \atop {1 \over 2}}} \right] (0)}{\vartheta \left[{\textstyle {{1 \over 2} \atop {1 \over 2}}} \right] (\frac{z_1}{2} - \frac{z_3}{2})}
        }\right)							\nn
        &\:&
        {\overset{{\rm eq.} (\ref{Riemannid})}{=}}	
        \frac{1}{2}
        \left({
        	\frac{1}{2}
        }\right)^3
        \frac{\vartheta' \left[{\textstyle {{1 \over 2} \atop {1 \over 2}}} \right] (0)^3}
             {\vartheta \left[{\textstyle {{1 \over 2} \atop {1 \over 2}}} \right] (\frac{z_1}{2} - \frac{z_2}{2}) \vartheta \left[{\textstyle {{1 \over 2} \atop {1 \over 2}}} \right] (\frac{z_2}{2} - \frac{z_3}{2}) \vartheta \left[{\textstyle {{1 \over 2} \atop {1 \over 2}}} \right] (\frac{z_1}{2} - \frac{z_3}{2})}
        				\nn
        &\:&	\qq\qq	%
        \times
        2 \vartheta \left[{\textstyle {{1 \over 2} \atop {1 \over 2}}} \right] \left({ \frac{z_1}{2} - \frac{z_3}{2} }\right)
           \vartheta \left[{\textstyle {{1 \over 2} \atop {1 \over 2}}} \right] \left({ \frac{z_3}{2} - \frac{z_2}{2} }\right)
            \vartheta \left[{\textstyle {{1 \over 2} \atop {1 \over 2}}} \right] \left({ \frac{z_2}{2} - \frac{z_1}{2} }\right)
             \vartheta \left[{\textstyle {{1 \over 2} \atop {1 \over 2}}} \right] (0)						\nn
	&\:&
        =	
        0						\:,
        	\label{theta(0)^4B^12B^23B^13}
\ea
 where
\ba
	x_1 &=& {1 \over 2} \left\{{ 0 + \left({ \frac{z_1}{2} - \frac{z_2}{2} }\right) + \left({ \frac{z_2}{2} - \frac{z_3}{2} }\right) + \left({ \frac{z_1}{2} - \frac{z_3}{2} }\right) }\right\} = \left({ \frac{z_1}{2} - \frac{z_3}{2} }\right)			\nn
        y_1 &=& {1 \over 2} \left\{{ 0 + \left({ \frac{z_1}{2} - \frac{z_2}{2} }\right) - \left({ \frac{z_2}{2} - \frac{z_3}{2} }\right) - \left({ \frac{z_1}{2} - \frac{z_3}{2} }\right) }\right\} = \left({ \frac{z_3}{2} - \frac{z_2}{2} }\right) 			\nn
        u_1 &=& {1 \over 2} \left\{{ 0 - \left({ \frac{z_1}{2} - \frac{z_2}{2} }\right) + \left({ \frac{z_2}{2} - \frac{z_3}{2} }\right) - \left({ \frac{z_1}{2} - \frac{z_3}{2} }\right) }\right\} = \left({ \frac{z_2}{2} - \frac{z_1}{2} }\right)			\nn
        v_1 &=& {1 \over 2} \left\{{ 0 - \left({ \frac{z_1}{2} - \frac{z_2}{2} }\right) - \left({ \frac{z_2}{2} - \frac{z_3}{2} }\right) + \left({ \frac{z_1}{2} - \frac{z_3}{2} }\right) }\right\} = 0			\:.
\ea

\subsubsection{case of non-maximal supersymmetry}

Finally, let us consider the case of type I superstring on $T^4/{\bf Z}_2$.
Among the summation over $\nu_{\rm f}$, only the part belonging to
 ${1 \over 2} {\cal J}_{{\rm I}, T^4}^{({\bf Z}_2)}$ in eq. (\ref{calJ_{I,T4/Z2}}) contributes to the $N=3$ amplitude.
So we will concentrate on this case.
The integrations of the exponential factor eq. (\ref{exp[...]}) over the grassmann coordinates
 $\displaystyle\prod_{J=1}^3 \zeta_J^{({\rm P})} \cdot \int \de \eta_J \int \de \theta_J \exp [\cdots]$ yield
\ba
	(2 \alpha')^3
        k_3 \cdot \zeta_1^{({\rm P})} k_1 \cdot \zeta_2^{({\rm P})} k_2 \cdot \zeta_3^{({\rm P})}
        &\:&
        \left\{{
        	- 2 B_{\nu_{\rm f}^6}^{12} B_{\nu_{\rm f}^6}^{23} B_{\nu_{\rm f}^6}^{31}
                + \left({ B_{\nu_{\rm f}^6}^{12} }\right)^2
                  \left({ C_{++}^{13} - C_{++}^{23} }\right)
	}\right.				\nn
        &\:&
        \left.{
                + \left({ B_{\nu_{\rm f}^6}^{23} }\right)^2
                  \left({ C_{++}^{21} - C_{++}^{31} }\right)
                + \left({ B_{\nu_{\rm f}^6}^{31} }\right)^2
                  \left({ C_{++}^{32} - C_{++}^{12} }\right)
        }\right\}					\:.
\ea

Coming back to eq. (\ref{calA':N,bullet}),
 we obtain the expression for ${\cal A}'_3{}^{T^4/{\bf Z}_2}$:
\ba
	{\cal A}'_3{}^{T^4/{\bf Z}_2}
        &=&	
        (2n)
        {\rm Tr} \left({ T^1 T^2 T^3 }\right)
        (2 \alpha')^3
        (k_3 \cdot \zeta_1^{({\rm P})}) (k_1 \cdot \zeta_2^{({\rm P})}) (k_2 \cdot \zeta_3^{({\rm P})})			\nn
        &\:&
        \times
        \int_0^\infty \frac{\de \tau_2}{\tau_2} \frac{1}{\tau_2^5}
        \left({
        	\prod_{I=1,2,3} \int \de z_I
        }\right)
        \sum_{\nu \in {\cal S}'}
        {1 \over 2} {\cal J}'_{\nu}{}^{({\bf Z}_2)}				\nn
        &\:&
        \times
        \left\{{
        	- 2 B_{\nu_{\rm f}^6}^{12} B_{\nu_{\rm f}^6}^{23} B_{\nu_{\rm f}^6}^{31}
                + \left({ B_{\nu_{\rm f}^6}^{12} }\right)^2
                  \left({ C_{++}^{13} - C_{++}^{23} }\right)
	}\right.				\nn
        &\:&	\qq\qq	%
        \left.{
                + \left({ B_{\nu_{\rm f}^6}^{23} }\right)^2
                  \left({ C_{++}^{21} - C_{++}^{31} }\right)
                + \left({ B_{\nu_{\rm f}^6}^{31} }\right)^2
                  \left({ C_{++}^{32} - C_{++}^{12} }\right)
        }\right\}			\:,
\ea
 where ${\cal J}'_{\nu}{}^{({\bf Z}_2)}$ has been introduced in eq. (\ref{calJ_{I,T4/Z2}})
  and $\nu_{\rm f}^6$ refers to the spacetime part (bin) of $\nu_{\rm f}$.
Let us recall from eqs. (\ref{box:+,+,rf,sf:concrete}), (\ref{box:rb,sb,rf,sf:concrete})
\ba
	\bn{+}{+}{-}{-}^2
        &=&	
        \frac{\vartheta \left[{\textstyle {0 \atop 0}} \right] (0)}{\eta^3}	\:,	\qq
        \bn{+}{-}{-}{+}^2
        =	
        \frac{2 \vartheta \left[{\textstyle {0 \atop {1 \over 2}}} \right] (0)}
             {\vartheta \left[{\textstyle {{1 \over 2} \atop 0}} \right] (0)}		\:,	\nn
	\bn{+}{+}{-}{+}^2
        &=&	
        \frac{\vartheta \left[{\textstyle {0 \atop {1 \over 2}}} \right] (0)}{\eta^3}	\:,	\qq
        \bn{+}{-}{-}{-}^2
        =	
        \frac{2 \vartheta \left[{\textstyle {0 \atop 0}} \right] (0)}
             {\vartheta \left[{\textstyle {{1 \over 2} \atop 0}} \right] (0)}		\:,	\:\:
        {\rm and \: therefore}
\ea
\be
	\bn{+}{+}{-}{-}^4
        \bn{+}{-}{-}{+}^4
        =	
        \bn{+}{+}{-}{+}^4
        \bn{+}{-}{-}{-}^4
        =	
        \frac{4}{\eta^6}
        \frac{
             	\vartheta \left[{\textstyle {0 \atop 0}} \right] (0)^2
                \vartheta \left[{\textstyle {0 \atop {1 \over 2}}} \right] (0)^2
             }
             {\vartheta \left[{\textstyle {{1 \over 2} \atop 0}} \right] (0)^2}
\ee
  as well as from eq. (\ref{DefABCE})
\ba
	B_{--}^{IJ}
        &=&	
        \frac{1}{2}
        \frac{
        	\vartheta \left[{\textstyle {0 \atop 0}} \right] \left({ \frac{z_I}{2} - \frac{z_J}{2} \left|{ \frac{i \tau}{2} }\right.}\right)
             }
             {
             	\vartheta \left[{\textstyle {0 \atop 0}} \right] \left({ 0 \left|{ \frac{i \tau}{2} }\right.}\right)
             }
        \frac{
        	\vartheta' \left[{\textstyle {{1 \over 2} \atop {1 \over 2}}} \right] \left({ 0 \left|{ \frac{i \tau}{2} }\right.}\right)
             }
             {
             	\vartheta \left[{\textstyle {{1 \over 2} \atop {1 \over 2}}} \right] \left({ \frac{z_I}{2} - \frac{z_J}{2} \left|{ \frac{i \tau}{2} }\right.}\right)
             }								\nn
	B_{-+}^{IJ}
        &=&	
        \frac{1}{2}
        \frac{
        	\vartheta \left[{\textstyle {0 \atop {1 \over 2}}} \right] \left({ \frac{z_I}{2} - \frac{z_J}{2} \left|{ \frac{i \tau}{2} }\right.}\right)
             }
             {
             	\vartheta \left[{\textstyle {0 \atop {1 \over 2}}} \right] \left({ 0 \left|{ \frac{i \tau}{2} }\right.}\right)
             }
        \frac{
        	\vartheta' \left[{\textstyle {{1 \over 2} \atop {1 \over 2}}} \right] \left({ 0 \left|{ \frac{i \tau}{2} }\right.}\right)
             }
             {
             	\vartheta \left[{\textstyle {{1 \over 2} \atop {1 \over 2}}} \right] \left({ \frac{z_I}{2} - \frac{z_J}{2} \left|{ \frac{i \tau}{2} }\right.}\right)
             }								\nn
        C_{++}^{IJ}
        &=&
        \frac{1}{2}
        \frac{\vartheta' \left[{\textstyle {{1 \over 2} \atop {1 \over 2}}} \right] (\frac{z_I}{2} - \frac{z_J}{2} | \frac{i \tau_2}{2})}{\vartheta \left[{\textstyle {{1 \over 2} \atop {1 \over 2}}} \right] (\frac{z_I}{2} - \frac{z_J}{2} | \frac{i \tau_2}{2})}
	 + \pi \frac{z_I - z_J}{\tau_2}					\:.
\ea
We obtain
\ba	
	&\:&
	{\cal A}'_3{}^{T^4/{\bf Z}_2}					\nn
        &\:&
        =	
        (2n)
        {\rm Tr} \left({ T^1 T^2 T^3 }\right)
        (2 \alpha')^3
        (k_3 \cdot \zeta_1^{({\rm P})}) (k_1 \cdot \zeta_2^{({\rm P})}) (k_2 \cdot \zeta_3^{({\rm P})})
        \int_0^\infty \frac{\de \tau_2}{\tau_2} \frac{1}{\tau_2^5}
        \left({
        	\prod_{I=1,2,3} \int \de z_I
        }\right)
        \left({ {1 \over 2} }\right)^3						\nn
        &\:&	%
        \left({ \prod_{I=5,6,7,8} (a_I \sqrt{\tau_2}) }\right)
        \frac{
             	\vartheta \left[{\textstyle {0 \atop 0}} \right]^2 (0)
                \vartheta \left[{\textstyle {0 \atop {1 \over 2}}} \right]^2 (0)
             }
             {
             	\vartheta \left[{\textstyle {{1 \over 2} \atop 0}} \right]^2 (0)
                \eta^6
             }
        \left[{
        	\frac{
                     	(-2)
                        \vartheta' \left[{\textstyle {{1 \over 2} \atop {1 \over 2}}} \right]^3 (0)
                     }
                     {
                     	\vartheta \left[{\textstyle {{1 \over 2} \atop {1 \over 2}}} \right] (1 - 2)
                        \vartheta \left[{\textstyle {{1 \over 2} \atop {1 \over 2}}} \right] (2 - 3)
                        \vartheta \left[{\textstyle {{1 \over 2} \atop {1 \over 2}}} \right] (3 - 1)
                     }
        }\right.				\nn
        &\:&	%
        \cdot
        \left\{{
        	\frac{
                	\vartheta \left[{\textstyle {0 \atop 0}} \right] (1 - 2)
                        \vartheta \left[{\textstyle {0 \atop 0}} \right] (2 - 3)
                        \vartheta \left[{\textstyle {0 \atop 0}} \right] (3 - 1)
                     }
                     {
                     	\vartheta \left[{\textstyle {0 \atop 0}} \right]^3 (0)
                     }
                -
                \frac{
                	\vartheta \left[{\textstyle {0 \atop {1 \over 2}}} \right] (1 - 2)
                        \vartheta \left[{\textstyle {0 \atop {1 \over 2}}} \right] (2 - 3)
                        \vartheta \left[{\textstyle {0 \atop {1 \over 2}}} \right] (3 - 1)
                     }
                     {
                     	\vartheta \left[{\textstyle {0 \atop {1 \over 2}}} \right]^3 (0)
                     }
        }\right\}							\nn
        &\:&	%
        +
        \left\{{
        	\left({
                	\frac{\vartheta \left[{\textstyle {0 \atop 0}} \right] (1 - 2)}
                             {\vartheta \left[{\textstyle {0 \atop 0}} \right] (0)}
                        \frac{\vartheta' \left[{\textstyle {{1 \over 2} \atop {1 \over 2}}} \right] (0)}
                             {\vartheta \left[{\textstyle {{1 \over 2} \atop {1 \over 2}}} \right] (1 - 2)}
                }\right)^2
                -
                \left({
                	\frac{\vartheta \left[{\textstyle {0 \atop {1 \over 2}}} \right] (1 - 2)}
                             {\vartheta \left[{\textstyle {0 \atop {1 \over 2}}} \right] (0)}
                        \frac{\vartheta' \left[{\textstyle {{1 \over 2} \atop {1 \over 2}}} \right] (0)}
                             {\vartheta \left[{\textstyle {{1 \over 2} \atop {1 \over 2}}} \right] (1 - 2)}
                }\right)^2
        }\right\}						\nn
        &\:&	\qq	%
        \times
        \left({
        	\frac{\vartheta' \left[{\textstyle {{1 \over 2} \atop {1 \over 2}}} \right] (1 - 3)}
                     {\vartheta \left[{\textstyle {{1 \over 2} \atop {1 \over 2}}} \right] (1 - 3)}
                -
                \frac{\vartheta' \left[{\textstyle {{1 \over 2} \atop {1 \over 2}}} \right] (2 - 3)}
                     {\vartheta \left[{\textstyle {{1 \over 2} \atop {1 \over 2}}} \right] (2 - 3)}
                +
                2 \pi (z_1 - z_2)
        }\right)						\nn
        &\:&
        +
        \left\{{
        	1 \rightarrow 2, 2 \rightarrow 3, 3 \rightarrow 1
        }\right\}
        \left\{{
        	1 \rightarrow 2, 2 \rightarrow 3, 3 \rightarrow 1
        }\right\}
        {\rm \: in \: the \: second \: term}			\nn
        &\:&
        \left.{\left.{
        	+
        	\left\{{
        		1 \rightarrow 2, 2 \rightarrow 3, 3 \rightarrow 1
        	}\right\}
        	\left\{{
        		1 \rightarrow 2, 2 \rightarrow 3, 3 \rightarrow 1
        	}\right\}
        	{\rm \: in \: the \: second \: term}
        }\right]}\right|_{\tau = \frac{i \tau_2}{2}}				\:,
        	\label{continue:{cal A}'_3^{T^4/Z_2}}
\ea
 where we have introduced shorthand notation
  $
	\vartheta \left[{\textstyle {\alpha \atop \beta}} \right] (I - J)
        \equiv	
        \vartheta \left[{\textstyle {\alpha \atop \beta}} \right] \left({ \frac{z_I}{2} - \frac{z_J}{2} }\right)
  $.
The second, third and fourth terms in eq. (\ref{continue:{cal A}'_3^{T^4/Z_2}}) can be further converted
 by using eq. (\ref{MumfordI:p.20,R6:+--+}):
\ba	
	&\:&
	\sum_{\nu \in {\cal S}'}
	{1 \over 2} {\cal J}'{}_{\nu, \: {\cal A}}^{({\bf Z}_2)}
        (B_{\nu_{\rm f}^6}^{IJ})^2							\nn
        &\:&
        =	
        {1 \over 2}
        \left({ \prod_{I=5,6,7,8} (a_I \sqrt{\tau_2}) }\right)
        {1 \over 2}
        \frac{4}{\eta^6}
        \left({
        	\frac{\vartheta \left[{\textstyle {0 \atop 0}} \right]^{2} (0) \vartheta \left[{\textstyle {0 \atop {1 \over 2}}} \right]^{2} (0)}
                     {\vartheta \left[{\textstyle {{1 \over 2} \atop 0}} \right]^{2} (0)}
        }\right)
        \left\{{
        	(B_{--}^{IJ})^2 - (B_{-+}^{IJ})^2
        }\right\}								\nn
        &\:&
        =	
        {1 \over 2}
        \left({ \prod_{I=5,6,7,8} (a_I \sqrt{\tau_2}) }\right)
        {1 \over 2}
        \frac{4}{\eta^6}
        \left({
        	\frac{\vartheta \left[{\textstyle {0 \atop 0}} \right]^{2} (0) \vartheta \left[{\textstyle {0 \atop {1 \over 2}}} \right]^{2} (0)}
                     {\vartheta \left[{\textstyle {{1 \over 2} \atop 0}} \right]^{2} (0)}
        }\right)								\nn
        &\:&		%
        \times
        \left\{{
        	\frac{1}{2^2}
        	\frac{
        		\vartheta \left[{\textstyle {0 \atop 0}} \right]^2 \left({ I - J \left|{ \frac{i \tau}{2} }\right.}\right)
        	     }
        	     {
        	     	\vartheta \left[{\textstyle {0 \atop 0}} \right]^2 \left({ 0 \left|{ \frac{i \tau}{2} }\right.}\right)
        	     }
        	\frac{
        		\vartheta' \left[{\textstyle {{1 \over 2} \atop {1 \over 2}}} \right]^2 \left({ 0 \left|{ \frac{i \tau}{2} }\right.}\right)
        	     }
        	     {
        	     	\vartheta \left[{\textstyle {{1 \over 2} \atop {1 \over 2}}} \right]^2 \left({ I - J \left|{ \frac{i \tau}{2} }\right.}\right)
        	     }
                -
                \frac{1}{2^2}
        	\frac{
        		\vartheta \left[{\textstyle {0 \atop {1 \over 2}}} \right]^2 \left({ I - J \left|{ \frac{i \tau}{2} }\right.}\right)
        	     }
        	     {
        	     	\vartheta \left[{\textstyle {0 \atop {1 \over 2}}} \right]^2 \left({ 0 \left|{ \frac{i \tau}{2} }\right.}\right)
        	     }
        	\frac{
        		\vartheta' \left[{\textstyle {{1 \over 2} \atop {1 \over 2}}} \right]^2 \left({ 0 \left|{ \frac{i \tau}{2} }\right.}\right)
        	     }
        	     {
        	     	\vartheta \left[{\textstyle {{1 \over 2} \atop {1 \over 2}}} \right]^2 \left({ I - J \left|{ \frac{i \tau}{2} }\right.}\right)
        	     }
        }\right\}									\nn
        &\:&
        {\overset{{\rm eq.} (\ref{MumfordI:p.20,R6:+--+})}{=}}	
        -
        \left({ \prod_{I=5,6,7,8} (a_I \sqrt{\tau_2}) }\right)
        \frac{1}{\eta^6}
        \frac{1}{2^2}
        \vartheta' \left[{\textstyle {{1 \over 2} \atop {1 \over 2}}} \right]^2 \left({ 0 \left|{ \frac{i \tau}{2} }\right.}\right)			\:,
		\label{sum{calJ}(B^IJ)^2}
\ea
 where in eq. (\ref{MumfordI:p.20,R6:+--+}),
\ba
	x_1 &=&	{1 \over 2} \left\{{ \left({ \frac{z_I}{2} - \frac{z_J}{2} }\right) + \left({ \frac{z_I}{2} - \frac{z_J}{2} }\right) + 0 + 0 }\right\}
             =  \frac{z_I}{2} - \frac{z_J}{2}						\nn
        y_1 &=&	{1 \over 2} \left\{{ \left({ \frac{z_I}{2} - \frac{z_J}{2} }\right) + \left({ \frac{z_I}{2} - \frac{z_J}{2} }\right) - 0 - 0 }\right\}
             =  \frac{z_I}{2} - \frac{z_J}{2}						\nn
        u_1 &=&	{1 \over 2} \left\{{ \left({ \frac{z_I}{2} - \frac{z_J}{2} }\right) - \left({ \frac{z_I}{2} - \frac{z_J}{2} }\right) + 0 - 0 }\right\}
             =  0								\nn
        v_1 &=&	{1 \over 2} \left\{{ \left({ \frac{z_I}{2} - \frac{z_J}{2} }\right) - \left({ \frac{z_I}{2} - \frac{z_J}{2} }\right) - 0 + 0 }\right\}
             =  0						\:.
\ea
Finally, we obtain
\ba	
	&\:&
	{\cal A}'_3{}^{T^4/{\bf Z}_2}					\nn
        &\:&
        =	
        (2n)
        {\rm Tr} \left({ T^1 T^2 T^3 }\right)
        (2 \alpha')^3
        (k_3 \cdot \zeta_1^{({\rm P})}) (k_1 \cdot \zeta_2^{({\rm P})}) (k_2 \cdot \zeta_3^{({\rm P})})
        \int_0^\infty \frac{\de \tau_2}{\tau_2} \frac{1}{\tau_2^5}
        \left({
        	\prod_{I=1,2,3} \int \de z_I
        }\right)
        \left({ {1 \over 2} }\right)^3						\nn
        &\:&	%
        \left({ \prod_{I=5,6,7,8} (a_I \sqrt{\tau_2}) }\right)
        \frac{1}{\eta^6}
        \left[{
        	        \frac{
        		     	\vartheta \left[{\textstyle {0 \atop 0}} \right]^2 (0)
        		        \vartheta \left[{\textstyle {0 \atop {1 \over 2}}} \right]^2 (0)
        		     }
        		     {
        		     	\vartheta \left[{\textstyle {{1 \over 2} \atop 0}} \right]^2 (0)
        		     }
        	\frac{
                     	(-2)
                        \vartheta' \left[{\textstyle {{1 \over 2} \atop {1 \over 2}}} \right]^3 (0)
                     }
                     {
                     	\vartheta \left[{\textstyle {{1 \over 2} \atop {1 \over 2}}} \right] (1 - 2)
                        \vartheta \left[{\textstyle {{1 \over 2} \atop {1 \over 2}}} \right] (2 - 3)
                        \vartheta \left[{\textstyle {{1 \over 2} \atop {1 \over 2}}} \right] (3 - 1)
                     }
        }\right.				\nn
        &\:&	%
        \cdot
        \left\{{
        	\frac{
                	\vartheta \left[{\textstyle {0 \atop 0}} \right] (1 - 2)
                        \vartheta \left[{\textstyle {0 \atop 0}} \right] (2 - 3)
                        \vartheta \left[{\textstyle {0 \atop 0}} \right] (3 - 1)
                     }
                     {
                     	\vartheta \left[{\textstyle {0 \atop 0}} \right]^3 (0)
                     }
                -
                \frac{
                	\vartheta \left[{\textstyle {0 \atop {1 \over 2}}} \right] (1 - 2)
                        \vartheta \left[{\textstyle {0 \atop {1 \over 2}}} \right] (2 - 3)
                        \vartheta \left[{\textstyle {0 \atop {1 \over 2}}} \right] (3 - 1)
                     }
                     {
                     	\vartheta \left[{\textstyle {0 \atop {1 \over 2}}} \right]^3 (0)
                     }
        }\right\}							\nn
        &\:&	%
        \left.{\left.{
        	+
                2
                \left({
        		\frac{\vartheta' \left[{\textstyle {{1 \over 2} \atop {1 \over 2}}} \right] (1 - 2)}{\vartheta \left[{\textstyle {{1 \over 2} \atop {1 \over 2}}} \right] (1 - 2)}
        		+
        		\frac{\vartheta' \left[{\textstyle {{1 \over 2} \atop {1 \over 2}}} \right] (2 - 3)}{\vartheta \left[{\textstyle {{1 \over 2} \atop {1 \over 2}}} \right] (2 - 3)}
        		+
        		\frac{\vartheta' \left[{\textstyle {{1 \over 2} \atop {1 \over 2}}} \right] (3 - 1)}{\vartheta \left[{\textstyle {{1 \over 2} \atop {1 \over 2}}} \right] (3 - 1)}
        	}\right)
                \vartheta' \left[{\textstyle {{1 \over 2} \atop {1 \over 2}}} \right]^2 \left({ 0 }\right)
        }\right]}\right|_{\tau = \frac{i \tau_2}{2}}								\:.		\nn
        	\label{continue2:{cal A}'_3^{T^4/Z_2}}
\ea

Unlike the case of maximal supersymmetry, after nationalizing, each term consists of the product of different $\vartheta$ functions
 and we do not find the use of the Riemann identity.

\section*{Acknowledgements}

We thank Takeshi Oota, Reiji Yoshioka, Nobuhito Maru and Takao Suyama for helpful discussions.
We also thank Takahiro Kubota for providing us with his useful lecture note.
The research is supported in part by the Grant-in-Aid for Scientific Research
 from the Ministry of Education, Science and Culture, Japan (15K05059).

\appendix

\section{Some of the notation	\label{notation:app}} 

The complex coordinates on the worldsheet torus are denoted by
\be
	z \equiv \sigma^1 + \tau \sigma^2			\:,	\quad
        {\bar z} = \sigma^1 + {\bar \tau} \sigma^2			\qq
        (0 \leq \sigma_1, \sigma_2 \leq 1)
\ee
 with modular parameter $\tau \equiv \tau_1 + i \tau_2$.

The Laplacian is defined by
\be
	\Delta \equiv 4 \partial_{z} \partial_{\bar z}			\:.
\ee

We introduce a real superfield by
\ba	
	{\bf{X}}^M (z, {\bar{z}}, \theta, {\bar{\theta}})
        &=&
        X^M (z, {\bar{z}}) + {\sqrt{1 \over 2}} \theta \psi^M (z, {\bar{z}}) + {\sqrt{1 \over 2}} {\bar{\psi}}^M (z, {\bar{z}}) {\bar{\theta}} + {1 \over 2} \theta {\bar{\theta}} F^M (z, {\bar{z}})		\nn
        &\:&	\qq\qq
        M = 0, 1, 2, \cdots, 9
         \label{defX}
\ea
 where $\theta$ and ${\bar \theta}$ are Grassmann numbers,
  $X^M (z, {\bar{z}})$ and $\psi^M (z, {\bar{z}})$ are bosonic and fermionic fields,
   and $F^M (z, {\bar{z}})$ is a auxiliary field.
The super-derivatives are defined by
\be	
	D = - \pardif{}{\theta} + i \theta \pardif{}{z}	\: , \:\:
        {\bar D} = \pardif{}{\bar{\theta}} - i {\bar{\theta}} \pardif{}{\bar{z}}			\:.
         \label{defDandbarD}
\ee

\section{Normalization in eq. (\ref{Phi})	\label{NoramalizationOfPhi}} 

Let us determine the normalization $N$ in
\be	
	\Phi_{n_1, n_2} \left[{\textstyle {\alpha \atop \beta}} \right] (\sigma^1, \sigma^2)
        =	
        N
        \ex^{2 \pi i (n_1 + \alpha) \sigma^1} \ex^{2 \pi i (n_2 + \beta) \sigma^2}		\:.
		\label{Phi:normtobedetermined}
\ee

The inner product with functions $f$, $g$ is defined
\ba
	(f, g)
        =	
        \int_0^1 \de \sigma^1
        \int_0^1 \de \sigma^2
        {\sqrt{\hat g}} f^* g				\:.
\ea
On a torus geometry
\be
	{\hat g}_{a b} (\tau)
        =	
        \left[
		\begin{array}{cc}
		 1 & \tau_1 \\
		 \tau_1 & \tau_1^2 + \tau_2^2
		\end{array}
	\right]						\qq
	{\rm and}	\qq
	{\sqrt{\hat g}}
        =	
        {\sqrt{{\rm det} \; {\hat g}_{a b}}}
        =	
        \tau_2			\:,
\ee
 the orthonormality of $\Phi_{n_1, n_2} \left[{\textstyle {\alpha \atop \beta}} \right] (\sigma^1, \sigma^2)$ implies
\be
	\delta_{m_1, n_1} \delta_{m_2, n_2}
        =	
        \left({
        	\Phi_{m_1, m_2} \left[{\textstyle {\alpha \atop \beta}} \right] (\sigma^1, \sigma^2),
                \Phi_{n_1, n_2} \left[{\textstyle {\alpha \atop \beta}} \right] (\sigma^1, \sigma^2)
        }\right)
        =	
        |N|^2
        \tau_2
        \delta_{m_1, n_1} \delta_{m_2, n_2}				\:.
\ee
Hence we take
\be
        N = \frac{1}{\sqrt{\tau_2}}			\:.
        	\label{normalization:torus}
\ee

\section{Some formulae} 

\subsection{Gauss hypergeometric function}

Gauss hypergeometric function is
\ba	
	F (\alpha, \beta, \gamma; z)
        =	%
        {}_2F_1 (\alpha, \beta; \gamma; z)
        &=&	
        \sum_{n=0}^\infty
        \frac{(\alpha)_n (\beta)_n}{(\gamma)_n}
        \frac{z^n}{n!}								\nn							\nn
        &=&	
        \frac{\Gamma(\gamma)}{\Gamma(\beta) \Gamma(\gamma-\beta)}
        \int_0^1
        	t^{\beta-1} (1-t)^{\gamma-\beta-1} (1-tz)^{-\alpha}
        \de t									\;, \qq
        	\label{GaussHGfunc}
\ea
 where
\be	
	(\alpha)_n = \alpha (\alpha+1) (\alpha+2) \cdots (\alpha+n-1)
		   = \frac{\Gamma(\alpha+n)}{\Gamma(\alpha)}		\;,\qq
        (\alpha)_0 = 1								\:.
\ee
In order to obtain the second line in eq. (\ref{GaussHGfunc}), we must have
\begin{itemize}
 \item ${\rm{Re}} \gamma > {\rm{Re}} \beta > 0$,
 \item $z$ can not be the real number which is greater than $1$,
 \item $(1-tz)^{-\alpha}$ takes the branch which goes to $1$ as $t \rightarrow 0$.
\end{itemize}

When $b=1$, $k=0$, the formula
\be	
	\sum_{n=0}^\infty
        \frac{x^n}{(a+nb) (a+nb+1) \cdots (a+nb+k)}
        =
        \frac{1}{k!}
        \int_0^1
        	\frac{t^{a-1} (1-t)^{k}}{1-xt^b}
        \de t								\;, \qq
        [a,b>0, \: |x|<1]
\ee
can be written as
\be	
	\sum_{n=0}^\infty
        \frac{x^n}{n+a}
        =	%
        \int_0^1
        	\frac{t^{a-1}}{1-tx}
        \de t
        =	%
        \frac{\Gamma(a)}{\Gamma(1+a)}
         F (1, a, 1+a; x)					\;.
        	\label{formula:b=1&k=0}
\ee
Thanks to eq. (\ref{formula:b=1&k=0}),
\ba	
	\sum_{n=0}^\infty
        \frac{1}{n+a}
        \frac{x^{n+a}}{1 - C y^{n+a}}
        &=&	
        \sum_{n=0}^\infty
        \frac{1}{n+a}
        x^{n+a}
        \sum_{m=0}^\infty
        \left({
        	C y^{n+a}
        }\right)^m
        =	%
        \sum_{m=0}^\infty
        x^a (C y^a)^m
        \sum_{n=0}^\infty
        \frac{(x y^m)^n}{n+a}						\nn
        &{\overset{(\ref{formula:b=1&k=0})}{=}}&	
        \sum_{m=0}^\infty
        x^a (C y^a)^m
        \frac{\Gamma(a)}{\Gamma(1+a)}
         F (1, a, 1+a; x y^m)						\nn
        &=&	
        \frac{x^a \Gamma(a)}{\Gamma(1+a)}
        \sum_{m=0}^\infty
        (C y^a)^m
         F (1, a, 1+a; x y^m)
        	\label{ExtendedGSWformula}
\ea
 with $a > 0$, $m \in \{0, {\bf{N}}\}$ and $|x y^m| < 1$.

\subsection{Jacobi theta function}

We define the Jacobi theta function as
\ba	
	\vartheta \left[{\textstyle {\alpha \atop \beta}} \right] (z|\tau)
        &=&	
        \sum_{n \in {\bold{Z}}}
        \ex^{\pi i (n + \alpha)^2 \tau}
        \ex^{2 \pi i (n + \alpha) (z + \beta)}			\nn
        &=&	
        \ex^{2 \pi i \alpha (z + \beta)} \ex^{\pi i \alpha^2 \tau}
        \prod_{n=1}^\infty
         (1 - \ex^{2 \pi i n \tau})
         (1 + \ex^{2 \pi i (n + \alpha - {1 \over 2}) \tau} \ex^{2 \pi i (z + \beta)})
         (1 + \ex^{2 \pi i (n - \alpha - {1 \over 2}) \tau} \ex^{- 2 \pi i (z + \beta)})		\;.			\nn
        	\label{Def.ofThetafunction}
\ea
We also use the following notation:
\be
	\vartheta_{--} \equiv \vartheta \left[{\textstyle {0 \atop 0}} \right]			\:,	\qq
	\vartheta_{-+} \equiv \vartheta \left[{\textstyle {0 \atop {1 \over 2}}} \right]	\:,	\qq
	\vartheta_{+-} \equiv \vartheta \left[{\textstyle {{1 \over 2} \atop 0}} \right]		\:.
\ee

This function has following properties:
\be	
	\vartheta \left[{\textstyle {\alpha+1 \atop \beta}} \right] (z|\tau)
         =	
         \vartheta \left[{\textstyle {\alpha \atop \beta}} \right] (z|\tau)			\:,	\qq
        \vartheta \left[{\textstyle {\alpha \atop \beta+1}} \right] (z|\tau)
         =	
         \ex^{2 \pi i \alpha}
         \vartheta \left[{\textstyle {\alpha \atop \beta}} \right] (z|\tau)			\:,
        	\label{alpha+1_beta+1}
\ee
\be	
	\overline{\vartheta \left[{\textstyle {\alpha \atop \beta}} \right] (z|\tau)}
         =	
         \vartheta \left[{\textstyle {\alpha \atop -\beta}} \right] (-{\bar z}| -{\bar \tau})				\:.
        	\label{bar_theta}
\ee

The theta function satisfies the heat equation:
\be	
	\frac{\partial^2}{\partial z^2}
        \vartheta \left[{\textstyle {\alpha \atop \beta}} \right] (z|\tau)
        =	
	4 \pi i
	\frac{\partial}{\partial \tau}
        \vartheta \left[{\textstyle {\alpha \atop \beta}} \right] (z|\tau)			\:.
        	\label{Heateq}
\ee

For $\alpha, \beta = 0, {1 \over 2}$, this function satisfies the Riemann identity \cite{Mumford}
\be	
	\sum_{\alpha, \beta = 0, {1 \over 2}} (-1)^{2\alpha + 2\beta}
        \vartheta \left[{\textstyle {\alpha \atop \beta}} \right] (x)
         \vartheta \left[{\textstyle {\alpha \atop \beta}} \right] (y)
          \vartheta \left[{\textstyle {\alpha \atop \beta}} \right] (u)
           \vartheta \left[{\textstyle {\alpha \atop \beta}} \right] (v)
	=
        2 \vartheta \left[{\textstyle {{1 \over 2} \atop {1 \over 2}}} \right] (x_1)
           \vartheta \left[{\textstyle {{1 \over 2} \atop {1 \over 2}}} \right] (y_1)
            \vartheta \left[{\textstyle {{1 \over 2} \atop {1 \over 2}}} \right] (u_1)
             \vartheta \left[{\textstyle {{1 \over 2} \atop {1 \over 2}}} \right] (v_1)
		\label{Riemannid}
\ee
 and also
\ba	
	&\:&
	\vartheta \left[{\textstyle {0 \atop 0}} \right] \left({ x \left|{ \tau }\right.}\right)
        \vartheta \left[{\textstyle {0 \atop 0}} \right] \left({ y \left|{ \tau }\right.}\right)
        \vartheta \left[{\textstyle {0 \atop {1 \over 2}}} \right] \left({ u \left|{ \tau }\right.}\right)
        \vartheta \left[{\textstyle {0 \atop {1 \over 2}}} \right] \left({ v \left|{ \tau }\right.}\right)			\nn
        &\:&	\qq
        -
        \vartheta \left[{\textstyle {0 \atop {1 \over 2}}} \right] \left({ x \left|{ \tau }\right.}\right)
        \vartheta \left[{\textstyle {0 \atop {1 \over 2}}} \right] \left({ y \left|{ \tau }\right.}\right)
        \vartheta \left[{\textstyle {0 \atop 0}} \right] \left({ u \left|{ \tau }\right.}\right)
        \vartheta \left[{\textstyle {0 \atop 0}} \right] \left({ v \left|{ \tau }\right.}\right)				\nn
        &\:&	\qq\qq
        -
        \vartheta \left[{\textstyle {{1 \over 2} \atop 0}} \right] \left({ x \left|{ \tau }\right.}\right)
        \vartheta \left[{\textstyle {{1 \over 2} \atop 0}} \right] \left({ y \left|{ \tau }\right.}\right)
        \vartheta \left[{\textstyle {{1 \over 2} \atop {1 \over 2}}} \right] \left({ u \left|{ \tau }\right.}\right)
        \vartheta \left[{\textstyle {{1 \over 2} \atop {1 \over 2}}} \right] \left({ v \left|{ \tau }\right.}\right)		\nn
        &\:&	\qq\qq\qq
        +
        \vartheta \left[{\textstyle {{1 \over 2} \atop {1 \over 2}}} \right] \left({ x \left|{ \tau }\right.}\right)
        \vartheta \left[{\textstyle {{1 \over 2} \atop {1 \over 2}}} \right] \left({ y \left|{ \tau }\right.}\right)
        \vartheta \left[{\textstyle {{1 \over 2} \atop 0}} \right] \left({ u \left|{ \tau }\right.}\right)
        \vartheta \left[{\textstyle {{1 \over 2} \atop 0}} \right] \left({ v \left|{ \tau }\right.}\right)			\nn
        &\:&
        =
        -2
        \vartheta \left[{\textstyle {{1 \over 2} \atop 0}} \right] \left({ x_1 \left|{ \tau }\right.}\right)
        \vartheta \left[{\textstyle {{1 \over 2} \atop 0}} \right] \left({ y_1 \left|{ \tau }\right.}\right)
        \vartheta \left[{\textstyle {{1 \over 2} \atop {1 \over 2}}} \right] \left({ u_1 \left|{ \tau }\right.}\right)
        \vartheta \left[{\textstyle {{1 \over 2} \atop {1 \over 2}}} \right] \left({ v_1 \left|{ \tau }\right.}\right)			\:,
        	\label{MumfordI:p.20,R6:+--+}
\ea
  where 
\ba	
	x_1 &=& {1 \over 2} (x + y + u + v)	\:,	\:\:
        y_1  =  {1 \over 2} (x + y - u - v)	\:,	\nn
        u_1 &=& {1 \over 2} (x - y + u - v)	\:,	\:\:
        v_1  =  {1 \over 2} (x - y - u + v)					\:.
\ea

\subsection{Dedekind eta function}

\be	
	\eta (\tau)
        =
        q^{\frac{1}{24}} \prod_{n=1}^\infty (1 - q^n)			\:.
        	\label{DedekintEta}
\ee
Using this,
\be	
	\vartheta' \left[{\textstyle {{1 \over 2} \atop {1 \over 2}}} \right] (0|\tau)
        =
        - 2 \pi \{ \eta (\tau) \}^3		\;.
        	\label{Theta'[1/2,1/2](0|tau)}
\ee

\subsection{Ramanujan's ${}_1\psi_1$ summation formula}

The Ramanujan's summation formula \cite{Ramanujan, Ramanujan2} is
\be	
	\sum_{n = - \infty}^\infty
        \frac{(a;q)_n}{(b;q)_n}
        z^n
        =
        \frac{(az;q)_\infty (q;q)_\infty (\frac{q}{az};q)_\infty (\frac{b}{a};q)_\infty}
             {(z;q)_\infty (b;q)_\infty (\frac{b}{az};q)_\infty (\frac{q}{a};q)_\infty}			\;,
        	\label{RamanajanSumFormula}
\ee
 with $|\frac{b}{a}| < |z| < 1$, $|q| < 1$.
Where, for $a, q \in {\bf{C}}$, $|q|<1$, $n \in {\bf{Z}}$,
\be	
	(a;q)_{\infty}
         \equiv
         \prod_{k=0}^\infty (1 - a q^k)			\:, \qq
        (a;q)_n
         \equiv
         \frac{(a;q)_\infty}{(a q^n;q)_\infty}
        	\label{qPochhammerSymbol}
\ee
 is $q$-Pochhammer symbol.

From eq. (\ref{RamanajanSumFormula}), one can derive
\be	
	\sum_{n=-\infty}^\infty
        \frac{z^n}{1 - a q^n}
        =
        \frac{(az;q)_\infty (\frac{q}{az};q)_\infty (q;q)_\infty^2}
             {(a;q)_\infty (z;q)_\infty (\frac{q}{z};q)_\infty (\frac{q}{a};q)_\infty}
        	\label{Ram:b=aq_divby1-a}
\ee
 with $|q| < |z| < 1$.
 
Proof: substituting $b=aq$ into eq. (\ref{RamanajanSumFormula}),
 the left hand side is
\ba	
        \frac{(a;q)_n}{(b;q)_n}
        &{\overset{b = aq}{=}}&
        \frac{(a;q)_n}{(aq;q)_n}
        =	%
        \frac{\displaystyle\prod_{k=0}^{n-1} (1 - a q^k)}{\displaystyle\prod_{k=0}^{n-1} (1 - a q^{k+1})}		\nn
        &=&	
        \frac{(1 - a) (1 - a q) (1 - a q^2) \cdots (1 - a q^{n-2}) (1 -a q^{n-1})}
             {(1 - a q) (1 - a q^2) \cdots (1 - a q^{n-2}) (1 - a q^{n-1}) (1 - a q^n)}			\nn
        &=&	
        \frac{1 - a}{1 - a q^n}
\ea
\be	
	\therefore		\qq
        \frac{1}{1 - a} \frac{(a;q)_n}{(aq;q)_n}
        =
        \frac{1}{1 - a q^n}				\;.
\ee
The right hand side is
\be	
	\frac{1}{1 - a}
        \frac{(az;q)_\infty (q;q)_\infty (\frac{q}{az};q)_\infty (\frac{b}{a};q)_\infty}
             {(z;q)_\infty (b;q)_\infty (\frac{b}{az};q)_\infty (\frac{q}{a};q)_\infty}
        {\overset{b=aq}{=}}	
        \frac{(az;q)_\infty (\frac{q}{az};q)_\infty (q;q)_\infty^2}
             {(z;q)_\infty (a;q)_\infty (\frac{q}{z};q)_\infty (\frac{q}{a};q)_\infty}			\;.
\ee
Therefore, substrituting $b=aq$ and dividing eq. (\ref{RamanajanSumFormula}) by $1 - a$ on the both sides,
 we obtain eq. (\ref{Ram:b=aq_divby1-a}). $\Box$

Using eq. (\ref{Ram:b=aq_divby1-a}), we find
\be	
	\sum_{n=-\infty}^\infty
        \frac{(\ex^{2 \pi i z})^n}{1 - (\ex^{- 2 \pi i \beta} q^\alpha) q^n}
        =
        \frac{i \ex^{- 2 \pi i \alpha z} }{2 \pi}
        \frac{
        	\vartheta \left[{\textstyle {\alpha - {1 \over 2}  \atop  {1 \over 2}- \beta}} \right] (z|\tau)
             }
             {
             	\vartheta \left[{\textstyle {\alpha - {1 \over 2}  \atop  {1 \over 2}- \beta}} \right] (0|\tau)
             }
        \frac{
        	\vartheta' \left[{\textstyle {{1 \over 2} \atop  {1 \over 2}}} \right] (0|\tau)
             }
             {
             	\vartheta \left[{\textstyle {{1 \over 2}  \atop  {1 \over 2}}} \right] (z|\tau)
             }								\;.
        	\label{b=aq,/1-aToTheta}
\ee

Proof:

With ${\zeta} = \ex^{2 \pi i z}$ and $a = \ex^{-2 \pi i \beta} q^{\alpha}$,
\ba	
	&\:&
	\sum_{n=-\infty}^\infty
        \frac{{\zeta}^n}{1 - a q^n}					\nn
        &\:&	\qq
        {\overset{{\rm eq.} (\ref{Ram:b=aq_divby1-a})}{=}}	
        \frac{(a{\zeta};q)_\infty (\frac{q}{a{\zeta}};q)_\infty (q;q)_\infty^2}
             {(a;q)_\infty ({\zeta};q)_\infty (\frac{q}{\zeta};q)_\infty (\frac{q}{a};q)_\infty}				\nn
        &\:&	\qq
        =	
        \frac{
             	\prod_{m=0}^\infty (1 + \ex^{2 \pi i \{z + ({1 \over 2} - \beta)\}} q^{m+\alpha}) (1 + \ex^{- 2 \pi i \{z + ({1 \over 2} - \beta)\}} q^{m-\alpha+1}) \prod_{m=1}^\infty (1 - q^{m})^2
             }
             {
             	\prod_{m=0}^\infty (1 + \ex^{2 \pi i \{0 + ({1 \over 2} - \beta)\}} q^{m+\alpha}) (1 + \ex^{2 \pi i (z + {1 \over 2})} q^m) (1 + \ex^{- 2 \pi i (z + {1 \over 2})} q^{m+1}) (1 + \ex^{2 \pi i \{0 + ({1 \over 2} - \beta)\}} q^{m-\alpha+1})
             }							\nn
        &\:&	\qq
        =	
        \frac{
        	\ex^{2 \pi i (\alpha - {1 \over 2}) (0 + ({1 \over 2} - \beta))} q^{\frac{(\alpha - {1 \over 2})^2}{2}} \prod_{m=1}^\infty (1 - q^m)
             }
             {
             	\ex^{2 \pi i (\alpha - {1 \over 2}) (z + ({1 \over 2} - \beta))} q^{\frac{(\alpha - {1 \over 2})^2}{2}} \prod_{m=1}^\infty (1 - q^m)
             }					\nn
        &\:&	\qq
        \times	%
        \frac{
        	\ex^{2 \pi i (\alpha - {1 \over 2}) (z + ({1 \over 2} - \beta))} q^{\frac{(\alpha - {1 \over 2})^2}{2}}
             }
             {
             	\ex^{2 \pi i (\alpha - {1 \over 2}) (0 + ({1 \over 2} - \beta))} q^{\frac{(\alpha - {1 \over 2})^2}{2}}
             }					\nn
        &\:&	\qq\qq
        \times
        \frac{
        	\prod_{m=1}^\infty (1 - q^m) (1 + \ex^{2 \pi i \{z + ({1 \over 2} - \beta)\}} q^{m + (\alpha - {1 \over 2}) - {1 \over 2}}) (1 + \ex^{- 2 \pi i \{z + ({1 \over 2} - \beta)\}} q^{m - (\alpha - {1 \over 2}) - {1 \over 2}})
             }
             {
             	\prod_{m=1}^\infty (1 - q^m) (1 + \ex^{2 \pi i \{0 + ({1 \over 2} - \beta)\}} q^{m + (\alpha - {1 \over 2}) - {1 \over 2}}) (1 + \ex^{2 \pi i \{0 + ({1 \over 2} - \beta)\}} q^{m - (\alpha - {1 \over 2}) - {1 \over 2}})
             }
        				\nn
        &\:&	\qq
        \times	%
        \frac{
        	\ex^{2 \pi i (- {1 \over 2}) (z + {1 \over 2})} q^{\frac{(-{1 \over 2})^2}{2}} \prod_{m=1}^\infty (1 - q^m)
             }
             {
             	(- 2 \pi) q^{\frac{3}{24}}
                \prod_{m=1}^\infty (1 - q^{m})
             }					\nn
        &\:&	\qq
        \times	%
        \frac{
        	(- 2 \pi)
        	\left\{{q^{\frac{1}{24}} \prod_{m=1}^\infty (1 - q^{m})}\right\}^3
             }
             {
             	\ex^{2 \pi i (- {1 \over 2}) (z + {1 \over 2})} q^{\frac{(- {1 \over 2})^2}{2}}
             	\prod_{m=1}^\infty (1 - q^m) (1 + \ex^{2 \pi i (z + {1 \over 2})} q^{m + (-{1 \over 2}) - {1 \over 2}}) (1 + \ex^{- 2 \pi i (z + {1 \over 2})} q^{m - (-{1 \over 2}) - {1 \over 2} })
             }					\nn
        &\:&	\qq
        {\overset{{\rm eqs.} (\ref{Def.ofThetafunction}), (\ref{DedekintEta})}{=}}	
        \frac{\ex^{- 2 \pi i (\alpha -{1 \over 2}) z} \ex^{- \pi i (z + {1 \over 2})} }{(- 2 \pi)}
        \frac{
        	\vartheta \left[{\textstyle {\alpha - {1 \over 2} \atop  {1 \over 2}- \beta}} \right] (z|\tau)
             }
             {
             	\vartheta \left[{\textstyle {\alpha - {1 \over 2} \atop  {1 \over 2}- \beta}} \right] (0|\tau)
             }
        \frac{
        	(- 2 \pi) \{ \eta (\tau) \}^3
             }
             {
             	\vartheta \left[{\textstyle {-{1 \over 2} \atop  +{1 \over 2}}} \right] (z|\tau)
             }					\nn
	&\:&	\qq
        {\overset{{\rm eqs.} (\ref{alpha+1_beta+1}), (\ref{Theta'[1/2,1/2](0|tau)})}{=}}	
        \frac{i \ex^{- 2 \pi i \alpha z} }{2 \pi}
        \frac{
        	\vartheta \left[{\textstyle {\alpha - {1 \over 2}  \atop  {1 \over 2}- \beta}} \right] (z|\tau)
             }
             {
             	\vartheta \left[{\textstyle {\alpha - {1 \over 2}  \atop  {1 \over 2}- \beta}} \right] (0|\tau)
             }
        \frac{
        	\vartheta' \left[{\textstyle {{1 \over 2} \atop  {1 \over 2}}} \right] (0|\tau)
             }
             {
             	\vartheta \left[{\textstyle {{1 \over 2}  \atop  {1 \over 2}}} \right] (z|\tau)
             }								\;.	\qq \Box
\ea

\subsection{Zeta function}

The zeta function is defined by
\be
	\zeta (z)
        =
        \sum_{n=1}^\infty
        \frac{1}{n^z}					\:.
        	\label{ZetaFunc}
\ee
For example,
\be
	\zeta (0) = - {1 \over 2}		\:,	\qq
        \zeta (2) = \frac{\pi^2}{6}		\:,	\qq
        \zeta (4) = \frac{\pi^4}{90}		\:.
        	\label{ZetaFunc:0,2,4}
\ee

In addition, the generalized zeta function is defined by
\be
	\zeta (z, a)
        =
        \sum_{n=0}^\infty
        \frac{1}{(a + n)^z}						\qq
        [a: \: {\rm const.}, \: {\rm Re}z > 1]			\:.
        	\label{GeneralizedZeta}
\ee
This function satisfies
\be
	\zeta (z, 1) = \zeta (z)	\:,	\qq
        \zeta \left({ z, {1 \over 2} }\right) = (2^z - 1) \zeta (z)			\:.
        	\label{GeneralizedZeta:1,1/2}
\ee

\section{$G_{++}$ and $G_{+-}$}  

\subsection{$G_{++} (z, {\bar{z}}|0, 0) = G \left[{\textstyle {0 \atop 0}} \right] (z, {\bar{z}}|0, 0)$	\label{app:G++}}

\ba	
        G_{++} (z, {\bar{z}}|0, 0)
        &\equiv&	
	G \left[{\textstyle {0 \atop 0}} \right] (z, {\bar{z}}|0, 0)			\nn
        &=&	
        \frac{1}{\tau_2}
        \sum_{\begin{subarray}{c}n_1, n_2 = - \infty \\ (n_1, n_2) \neq (0,0) \end{subarray}}^{\infty}
        \frac{1}{\frac{4 (2 \pi)^2}{(\tau - {\bar \tau})^2} |n_2 - n_1 \tau|^2}
        \ex^{2 \pi i n_1 \sigma^1} \ex^{2 \pi i n_2 \sigma^2}				\nn
        &=&	
        \frac{1}{\tau_2}
        \sum_{\begin{subarray}{c}n_1, n_2 = - \infty \\ n_1 \neq 0 \end{subarray}}^{\infty}
        \frac{1}{\frac{4 (2 \pi)^2}{(\tau - {\bar \tau})^2} |n_2 - n_1 \tau|^2}
        \ex^{2 \pi i n_1 \sigma^1} \ex^{2 \pi i n_2 \sigma^2}
        +
        \frac{1}{\tau_2}
        \sum_{\begin{subarray}{c} n_2 = - \infty \\ n_2 \neq 0 \end{subarray}}^{\infty}
        \frac{1}{\frac{4 (2 \pi)^2}{(\tau - {\bar \tau})^2} |n_2|^2}
        \ex^{2 \pi i n_2 \sigma^2}						\:.		\nn
        	\label{G_++}
\ea
Using
\be	
	\sum_{n=1}^\infty
        {1 \over n} \frac{x^n}{1 - q^n}
        =	
        - \sum_{m=0}^\infty \ln (1 - x q^m)			\:,
        	\label{GSWformula}
\ee
 the first term in the last line of eq. (\ref{G_++}) can be computed as
\ba	
	&\:&
	\frac{1}{\tau_2}
        \sum_{\begin{subarray}{c}n_1, n_2 = - \infty \\ n_1 \neq 0 \end{subarray}}^{\infty}
        \frac{1}{\frac{4 (2 \pi)^2}{(\tau - {\bar \tau})^2} |n_2 - n_1 \tau|^2}
        \ex^{2 \pi i n_1 \sigma^1} \ex^{2 \pi i n_2 \sigma^2}				\nn
        &\:&
        =	
        \frac{1}{\tau_2}
        \sum_{\begin{subarray}{c}n_1, n_2 = - \infty \\ n_1 \neq 0 \end{subarray}}^{\infty}
        \frac{\tau - {\bar{\tau}}}{4 (2 \pi)^2}
        \frac{1}{n_1}
        \left\{{
        	\frac{1}{n_2 - n_1 \tau}
                 - \frac{1}{n_2 - n_1 {\bar{\tau}}}
        }\right\}
        \ex^{2 \pi i n_1 \sigma^1} \ex^{2 \pi i n_2 \sigma^2}				\nn
        &\:&
        =	
        \frac{1}{\tau_2}
        \sum_{\begin{subarray}{c}n_1, n_2 = - \infty \\ n_1 \neq 0 \end{subarray}}^{\infty}
        \frac{i(\tau - {\bar{\tau}})}{4 (2 \pi)}
        \frac{1}{n_1}
        \left[{
        	\frac{
                     	\int_0^1 \de \sigma
        		\ex^{- 2 \pi i (n_2 - n_1 \tau) \sigma}
                     }
                     {
                     	1 -  q^{n_1}
                     }
        -
        	\frac{
                     	\int_0^1 \de \sigma
        		\ex^{- 2 \pi i (n_2 - n_1 {\bar{\tau}}) \sigma}
                     }
                     {
                     	1 - {\bar{q}}^{-n_1}
                     }
        }\right]
        \ex^{2 \pi i n_1 \sigma^1} \ex^{2 \pi i n_2 \sigma^2}						\nn
        &\:&
        =	
        \frac{-2}{4 (2 \pi)}
        \sum_{\begin{subarray}{c}n_1 = - \infty \\ n_1 \neq 0 \end{subarray}}^{\infty}
        \frac{1}{n_1}
        \left[{
        	\frac{
                     	\int_0^1 \de \sigma
                        \delta (\sigma^2 - \sigma)
                        \ex^{2 \pi i n_1 (\sigma^1 + \tau \sigma)}
                     }
                     {
                     	1 -  q^{n_1}
                     }
        -
        	\frac{
                     	\int_0^1 \de \sigma
                        \delta (\sigma^2 - \sigma)
                        \ex^{2 \pi i n_1 (\sigma^1 + {\bar{\tau}} \sigma)}
                     }
                     {
                     	1 - {\bar{q}}^{-n_1}
                     }
        }\right]						\nn
        &\:&
        =	
        \frac{-2}{4 (2 \pi)}
        \sum_{\begin{subarray}{c}n_1 = - \infty \\ n_1 \neq 0 \end{subarray}}^{\infty}
        \left[{
        	\frac{1}{n_1}
        	\frac{
                        \zeta^{n_1}
                     }
                     {
                     	1 -  q^{n_1}
                     }
        -
        	\frac{1}{n_1}
        	\frac{
                        {\bar \zeta}^{- n_1}
                     }
                     {
                     	1 - {\bar{q}}^{-n_1}
                     }
        }\right]						\nn
        &\:&
        =	
        \frac{-2}{4 (2 \pi)}
        \sum_{n_1 = 1}^{\infty}
        \left[{
        	\frac{1}{n_1}
        	\frac{
                        \zeta^{n_1}
                     }
                     {
                     	1 -  q^{n_1}
                     }
        +
        	\frac{1}{n_1}
        	\frac{
                        \left({ \frac{\bar q}{\bar \zeta} }\right)^{n_1}
                     }
                     {
                     	1 - {\bar{q}}^{n_1}
                     }
        +
        	\frac{1}{n_1}
        	\frac{
                        \left({ \frac{q}{\zeta} }\right)^{n_1}
                     }
                     {
                     	1 -  q^{n_1}
                     }
        +
        	\frac{1}{n_1}
        	\frac{
                        {\bar \zeta}^{n_1}
                     }
                     {
                     	1 - {\bar{q}}^{n_1}
                     }
        }\right]						\nn
        &\:&
        {\overset{{\rm eq.} (\ref{GSWformula})}{=}}	
        \frac{-2}{4 (2 \pi)}
        \sum_{m = 0}^{\infty}
        \left[{
        	- \ln \left({ 1 - \zeta q^m }\right)
                - \ln \left({ 1 - \frac{\bar q}{\bar \zeta} {\bar q}^m }\right)
                - \ln \left({ 1 - \frac{q}{\zeta} {\bar q}^m }\right)
                - \ln \left({ 1 - {\bar \zeta} {\bar q}^m }\right)
        }\right]						\nn
        &\:&
        =	
        \frac{2}{4 (2 \pi)}
        \left[{
        	  \ln \left|{ \zeta - 1 }\right|^2
                + \sum_{m = 1}^{\infty}
        	  \ln \left|{ 1 - \zeta q^m }\right|^2
                      \left|{ 1 - \frac{q^m}{\zeta} }\right|^2
        }\right]						\:.
        	\label{continue:n1/=0}
\ea
Next, turning to the second term,
\ba	
	\frac{1}{\tau_2}
        \sum_{\begin{subarray}{c} n_2 = - \infty \\ n_2 \neq 0 \end{subarray}}^{\infty}
        \frac{1}{\frac{4 (2 \pi)^2}{(\tau - {\bar \tau})^2} |n_2|^2}
        \ex^{2 \pi i n_2 \sigma^2}
        &=&	
        \frac{1}{\tau_2}
        \frac{(\tau - {\bar \tau})^2}{4 (2 \pi)^2}
        \left({
        	\sum_{n_2 = 1}^{\infty}
        	\frac{1}{n_2^2}
                \ex^{2 \pi i n_2 \sigma^2}
                +
                \sum_{n_2 = 1}^{\infty}
        	\frac{1}{n_2^2}
                \ex^{- 2 \pi i n_2 \sigma^2}
        }\right)						\nn
        &\equiv&	
        \frac{1}{\tau_2}
        \frac{(\tau - {\bar \tau})^2}{4 (2 \pi)^2}
        F (\sigma^2)					\:.
\ea
Now
\ba	
	\frac{1}{(2 \pi i)^2}
        \frac{\de^2 F (\sigma^2)}{\de (\sigma^2)^2}
        &=&	
        \sum_{n_2 = 1}^{\infty}
        \ex^{2 \pi i n_2 \sigma^2}
        +
        \sum_{n_2 = 1}^{\infty}
        \ex^{- 2 \pi i n_2 \sigma^2}
        =	
        \frac{\ex^{2 \pi i \sigma^2 - \varepsilon_{+}}}{1 - \ex^{2 \pi i \sigma^2 - \varepsilon_{+}}}
        -
        \frac{1}{1 - \ex^{2 \pi i \sigma^2 - \varepsilon_{-}}}			\nn
        &=&	
        -1	\qq	{\rm modulo} \:\: \delta (\sigma^2)			\:.
\ea
\be
	\therefore	\qq
	F (\sigma^2)
        =
        (2 \pi i)^2
        \left({
        	- {1 \over 2} (\sigma^2)^2 + A (\sigma^2) + B
        }\right)							\:.
\ee
Using
\be
	A
        =	
        F' (0)
        =
        0					\:,	\qq
	B
        =	
        F (0)
        =	%
        2
        \sum_{n_2 = 1}^{\infty}
        \frac{1}{n_2^2}
        =	%
        2 \zeta (2)
        =	%
        2 \cdot \frac{\pi^2}{6}			\:,
\ee
we obtain
\be	
	\frac{1}{\tau_2}
        \sum_{\begin{subarray}{c} n_2 = - \infty \\ n_2 \neq 0 \end{subarray}}^{\infty}
        \frac{1}{\frac{4 (2 \pi)^2}{(\tau - {\bar \tau})^2} |n_2|^2}
        \ex^{2 \pi i n_2 \sigma^2}
        =	
        \frac{1}{\tau_2}
        \frac{(\tau - {\bar \tau})^2}{4 (2 \pi)^2}
        (2 \pi i)^2
        \left({
        	- {1 \over 2} (\sigma^2)^2 + 2 \cdot \frac{\pi^2}{6}
        }\right)
        =	
        -
        {1 \over 2}
        \frac{\left({ {\rm Im} z }\right)^2}{\tau_2}
        +
        2\tau_2 \cdot \frac{\pi^2}{6}				\:.
        	\label{continue:n1=0}
\ee
As a result of eqs. (\ref{G_++}), (\ref{continue:n1/=0}) and (\ref{continue:n1=0}),
\be	
        G_{++} (z, {\bar{z}}|0, 0)
        =	
        \frac{2}{4 (2 \pi)}
        \left[{
        	  \ln \left|{ \zeta - 1 }\right|^2
                + \sum_{m = 1}^{\infty}
        	  \ln \left|{ 1 - \zeta q^m }\right|^2
                      \left|{ 1 - \frac{q^m}{\zeta} }\right|^2
        }\right]
        -
        {1 \over 2}
        \frac{\left({ {\rm Im} z }\right)^2}{\tau_2}
        +
        2\tau_2 \cdot \frac{\pi^2}{6}					\:.
        	\label{summary:G_++}
\ee
Due to
\ba	
	&\:&
	\prod_{m=1}^\infty
        \left({ 1 - \zeta q^m }\right)
        \left({ 1 - \frac{q^m}{\zeta} }\right)					\nn
        &\:&
        =	
        \frac{
        	i q^{\frac{1}{8}}
        	\ex^{\pi i z} (1 - \ex^{- 2 \pi i z})
        	\prod_{n=1}^\infty
        	 (1 - q^n)
             }
             {
             	i q^{\frac{1}{8}}
        	\ex^{\pi i z} (1 - \ex^{- 2 \pi i z})
        	\prod_{n=1}^\infty
        	 (1 - q^n)
             }
        \frac{
        	(- 2 \pi)
        	\left\{{
        		q^{\frac{1}{24}}
                	\prod_{n=1}^\infty (1 - q^n)
        	}\right\}^3
             }
             {
             	(- 2 \pi)
        	\left\{{
        		q^{\frac{1}{24}}
                	\prod_{n=1}^\infty (1 - q^n)
        	}\right\}^3
             }
        \prod_{m=1}^\infty
        \left({ 1 - \ex^{2 \pi i z} q^m }\right)
        \left({ 1 - \ex^{- 2 \pi i z} q^m }\right)					\nn
        &\:&
        {\overset{{\rm eqs.} (\ref{DedekintEta}), (\ref{Theta'[1/2,1/2](0|tau)})}{=}}	
        \frac{
        	(2 \pi i)
                \prod_{n=1}^\infty (1 - q^n)^2
             }
             {
        	\ex^{\pi i z} (1 - \ex^{- 2 \pi i z})
             }
        \frac{\vartheta \left[{\textstyle {{1 \over 2} \atop {1 \over 2}}} \right] (z)}
             {\vartheta' \left[{\textstyle {{1 \over 2} \atop {1 \over 2}}} \right] (0)}			\:,
        	\label{prod&theta[1/2,1/2]}
\ea
 the terms in the box brackets of eq. (\ref{summary:G_++}) can be recast as follows:
\ba	
	&\:&
	\ln \left|{ \zeta - 1 }\right|^2
        +
        \sum_{m = 1}^{\infty}
        \ln \left|{ 1 - \zeta q^m }\right|^2
        \left|{ 1 - \frac{q^m}{\zeta} }\right|^2			\nn
        &\:&
        {\overset{{\rm eq.} (\ref{prod&theta[1/2,1/2]})}{=}}	
        \ln{
        	\left|{ \ex^{2 \pi i z} - 1 }\right|^2
                \left|{
        		\frac{
        			(2 \pi i)
                		\prod_{n=1}^\infty (1 - q^n)^2
             		     }
             		     {
        			\ex^{\pi i z} (1 - \ex^{- 2 \pi i z})
             		     }
        		\frac{\vartheta \left[{\textstyle {{1 \over 2} \atop {1 \over 2}}} \right] (z)}
             		     {\vartheta' \left[{\textstyle {{1 \over 2} \atop {1 \over 2}}} \right] (0)}
                }\right|^2
           }								\nn
        &\:&
        =	
        \ln \left|{
        		\frac{\vartheta \left[{\textstyle {{1 \over 2} \atop {1 \over 2}}} \right] (z)}
             		     {\vartheta' \left[{\textstyle {{1 \over 2} \atop {1 \over 2}}} \right] (0)}
            	  }\right|^2
        +
        2
        \sum_{n=1}^\infty
        \ln |1 - q^n|^2
        -
        2 \pi ({\rm Im} \: z)
        +
        2
        \ln (2 \pi)						\:.
        	\label{connecttotheta}
\ea
Hence we obtain
\ba	
	&\:&
        G_{++} (z, {\bar{z}}|0, 0)							\nn
        &\:&
        {\overset{{\rm eqs.} (\ref{summary:G_++}), (\ref{connecttotheta})}{=}}	
        \frac{1}{2 \pi}
        \ln \left|{
        		\frac{\vartheta \left[{\textstyle {{1 \over 2} \atop {1 \over 2}}} \right] (z)}
             		     {\vartheta' \left[{\textstyle {{1 \over 2} \atop {1 \over 2}}} \right] (0)}
            	  }\right|
        -
        {1 \over 2}
        \frac{\left({ {\rm Im} \: z }\right)^2}{\tau_2}						\nn
        &\:&	\qq\qq\qq	%
        +
        \left[{
        	\frac{1}{2 \pi}
                2
        	\sum_{n=1}^\infty
        	\ln |1 - q^n|
                -
                \frac{1}{2}
                ({\rm Im} \: z)
                +
                \frac{1}{2 \pi}
                \ln (2 \pi)
                +
                2 \tau_2 \cdot \frac{\pi^2}{6}
        }\right]						\:.
        	\label{summary:G_++calc.App}
\ea
The terms in the brackets on the last line vanish when acting $\Delta = 4 \partial_{z} \partial_{\bar z}$.

\subsection{$G_{+-} (z, {\bar{z}}|0, 0) = G \left[{\textstyle {0 \atop {1 \over 2}}} \right] (z, {\bar{z}}|0, 0)$	\label{app:G+-}}

\ba
	&\:&
        G_{+-} (z, {\bar{z}}|0, 0)								\nn
        &\:&
        \equiv	
	G \left[{\textstyle {0 \atop {1 \over 2}}} \right] (z, {\bar{z}}|0, 0)			\nn
        &\:&
        =	
        \frac{1}{\tau_2}
        \sum_{n_1, n_2 = - \infty}^{\infty}
        \frac{1}{\frac{4 (2 \pi)^2}{(\tau - {\bar \tau})^2} \left|{ \left({ n_2 + {1 \over 2} }\right) - n_1 \tau }\right|^2}
        \ex^{2 \pi i n_1 \sigma^1} \ex^{2 \pi i \left({ n_2 + {1 \over 2} }\right) \sigma^2}				\nn
        &\:&
        =	
        \frac{1}{\tau_2}
        \sum_{\begin{subarray}{c}n_1, n_2 = - \infty \\ n_1 \neq 0 \end{subarray}}^{\infty}
        \frac{1}{\frac{4 (2 \pi)^2}{(\tau - {\bar \tau})^2} \left|{ \left({ n_2 + {1 \over 2} }\right) - n_1 \tau }\right|^2}
        \ex^{2 \pi i n_1 \sigma^1} \ex^{2 \pi i \left({ n_2 + {1 \over 2} }\right) \sigma^2}
        +
        \frac{1}{\tau_2}
        \sum_{n_2 = - \infty}^{\infty}
        \frac{1}{\frac{4 (2 \pi)^2}{(\tau - {\bar \tau})^2} \left|{ n_2 + {1 \over 2} }\right|^2}
        \ex^{2 \pi i \left({ n_2 + {1 \over 2} }\right) \sigma^2}			\:.			\nn
        	\label{G_{+-}:def}
\ea

The first term in the last line of eq. (\ref{G_{+-}:def}) is
\ba	
	&\:&
        \frac{1}{\tau_2}
        \sum_{\begin{subarray}{c}n_1, n_2 = - \infty \\ n_1 \neq 0 \end{subarray}}^{\infty}
        \frac{1}{\frac{4 (2 \pi)^2}{(\tau - {\bar \tau})^2} \left|{ \left({ n_2 + {1 \over 2} }\right) - n_1 \tau }\right|^2}
        \ex^{2 \pi i n_1 \sigma^1} \ex^{2 \pi i \left({ n_2 + {1 \over 2} }\right) \sigma^2}						\nn
        &\:&
        =	
        \frac{1}{\tau_2}
        \sum_{\begin{subarray}{c}n_1, n_2 = - \infty \\ n_1 \neq 0 \end{subarray}}^{\infty}
        \frac{\tau - {\bar \tau}}{4 (2 \pi)^2}
        \frac{1}{n_1}
        \left\{{
        	\frac{1}{\left({ n_2 + {1 \over 2} }\right) - n_1 \tau}
                -
                \frac{1}{\left({ n_2 + {1 \over 2} }\right) - n_1 {\bar \tau}}
        }\right\}
        \ex^{2 \pi i n_1 \sigma^1} \ex^{2 \pi i \left({ n_2 + {1 \over 2} }\right) \sigma^2}						\nn
        &\:&
        =	
        \frac{1}{\tau_2}
        \sum_{\begin{subarray}{c}n_1, n_2 = - \infty \\ n_1 \neq 0 \end{subarray}}^{\infty}
        \frac{\tau - {\bar \tau}}{4 (2 \pi)^2}
        \frac{1}{n_1}
        \left[{
        	\frac{2 \pi i}
                     {
                     	1 - 
                	\ex^{- 2 \pi i {1 \over 2}}
                	q^{n_1}
                     }
                \int_0^1 \de \sigma
        	\ex^{
        		- 2 \pi i
                	\left\{{
                		\left({ n_2 + {1 \over 2} }\right) - n_1 \tau
                	}\right\}
                	 \sigma
            	    }
        }\right.										\nn
        &\:&	\qq\qq\qq\qq\qq\qq	%
        \left.{
                -
                \frac{2 \pi i}
                     {
                     	1 - 
                	\ex^{2 \pi i {1 \over 2}}
                	{\bar{q}}^{- n_1}
                     }
                \int_0^1 \de \sigma
        	\ex^{
        		- 2 \pi i
                	\left\{{
                		\left({ n_2 + {1 \over 2} }\right) - n_1 {\bar{\tau}}
                	}\right\}
                	 \sigma
            	    }
        }\right]
        \ex^{2 \pi i n_1 \sigma^1} \ex^{2 \pi i \left({ n_2 + {1 \over 2} }\right) \sigma^2}						\nn
        &\:&
        =	
        \frac{-2}{4 (2 \pi)}
        \sum_{\begin{subarray}{c}n_1 = - \infty \\ n_1 \neq 0 \end{subarray}}^{\infty}
        \frac{1}{n_1}
        \left[{
        	\frac{
                	\int_0^1 \de \sigma
        		\left({
                        	\delta (\sigma^2 - \sigma)
                        }\right)
                        \ex^{2 \pi i {1 \over 2} (\sigma^2 - \sigma)}
        		\ex^{2 \pi i n_1 (\sigma^1 + \tau \sigma)}
                     }
                     {
                     	1 - (-q^{n_1})
                     }
        }\right.										\nn
        &\:&	\qq\qq\qq\qq\qq\qq	%
        \left.{
                -
                \frac{
                	\int_0^1 \de \sigma
                        \left({
                        	\delta (\sigma^2 - \sigma)
                        }\right)
                        \ex^{2 \pi i {1 \over 2} (\sigma^2 - \sigma)}
        		\ex^{2 \pi i n_1 (\sigma^1 + {\bar \tau} \sigma)}
                     }
                     {
                     	1 - (-{\bar{q}}^{- n_1})
                     }
        }\right]										\nn
        &\:&
        =	
        \frac{-2}{4 (2 \pi)}
        \sum_{\begin{subarray}{c}n_1 = - \infty \\ n_1 \neq 0 \end{subarray}}^{\infty}
        \frac{1}{n_1}
        \left[{
        	\frac{
        		\zeta^{n_1}
                     }
                     {
                     	1 - (-q^{n_1})
                     }
                -
                \frac{
        		{\bar \zeta}^{- n_1}
                     }
                     {
                     	1 - (-{\bar{q}}^{- n_1})
                     }
        }\right]										\nn
        &\:&
        =	
        \frac{-2}{4 (2 \pi)}
        \sum_{n_1 = 1}^{\infty}
        \left[{
        	\frac{1}{n_1}
        	\frac{
        		\zeta^{n_1}
                     }
                     {
                     	1 - (-q^{n_1})
                     }
                -
                \frac{1}{n_1}
                \frac{
                	\left({
                        	\frac{\bar{q}}{\bar \zeta}
                        }\right)^{n_1}
                     }
                     {
                     	1 - (-{\bar{q}}^{n_1})
                     }
                -
                \frac{1}{n_1}
        	\frac{
                	\left({
                        	\frac{q}{\zeta}
                        }\right)^{n_1}
                     }
                     {
                     	1 - (-q^{n_1})
                     }
                +
                \frac{1}{n_1}
                \frac{
        		{\bar \zeta}^{+ n_1}
                     }
                     {
                     	1 - (-{\bar{q}}^{+ n_1})
                     }
        }\right]										\nn
        &\:&
        {\overset{{\rm eq.} (\ref{GSWformula:deformed})}{=}}	
        \frac{-2}{4 (2 \pi)}
        \sum_{m=0}^\infty
        \left[{
        	(-1)^{m+1} \ln (1 - \zeta q^m)
                -
                (-1)^{m+1} \ln \left({ 1 - \frac{\bar q}{\bar \zeta} {\bar q}^m }\right)
        }\right.										\nn
        &\:&	\qq\qq\qq\qq\qq	%
        \left.{
                -
                (-1)^{m+1} \ln \left({ 1 - \frac{q}{\zeta} q^m }\right)
                +
                (-1)^{m+1} \ln (1 - {\bar \zeta} {\bar q}^m)
        }\right]										\nn
        &\:&
        =	
        \frac{2}{4 (2 \pi)}
        \left[{
        	\ln |1 - \zeta|^2
                +
        	\sum_{m=1}^\infty
        	(-1)^{m} \ln |1 - \zeta q^m|^2 \left|{ 1 - \frac{q^{m}}{\zeta} }\right|^2
        }\right]										\:,
        	\label{G_+-:term1}
\ea
 where we have used
\ba	
	\sum_{n=1}^\infty
        {1 \over n} \frac{x^n}{1 - (-y^n)}
        &=&	
        \sum_{n=1}^\infty {1 \over n}
        \sum_{m=0}^\infty x^n (-y^{n})^m
        =	
        \sum_{m=0}^\infty (-1)^m
        \sum_{n=1}^\infty
        {1 \over n} (x y^m)^n							\nn
        &=&	
        \sum_{m=0}^\infty (-1)^m \{ - \ln (1 - x y^m)\}
        =	
        \sum_{m=0}^\infty (-1)^{m+1} \ln (1 - x y^m)			\:.
        	\label{GSWformula:deformed}
\ea

The second term in eq. (\ref{G_{+-}:def}) is
\ba	
	&\:&
	\frac{1}{\tau_2}
        \sum_{n_2 = - \infty}^{\infty}
        \frac{1}{\frac{4 (2 \pi)^2}{(\tau - {\bar \tau})^2} \left|{ n_2 + {1 \over 2} }\right|^2}
        \ex^{2 \pi i \left({ n_2 + {1 \over 2} }\right) \sigma^2}						\nn
        &\:&
        =	
        \frac{1}{\tau_2}
        \frac{(\tau - {\bar \tau})^2}{4 (2 \pi)^2}
        \left({
        	\sum_{n_2 = 0}^{\infty}
        	\frac{1}{\left({ n_2 + {1 \over 2} }\right)^2}
        	\ex^{2 \pi i \left({ n_2 + {1 \over 2} }\right) \sigma^2}
                +
                \sum_{n_2 = 1}^{\infty}
        	\frac{1}{\left({ n_2 - {1 \over 2} }\right)^2}
        	\ex^{- 2 \pi i \left({ n_2 - {1 \over 2} }\right) \sigma^2}
        }\right)											\nn
        &\:&
        \equiv 
        \frac{1}{\tau_2}
        \frac{(\tau - {\bar \tau})^2}{4 (2 \pi)^2}
        {\tilde F} (\sigma^2)							\:.
\ea
Then
\ba	
	\frac{1}{(2 \pi)^2}
        \frac{\de^2 {\tilde F} (\sigma^2)}{\de (\sigma^2)^2}
        &=&	
        \sum_{n_2 = 0}^{\infty}
        \ex^{2 \pi i \left({ n_2 + {1 \over 2} }\right) \sigma^2}
        +
        \sum_{n_2 = 1}^{\infty}
        \ex^{- 2 \pi i \left({ n_2 - {1 \over 2} }\right) \sigma^2}					\nn
        &=&	
        \ex^{2 \pi i {1 \over 2} \sigma^2}
        \left({
        	\frac{1}{1 - \ex^{2 \pi i \sigma^2 - \varepsilon_{+}}}
                +
                \frac{\ex^{- 2 \pi i \sigma^2 - \varepsilon_{-}}}{1 - \ex^{- 2 \pi i \sigma^2 - \varepsilon_{-}}}
        }\right)
        =	
        0	\qq	{\rm modulo} \:\: \delta (\sigma^2)			\:.
\ea
\be	
	\therefore	\qq
        {\tilde F} (\sigma^2)
        =	
        (2 \pi)^2
        \left({
        	A \sigma^2 + B
        }\right)
\ee
 and
\ba	
	B
        &=&	
        {\tilde F} (0)
        =	%
        \sum_{n_2 = 0}^{\infty}
        \frac{1}{\left({ n_2 + {1 \over 2} }\right)^2}
        +
        \sum_{n_2 = 1}^{\infty}
        \frac{1}{\left({ n_2 - {1 \over 2} }\right)^2}
        {\overset{{\rm eq.} (\ref{GeneralizedZeta})}{=}}	%
        \zeta \left({ 2, {1 \over 2} }\right)
        +
        \left\{{
        	\zeta \left({ 2, - {1 \over 2} }\right)
                -
                \frac{1}{\left({ 0 - {1 \over 2} }\right)^2}
        }\right\}									\nn
        &{\overset{{\rm eqs.} (\ref{ZetaFunc:0,2,4}), (\ref{GeneralizedZeta:1,1/2})}{=}}&	
        \frac{\pi^2}{2}
        +
        \left\{{
        	\left({ 4 + \frac{\pi^2}{2} }\right)
                -
                4
        }\right\}
        =	%
        2 \cdot \frac{\pi^2}{2}
        =	%
        \pi^2			\:.
\ea
Using the formula
\ba	
	\sum_{n=1}^\infty
        \frac{1}{n^2 + a^2}
        =
        - \frac{1}{2 a^2}
        + \frac{\pi}{2 a} \coth (a \pi)					\:,
        	\label{iwanamiI:p48line2}
\ea
\ba	
	A
        &=&	
        {\tilde F}' (0)
        =	%
        \sum_{n_2 = 0}^{\infty}
        \frac{2 \pi i}{n_2 + {1 \over 2}}
        +
        \sum_{n_2 = 1}^{\infty}
        \frac{- 2 \pi i}{n_2 - {1 \over 2}}
        =	%
        2 \pi i
        \left[{
        	\frac{1}{0 + {1 \over 2}}
                +
                \sum_{n_2 = 1}^{\infty}
                \left\{{
                	\frac{1}{n_2 + {1 \over 2}}
                        -
                        \frac{1}{n_2 - {1 \over 2}}
                }\right\}
        }\right]							\nn
        &=&	
        2 \pi i
        \left[{
        	2
                -
                \sum_{n_2 = 1}^{\infty}
                \frac{1}{n_2^2 + \left({ \frac{i}{2} }\right)^2}
        }\right]
        {\overset{{\rm eq.} (\ref{iwanamiI:p48line2})}{=}}	
        2 \pi i
        \left[{
        	2
                -
                \left\{{
                	- \frac{1}{2 \left({ \frac{i}{2} }\right)^2}
                        + \frac{\pi}{2 \frac{i}{2}} \coth \left({ \frac{i}{2} \pi }\right)
                }\right\}
        }\right]
        =	%
        0				\:.
\ea
\be
	\therefore	\qq
        {\tilde F} (\sigma^2)
        =	
        (2 \pi)^2 \pi^2					\:.
\ee
Therefore
\be	
	\frac{1}{\tau_2}
        \sum_{n_2 = - \infty}^{\infty}
        \frac{1}{\frac{4 (2 \pi)^2}{(\tau - {\bar \tau})^2} \left|{ n_2 + {1 \over 2} }\right|^2}
        \ex^{2 \pi i \left({ n_2 + {1 \over 2} }\right) \sigma^2}
        \equiv 
        \frac{1}{\tau_2}
        \frac{(\tau - {\bar \tau})^2}{4 (2 \pi)^2}
        {\tilde F} (\sigma^2)
        =	%
        \frac{1}{\tau_2}
        \frac{(2 i \tau_2)^2}{4 (2 \pi)^2}
        (2 \pi)^2 \pi^2
        =	
        - \pi^2 \tau_2				\:.
        	\label{continue:G_+-:term2}
\ee

Finally,
\be	
        G_{+-} (z, {\bar{z}}|0, 0)
        {\overset{{\rm eqs.} (\ref{G_+-:term1}), (\ref{continue:G_+-:term2})}{=}}	
        \frac{1}{2 \pi}
        \left[{
        	\ln |1 - \zeta|
                +
        	\sum_{m=1}^\infty
        	(-1)^{m} \ln |1 - \zeta q^m| \left|{ 1 - \frac{q^{m}}{\zeta} }\right|
        }\right]
        -
        \pi^2 \tau_2					\:.
        	\label{summary:G_+-}
\ee

\section{supertorus Green function and supersphere Green function	\label{G_STandGSS}}  

Since
\be	
	\frac{\vartheta \left[{\textstyle {{1 \over 2} \atop {1 \over 2}}} \right] (z|\tau)}
             {\vartheta' \left[{\textstyle {{1 \over 2} \atop {1 \over 2}}} \right] (0|\tau)}
        {\overset{z \sim 0}{\sim}}	
        z				\:,		\qq
        \ln \left|{ 1 - \ex^{2 \pi iz} }\right|
        {\overset{z \sim 0}{\sim}}	
        \ln |z|
		\label{Theta[1/2,1/2](z)/Theta'[1/2,1/2](0):z->0}
\ee
\be	
	G_{+ \pm} (z, {\bar{z}}|0, 0)
        {\overset{z \sim 0}{\sim}}	
        \frac{1}{2 \pi} \ln |z|
        	\label{G_{++}:z->0}
\ee
 and
\ba	
	{\cal{S}}_{\nu_{\rm f}} (z, {\bar{z}}|0, 0)
        &{\overset{z \sim 0}{\sim}}&
        \frac{i}{\pi} \frac{1}{z}			\:,	\qq
        {\overline{\cal{S}}_{\nu_{\rm f}}} (z, {\bar{z}}|0, 0)
        {\overset{{\bar{z}} \sim 0}{\sim}}
        - \frac{i}{\pi} \frac{1}{\bar z}				\:,
        	\label{S_{--},S_{-+},S_{+-}:z->0}
\ea
 where $\nu_{\rm f} = (--), (-+), (+-)$.
Using eqs. (\ref{G_{++}:z->0}) and (\ref{S_{--},S_{-+},S_{+-}:z->0}),
\ba	
	{\bf G}_{\shortstack{$+ \pm$ \\ $\nu_{\rm f}$}}^{\rm supertorus} (z_I, {\bar{z}}_I|z_J, {\bar{z}}_J)
        &\equiv&	
	G_{+ \pm} (z_I, {\bar{z}}_I|z_J, {\bar{z}}_J)
        +
        \frac{\theta_I \theta_J}{4}
        {\cal{S}}_{\nu_{\rm f}} (z_I, {\bar{z}}_I|z_J, {\bar{z}}_J)
        -
        \frac{{\bar \theta}_I {\bar \theta}_J}{4}
        {\overline{\cal{S}}_{\nu_{\rm f}}} (z_I, {\bar{z}}_I|z_J, {\bar{z}}_J)				\nn
        &{\overset{z_I \sim z_J}{\sim}}&	
        \frac{1}{2 \pi} \ln |z_I - z_J|
        +
        \frac{\theta_I \theta_J}{4}
        \left({
        	\frac{i}{\pi} \frac{1}{z_I - z_J}
        }\right)
        -
        \frac{{\bar \theta}_I {\bar \theta}_J}{4}
        \left({
        	- \frac{i}{\pi} \frac{1}{{\bar z}_I - {\bar z}_J}
        }\right)										\nn
        &=&	
        \frac{1}{4 \pi}
        \ln (z_I - z_J + i \theta_I \theta_J)
        +
        \frac{1}{4 \pi}
        \ln ({\bar z}_I - {\bar z}_J + i {\bar \theta}_I {\bar \theta}_J)			\nn
        &=&	
        \frac{1}{2 \pi}
        \ln |z_I - z_J + i \theta_I \theta_J|
        =	%
        {\bf G}^{\rm supersphere} (z_I, {\bar{z}}_I|z_J, {\bar{z}}_J)				\:.
\ea

\section{Path integral of a fermionic string at one-loop	\label{App:pathint:review}}  

 In the appendix, the path integral of a fermionic string \cite{Polyakov, Pol, CKT} is briefly recalled.
 We present formulas in the flat ten dimensional case but they can be easily adapted to the other cases such as $T^4/{\cal Z}_2$ orbifold given in the text.
 The basic variables are the bosonic coordinates ($2$d massless scalar fields) $X^\mu$,
  the fermionic ones ($2$d massless two-component Majorana spinor fields) $\psi_{\rm Maj \:}{}^\mu_\alpha$,
   zwiebein $e_a{}^m$ (or $e_m{}^a$ such that $g_{mn} = e_m{}^a e_n{}^b \delta_{ab}$) and the Rarita - Schwinger field $\chi_\alpha{}^m$,
    we denoted by $\mu, \nu, ...$, $\alpha, \beta, ...$, $m, n, ...$ and $a, b, ...$ ten dimensional vector indices, two dimensional spinor indices, two dimensional worldsheet indices,
     and two dimensional local Lorentz indices respectively.
The action \cite{DiVecchia} is
\ba
	S
        &=&	
        \frac{1}{2 \pi \alpha'} \int d^2\sigma \sqrt{g}
        \left\{{
        	{1 \over 2} g^{m n} \partial_m X^\mu \partial_n X_\mu
                  - {i \over 2} \psi_{\rm Maj \:}{}^\mu \gamma^a \nabla_a \psi_{\rm Maj \:}{}_\mu
        }\right.				\nn
        &\:&	\qq\qq\qq\qq	%
        \left.{
                - {1 \over 2}
                 \left({
                 	\psi_{\rm Maj \:}{}^\mu \gamma^a \gamma^b \chi_a
                 }\right)
                 \left({
                 	\partial_b X_\mu - {1 \over 4} \chi_b \psi_{\rm Maj \:}{}_\mu
                 }\right)
        }\right\},
\ea
 where
 \ba
 	\chi_a &=& e_a{}^m \chi_m, \;\; \partial_b = e_b{}^m \partial_m		\nn
        \nabla_a &=& e_a{}^m \left({ \partial_m - \omega_m {1 \over 2} \gamma^5 }\right)			\nn
        \omega_m &=& e_m{}^a \varepsilon^{pq} \partial_p e_q{}^b \delta_{ab}			\:.
 \ea

 The superstring scattering amplitudes are given in general by the functional integrals
  with the appropriately chosen vertex operators $\displaystyle\prod_{I} O_I$ over these worldsheet fields
   with respect to this action modulo the local symmetries
\be
	\langle \prod_I O_I \rangle
        =
        \sum_{\rm{top.}} \sum_{\rm{s.s.}}
         \int \frac{{\cal{D}} e_m{}^a}{\Omega (\rm{D}) \Omega (\rm{W}) \Omega (\rm{L})}
         \int \frac{{\cal{D}} \chi_m{}^a}{\Omega (\rm{S}) \Omega (\rm{SW})}
        \int {\cal{D}} X^\mu \int {\cal{D}}\psi_{\rm Maj \:}{}^\mu
         e^{-S} \prod_I O_I			\:.
\ee
Here we have denoted by top. and by s.s. the summation over the worldsheet topology
 and the summation over the spin structure respectively.
The functional integrals are (at least formally) defined through inner product
 and are modded out by the volumes of the two dimensional diffeomorphisms,
  the local Weyl symmetry, the local Lorentz invariance, the two dimensional local symmetry
   and the local Weyl symmetry denoted respectively by $\Omega (\rm{D})$, $\Omega (\rm{W})$, $\Omega (\rm{L})$, $\Omega (\rm{S})$ and $\Omega (\rm{SW})$.

 The functional integral measure is first decomposed by noting general variation of vierbeins
  and gravitino variables exploiting the local transformations and the orthogonal decomposition:
\be
	\delta e_m{}^a
        =
        \delta \sigma e_m{}^a + (P_1 \delta \eta)_m{}^a + \delta \ell \varepsilon^{ab} e_m{}^b + \sum_i \delta c_i \psi^i_m{}^a .
\ee
Here
 \ba
 	P_1 &:& (P_1 \delta \eta)_m{}^a
        =
        \left\{{
        	\delta \eta^n \partial_n e_m{}^a + e_n{}^a \partial_m \delta \eta^n
        }\right\}		\nn
        &\;& \qquad \qquad \qquad
        - {1 \over 2} e_m{}^a e_b{}^n
         \left\{{
         	\delta \eta^\ell \partial_\ell e_n{}^b + e_\ell{}^b \partial_n \delta \eta^\ell
         }\right\}		\nn
        &\;& \qquad \qquad \qquad
        - {1 \over 2} \varepsilon^{ab} e_m{}^b \varepsilon_{dc} e_c{}^n
         \left\{{
         	\delta \eta^\ell \partial_\ell e_n{}^d + e_\ell{}^d \partial_n \delta \eta^\ell
         }\right\}		\nn
        \psi^i &\in& {\rm{Ker}} P_1^\dag		\nonumber
 \ea
\be
	\delta \chi_m
        =
        \gamma_m \delta \rho + (P_{1/2} \delta \zeta)_m + \sum_i \delta \epsilon_i \Psi^i_m
\ee
 \ba
 	(P_{1/2} \delta \zeta)_m
        &=&
        2 \nabla_m \delta \zeta - \gamma_m \gamma^n \nabla_n \delta \zeta	\nn
        \Psi^i_m &\in& {\rm{Ker}} P_{1/2}^\dag	.	\nonumber
 \ea
Hence
\ba
	\langle \prod_I O_I \rangle
        &=&
        \int \frac{{\cal{D}} \sigma {\cal{D}}(P_1 \eta) {\cal{D}} \ell}{\Omega (\rm{D}) \Omega (\rm{W}) \Omega (\rm{L})}
        \prod_i \de c_i {\rm{det}} \langle \psi^i | \psi^j \rangle^{1 \over 2}		\nn
        &\;& \qq
        \int \frac{{\cal{D}} \rho {\cal{D}}(P_{1/2} \zeta)}{\Omega (\rm{S}) \Omega (\rm{SW})}
        \prod_i \de \epsilon_i {\rm{det}} \langle \Psi_i | \Psi_j \rangle^{- {1 \over 2}}
        \prod_I O_I .
\ea

 This expression is further converted by referring to the global gauge slice
  and moduli and supermoduli parameters
  \ba
  	e_m{}^a &=& e^\Lambda ({\rm{Diff}} {\hat{e}})_m{}^a (\tau_i)		\nn
        \chi_m &=& \gamma_m \lambda + \sum_i a_i \Phi^i{}_m .
  \ea
The vierbein variation reads
\be
	\delta e_m{}^a
        =
        \delta \Lambda e_m{}^a + \delta \eta^n \partial_n e_m{}^a + e_n{}^a \partial_m \delta \eta^n
         + \delta L \varepsilon^{ab} e_m{}^b + \sum_i \delta \tau_i \left({ \partial e_m{}^a / \partial \tau_i }\right)		\nonumber
\ee
\be
	\left[
		\begin{array}{c}
		 \de \sigma \\
		 \de \ell \\
		 P_1 \delta \eta \\
 		 \de c_i
		\end{array}
	\right]
        =
        \left[
		\begin{array}{cccc}
		 1 & 0 & * & 0 \\
		 0 & 1 & * & 0 \\
		 0 & 0 & P_1 & 0 \\
		 0 & 0 & 0 & T_{ij}
		\end{array}
	\right]
	\left[
		\begin{array}{c}
		 \delta \Lambda \\
		 \delta L \\
		 \delta \eta \\
 		 \delta T_{ij}
		\end{array}
	\right]		\nonumber
\ee
\be
	{\rm{det}} T
        =
        {\rm{det}} \langle \psi_i | \psi_j \rangle^{-1}
        {\rm{det}} \langle \psi_i | \partial e_m{}^a / \partial \tau_i \rangle	.	\nonumber
\ee
Hence
\ba
	\frac{{\cal{D}} e_m{}^a}{\Omega (\rm{D}) \Omega (\rm{W}) \Omega (\rm{L})}
        &=&
        \frac{{\cal{D}} \Lambda {\cal{D}} L {\cal{D}}' \eta}{\Omega (\rm{D}) \Omega (\rm{W}) \Omega (\rm{L})}
        \prod_i \de \tau_i {\rm{det}}' (P_1^\dag P_1)^{1 \over 2}		\nn
        &\;& \qq
        {\rm{det}} \langle \psi_i | \psi_j \rangle^{-{1 \over 2}}
        {\rm{det}} \left\langle{ \psi_i \left|{ \pardif{e_m{}^a}{\tau_i} }\right. }\right\rangle		\nn
        &=&
        \prod_i \de \tau_i \frac{1}{\Omega({\rm{CK}})}
        {\rm{det}}' (P_1^\dag P_1)^{1 \over 2}
        {\rm{det}} \langle \psi_i | \psi_j \rangle^{-{1 \over 2}}
        {\rm{det}} \left\langle{ \psi_i \left|{ \pardif{e_m{}^a}{\tau_i} }\right. }\right\rangle		\nonumber
\ea
 where $\Omega({\rm{CK}})$ is the volume of conformal killing vectors
  and ${\cal{D}}' \eta$ indicates that ${\rm{ker}}P_1$ has been excluded.
The variation of the gravitinos reads
\be
	\delta \chi_m
        =
        \gamma_m \delta \lambda + 2 \nabla_m \delta \zeta + \sum_i \de a_i \Phi_i		\nonumber
\ee
\be
	\left[
		\begin{array}{c}
		 \delta \rho \\
		 P_{1/2} \delta \zeta \\
		 \delta \epsilon_i
		\end{array}
	\right]
        =
        \left[
		\begin{array}{ccc}
		 1 & * & 0 \\
		 0 & P_{1/2} & 0 \\
		 0 & 0 & S_{ij}
		\end{array}
	\right]
	\left[
		\begin{array}{c}
		 \delta \lambda \\
		 \delta \zeta \\
		 \de a_j
		\end{array}
	\right]		\nonumber
\ee
\be
	{\rm{det}} S
        =
        {\rm{det}} \langle \Psi_i | \Psi_k \rangle^{-1}
        {\rm{det}} \langle \Psi_k | \Phi_j \rangle .		\nonumber
\ee
Hence
\ba
	\frac{{\cal{D}} \chi_m}{\Omega (\rm{S}) \Omega (\rm{SW})}
        &=&
        \frac{{\cal{D}} \lambda {\cal{D}}' \zeta}{\Omega (\rm{S}) \Omega (\rm{SW})}
        {\rm{det}}' (P_{1/2}^\dag P_{1/2})^{- {1 \over 2}}
        \prod_i \de a_i
        {\rm{det}} S^{-1}
        {\rm{det}} \langle \Psi | \Psi \rangle^{-{1 \over 2}}		\nn
        &=&
        \prod_i \de a_i \frac{1}{\Omega({\rm{CKS}})}
        {\rm{det}}' (P_{1/2}^\dag P_{1/2})^{- {1 \over 2}}
        {\rm{det}} \langle \Psi_i | \Psi_k \rangle^{1 \over 2}
        {\rm{det}} \langle \Psi_k | \Phi_j \rangle^{-1}		\nonumber
\ea
 where $\Omega({\rm{CKS}})$ is the volume of the conformal killing spinor
  and ${\cal{D}}' \zeta$ indicates that ${\rm{ker}} P_{1/2}$ has been excluded.
The final formula is
\ba	
	\langle \prod_I O_I \rangle
        &=&
        \sum_{\rm{top.}} \sum_{\rm{s.s.}}
        \int \prod_i \de \tau_i \frac{1}{\Omega({\rm{CKV}})}
        {\rm{det}}' (P_1^\dag P_1)^{1 \over 2}
        {\rm{det}} \langle \psi_i | \psi_j \rangle^{-{1 \over 2}}
        {\rm{det}} \left\langle{ \psi_i \left|{ \pardif{e_m{}^a}{\tau_i} }\right. }\right\rangle	\nn
        &\;&	\qq
        \int \prod_i \de a_i \frac{1}{\Omega({\rm{CKS}})}
        {\rm{det}}' (P_{1/2}^\dag P_{1/2})^{-{1 \over 2}}
        {\rm{det}} \langle \Psi_i | \Psi_j \rangle^{1 \over 2}
        {\rm{det}} \langle \Psi_i | \Phi_j \rangle^{-1}				\nn
        &\:&	\qq\qq
        \int {\cal{D}} X^\mu \int {\cal{D}}\psi_{\rm Maj \:}{}^\mu
         e^{-S} \prod_I O_I				\:.
        	\label{pathintformula}
\ea

\section{Superannulus	\label{App:imagemethod:SA}}  

In this appendix, we apply the method of images in superspace to superannulus \cite{ItoMox, IKK}.

Let the conjugate point of $(z, \theta)$ be $({\tilde z}, {\tilde \theta})$.
The involution acting on $f (z, \theta)$ associated with $(z, \theta) \rightarrow ({\tilde z}, {\tilde \theta})$ is denoted by
\be
	{\hat i} f (z, \theta)
        =	
        f ({\tilde z}, {\tilde \theta})
        =	
        {\rm fn} ({\bar z}, \: {\rm and} \: {\bar \theta} \: {\rm only})	\:.
\ee
Let the supersymmetry transformation of $f (z, \theta)$ be
\be
	\delta f
        =
        (\varepsilon Q - {\bar \varepsilon} {\bar Q}) f (z, \theta)		\:.
\ee
We require
\ba
	{\hat i} \delta f
        &=&	
        \delta {\hat i} f (z, \theta)
        =	%
        \delta f ({\tilde z}, {\tilde \theta})
        =	%
        - {\bar \varepsilon} {\bar Q} f ({\tilde z}, {\tilde \theta})			\nn
        =	
        {\hat i} \varepsilon Q f
        &=&	
        {\hat i} \varepsilon
        \left({
        	i \theta \pardif{}{z} + \pardif{}{\theta}
        }\right)
        f (z, \theta)
        =	%
        {\tilde \varepsilon} {\tilde Q} f ({\tilde z}, {\tilde \theta})			\:.
\ea
So we conclude
\be
	{\tilde \varepsilon} {\tilde Q}
        =	
        - {\bar \varepsilon} {\bar Q}
\ee
\be
	\delta {\tilde \theta}
        \left({
        	i {\tilde \theta} \pardif{}{\tilde z} + \pardif{}{\tilde \theta}
        }\right)
        =	
        -
        \delta {\bar \theta}
        \left({
        	- i {\bar \theta} \pardif{}{\bar z} - \pardif{}{\bar \theta}
        }\right)			\:.
\ee
Therefore,
\ba
	{\rm if} \:\:\:
        {\tilde z} &=& {\bar z}, \:\:\:\:\: {\rm then} \:\: {\tilde \theta} = \pm {\bar \theta} \:\:\:\:\: ({\rm UHP})		\:,		\nn
        {\tilde z} &=& - {\bar z}, \:\: {\rm then} \:\: {\tilde \theta} = \pm i {\bar \theta} \:\:\: ({\rm annulus})		\:,		\nn
        {\rm and}	\:\:\:
        {\tilde z} &=& \frac{1}{\bar z}, \:\:\:\:\: {\rm then} \:\: {\tilde \theta} = \pm \frac{i {\bar \theta}}{\bar z} \:\: ({\rm disk})		\:.
\ea

\section{Supplement to $N_{+ \pm}^{IJ}, B_{\nu_{\rm f}}^{IJ}, C_{+ \pm}^{IJ}, E_{+ \pm}^{IJ}$	\label{app:propertyofNBCE}}  

\subsection{properties under $I \leftrightarrow J$}

Here we check the properties under $I \leftrightarrow J$.

Due to the even/odd properties for theta functions,
\be	
        \frac{\vartheta \left[{\textstyle {{1 \over 2} \atop {1 \over 2}}} \right] (-z)}{\vartheta' \left[{\textstyle {{1 \over 2} \atop {1 \over 2}}} \right] (0)}
        =	%
        \frac{- \vartheta \left[{\textstyle {{1 \over 2} \atop {1 \over 2}}} \right] (z)}{\vartheta' \left[{\textstyle {{1 \over 2} \atop {1 \over 2}}} \right] (0)}
        	\label{Theta [1/2,1/2]:even_or_odd}
\ee
\be	
        {\cal{S}}_{\nu_{\rm f}}	(-z)	
        =	
        \frac{i}{\pi}
        \frac{\vartheta_{\nu_{\rm f}} (-z)}{\vartheta_{\nu_{\rm f}} (0)} \frac{\vartheta' \left[{\textstyle {{1 \over 2} \atop {1 \over 2}}} \right] (0)}{\vartheta \left[{\textstyle {{1 \over 2} \atop {1 \over 2}}} \right] (-z)}
        =	%
        \frac{i}{\pi}
        \left({
        	\frac{+ \vartheta_{\nu_{\rm f}} (z)}{\vartheta_{\nu_{\rm f}} (0)}
        }\right)
        \left({
        	\frac{\vartheta' \left[{\textstyle {{1 \over 2} \atop {1 \over 2}}} \right] (0)}{- \vartheta \left[{\textstyle {{1 \over 2} \atop {1 \over 2}}} \right] (z)}
        }\right)
        =	%
        - {\cal{S}}_{\nu_{\rm f}} (z)					\:.
        	\label{S_nu_evenodd}
\ee
In addition, by the fact that the derivative of a even/odd function becomes odd/even,
\ba	
	\frac{\vartheta' \left[{\textstyle {{1 \over 2} \atop {1 \over 2}}} \right] (-z)}{\vartheta \left[{\textstyle {{1 \over 2} \atop {1 \over 2}}} \right] (-z)}	
        &=&	
        \frac{+ \vartheta' \left[{\textstyle {{1 \over 2} \atop {1 \over 2}}} \right] (z)}{- \vartheta \left[{\textstyle {{1 \over 2} \atop {1 \over 2}}} \right] (z)}
        =	%
        - \frac{\vartheta' \left[{\textstyle {{1 \over 2} \atop {1 \over 2}}} \right] (z)}{\vartheta \left[{\textstyle {{1 \over 2} \atop {1 \over 2}}} \right] (z)}
        	\label{fracTheta1stder}			\\
        \frac{\vartheta'' \left[{\textstyle {{1 \over 2} \atop {1 \over 2}}} \right] (-z)}			
               {\vartheta \left[{\textstyle {{1 \over 2} \atop {1 \over 2}}} \right] (-z)}
        &=&	
        \frac{- \vartheta'' \left[{\textstyle {{1 \over 2} \atop {1 \over 2}}} \right] (z)}
             {- \vartheta \left[{\textstyle {{1 \over 2} \atop {1 \over 2}}} \right] (z)}
	=	%
        + \frac{\vartheta'' \left[{\textstyle {{1 \over 2} \atop {1 \over 2}}} \right] (z)}
               {\vartheta \left[{\textstyle {{1 \over 2} \atop {1 \over 2}}} \right] (z)}				\:.
        	\label{fracTheta2ndder}
\ea
Using these,
\ba	
	\pi N_{++}^{JI}	
        &{\overset{{\rm eqs.} (\ref{Theta [1/2,1/2]:even_or_odd}), (\ref{S_nu_evenodd})}{=}}&
        \ln \left|{
        	-
                \frac{\vartheta \left[{\textstyle {{1 \over 2} \atop {1 \over 2}}} \right] \left({ \frac{z_I}{2} - \frac{z_J}{2} \left|{\frac{i\tau_2}{2}}\right. }\right)}
        	     		     {\vartheta' \left[{\textstyle {{1 \over 2} \atop {1 \over 2}}} \right] \left({ 0 \left|{\frac{i\tau_2}{2}}\right. }\right)}
            }\right|
         + {\pi \over 2} \frac{(z_I - z_J)^2}{\tau_2}
        = \pi N_{++}^{IJ}						\nn
        B_{\nu_{\rm f}}^{JI}	
        &{\overset{{\rm eq.} (\ref{S_nu_evenodd})}{=}}&
        \frac{1}{2} \frac{\pi}{i} (-1) {\cal{S}}_{\nu_{\rm f}} \left({ \frac{z_I}{2} - \frac{z_J}{2} \left|{ \frac{i \tau_2}{2} }\right.}\right)
        = - B_{\nu_{\rm f}}^{IJ}						\nn
        C_{++}^{JI}	
        &{\overset{{\rm eq.} (\ref{fracTheta1stder})}{=}}&
        \frac{1}{2} (-1)
        \frac{\vartheta' \left[{\textstyle {{1 \over 2} \atop {1 \over 2}}} \right] (\frac{z_I}{2} - \frac{z_J}{2} | \frac{i \tau_2}{2})}{\vartheta \left[{\textstyle {{1 \over 2} \atop {1 \over 2}}} \right] (\frac{z_I}{2} - \frac{z_J}{2} | \frac{i \tau_2}{2})}
	 + (-1) \pi \frac{z_I - z_J}{\tau_2}
        = - C_{++}^{IJ}						\nn
        E_{++}^{JI}	
        &{\overset{{\rm eqs.} (\ref{fracTheta1stder}) ,\: (\ref{fracTheta2ndder})}{=}}&
        \frac{1}{4}
        \left\{{
          + \frac{\vartheta'' \left[{\textstyle {{1 \over 2} \atop {1 \over 2}}} \right] \left({ \frac{z_I}{2} - \frac{z_J}{2} \left|{\frac{i\tau_2}{2}}\right. }\right)}
               {\vartheta \left[{\textstyle {{1 \over 2} \atop {1 \over 2}}} \right] \left({ \frac{z_I}{2} - \frac{z_J}{2} \left|{\frac{i\tau_2}{2}}\right. }\right)}
          - \left({
        	- \frac{\vartheta' \left[{\textstyle {{1 \over 2} \atop {1 \over 2}}} \right] \left({ \frac{z_I}{2} - \frac{z_J}{2} \left|{\frac{i\tau_2}{2}}\right. }\right)}
        	       {\vartheta \left[{\textstyle {{1 \over 2} \atop {1 \over 2}}} \right] \left({ \frac{z_I}{2} - \frac{z_J}{2} \left|{\frac{i\tau_2}{2}}\right. }\right)}
            }\right)^2
        }\right\}
         + \frac{\pi}{\tau_2}
        = E_{++}^{IJ}						\:.		\nn
        	\label{A,B,C,E:IJandJI}
\ea
From eq. (\ref{Def:N,C,E:+-}),
\ba
	\pi N_{+-}^{JI} &=& + \pi N_{+-}^{IJ}						\nn
	B_{\nu_{\rm f}}^{JI} &=& - B_{\nu_{\rm f}}^{IJ}					\nn
        C_{+-}^{JI} &=& - C_{+-}^{IJ}							\nn
        E_{+-}^{JI} &=& + E_{+-}^{IJ}				\:.
        	\label{NBCE:+-:IJ<->JI}
\ea

\subsection{singularity at $z_I \sim z_J$}

Let us look at a singularity of $\pi N_{+ \pm}^{IJ}$, $B_{\nu_{\rm f}}^{IJ}$, $C_{+ \pm}^{IJ}$ and $E_{+ \pm}^{IJ}$ at $z_I \sim z_J$.

By eq. (\ref{Theta[1/2,1/2](z)/Theta'[1/2,1/2](0):z->0}),
\ba	
	\pi N_{+ \pm}^{IJ}
        &{\overset{z_J \sim z_I}{\sim}}&	
         \ln \left|{ \frac{z_I}{2} - \frac{z_I}{2} }\right|			\nn
        B_{\nu_{\rm f}}^{IJ}
        &{\overset{z_J \sim z_I}{\sim}}&	
         \frac{1}{2} \frac{1}{\frac{z_I}{2} - \frac{z_I}{2}}		\nn
        C_{+ \pm}^{IJ}
        &{\overset{z_J \sim z_I}{\sim}}&	
         \frac{1}{2} \frac{1}{\frac{z_I}{2} - \frac{z_I}{2}}		\nn
        E_{+ \pm}^{IJ}
        &{\overset{z_J \sim z_I}{\sim}}&	
        - \frac{1}{4}
           \frac{1}{\left({\frac{z_I}{2} - \frac{z_I}{2}}\right)^2}				\:,
        	\label{singularityofA,B,C,E}
\ea
 where we have also used
\ba	
	\lim_{z \rightarrow 0} \frac{\vartheta'' \left[{\textstyle {{1 \over 2} \atop {1 \over 2}}} \right] (z|\tau)}{\vartheta \left[{\textstyle {{1 \over 2} \atop {1 \over 2}}} \right] (z|\tau)}
        &{\overset{\rm L'Hopital's \: rule}{=}}&	
        \lim_{z \rightarrow 0} \frac{\vartheta''' \left[{\textstyle {{1 \over 2} \atop {1 \over 2}}} \right] (z|\tau)}{\vartheta' \left[{\textstyle {{1 \over 2} \atop {1 \over 2}}} \right] (z|\tau)}
        {\overset{{\rm eq.} (\ref{Heateq})}{=}}	%
        \lim_{z \rightarrow 0} \frac{\pardif{}{z} \left\{{ 4 \pi i \pardif{}{\tau} \vartheta \left[{\textstyle {{1 \over 2} \atop {1 \over 2}}} \right] (z|\tau) }\right\}}
        			    {\vartheta' \left[{\textstyle {{1 \over 2} \atop {1 \over 2}}} \right] (z|\tau)}		\nn
        &=&	
        4 \pi i
        \lim_{z \rightarrow 0} \frac{\pardif{}{\tau} \left\{{ \vartheta' \left[{\textstyle {{1 \over 2} \atop {1 \over 2}}} \right] (z|\tau) }\right\}}
        			    {\vartheta' \left[{\textstyle {{1 \over 2} \atop {1 \over 2}}} \right] (z|\tau)}
        =	%
        4 \pi i
        \frac{\pardif{}{\tau} \left\{{ \lim_{z \rightarrow 0} \vartheta' \left[{\textstyle {{1 \over 2} \atop {1 \over 2}}} \right] (z|\tau) }\right\}}
             {\lim_{z \rightarrow 0} \vartheta' \left[{\textstyle {{1 \over 2} \atop {1 \over 2}}} \right] (z|\tau)}			\nn
	&=&	
        4 \pi i
        \frac{\pardif{}{\tau} \left[{- 2 \pi \{{\eta (\tau)}\}^3 }\right]}{- 2 \pi \{{\eta (\tau)}\}^3}
        =	%
        4 \pi i
        \frac{\pardif{}{\tau} \left[{\{{\eta (\tau)}\}^3 }\right]}{\{{\eta (\tau)}\}^3}			\nn
        &=&	
        4 \pi i
        \pardif{}{\tau}
        \ln{ \{{\eta (\tau)}\}^3}
        =	%
        3 \cdot 4 \pi i
        \pardif{}{\tau} \ln{\eta (\tau)}
        	\label{Theta''(0|tau)/Theta(0|tau)}
\ea
  to evaluate $E_{++}^{IJ}$.

\subsection{eq. (\ref{exp[...]}) at $z_I \sim z_J$ in case of maximal supersymmetry}

Let us check that eq. (\ref{exp[...]}) reduces to that of \cite{ItoMox} at $z_I \sim z_J$ in case of maximal supersymmetry.

According to eq. (\ref{singularityofA,B,C,E}),
\ba	
	&\:&	
        \exp \left[{
        	2 \alpha' \sum_{1 \leq I < J \leq N}
        	k_I \cdot k_J \pi N_{++}^{IJ}
             }\right]					\nn
        &\:&	
        \exp \left[{
        	2 \alpha' \sum_{1 \leq I < J \leq N}
                \left\{{
        		i k_I \cdot k_J \theta_I \theta_J B_{\nu_{\rm f}}^{IJ}
                }\right.
             }\right.					\nn
        &\:&	\qq\qq	
        	\left.{
                	+ {
                		(k_I \cdot \eta_J \theta_I - k_J \cdot \eta_I \theta_J) B_{\nu_{\rm f}}^{IJ}
                		+ (k_J \cdot \eta_I \theta_I - k_I \cdot \eta_J \theta_J) C_{++}^{IJ}
                	  }
             	}\right.					\nn
        &\:&	\qq\qq	
        \left.{
        	\left.{
        		- i \eta_I \cdot \eta_J	B_{\nu_{\rm f}}^{IJ}
        		+ \eta_I \cdot \eta_J \theta_I \theta_J E_{++}^{IJ}
                }\right\}
             }\right]						\nn
        &\:&
        {\overset{z_I \sim z_J}{\sim}}	
        \exp \left[{
        	2 \alpha' \sum_{1 \leq I < J \leq N}
        	k_I \cdot k_J \ln \left|{ \frac{z_I}{2} - \frac{z_J}{2} }\right|
             }\right]					\nn
        &\:&	
        \exp \left[{
        	2 \alpha' \sum_{1 \leq I < J \leq N}
                \left\{{
        		i k_I \cdot k_J \theta_I \theta_J \frac{1}{2} \frac{1}{\frac{z_I}{2} - \frac{z_J}{2}}
                }\right.
             }\right.					\nn
        &\:&	\qq\qq	
        	\left.{
                	+ \left({
                		(k_I \cdot \eta_J \theta_I - k_J \cdot \eta_I \theta_J) \frac{1}{2} \frac{1}{\frac{z_I}{2} - \frac{z_J}{2}}
                		+ (k_J \cdot \eta_I \theta_I - k_I \cdot \eta_J \theta_J) \frac{1}{2} \frac{1}{\frac{z_I}{2} - \frac{z_J}{2}}
                	  }\right)
             	}\right.					\nn
        &\:&	\qq\qq	
        \left.{
        	\left.{
        		- i \eta_I \cdot \eta_J	\frac{1}{2} \frac{1}{\frac{z_I}{2} - \frac{z_J}{2}}
        		+ \eta_I \cdot \eta_J \theta_I \theta_J \frac{(-1)}{4} \frac{1}{\left({ \frac{z_I}{2} - \frac{z_J}{2} }\right)^2}
                }\right\}
             }\right]						\nn
        &\:&
        =	
        \prod_{1 \leq I < J \leq N}
        \left|{ \frac{z_I}{2} - \frac{z_J}{2} }\right|^{2 \alpha' k_I \cdot k_J}		\nn
        &\:&	\qq	
        \exp \left[{
        	\sum_{1 \leq I < J \leq N}
                \left\{{
        		2 \alpha' i k_I \cdot k_J \theta_I \theta_J \frac{1}{2} \frac{1}{\frac{z_I}{2} - \frac{z_J}{2}}
                }\right.
             }\right.
                	+ 2 \alpha' (k_I \cdot \eta_J + k_J \cdot \eta_I)
                          \frac{1}{2}
                          \frac{\theta_I - \theta_J}{\frac{z_I}{2} - \frac{z_J}{2}}
                        				\nn
        &\:&	\qq\qq	
        \left.{
        	\left.{
        		- 2 \alpha' i \eta_I \cdot \eta_J \frac{1}{2} \frac{1}{\frac{z_I}{2} - \frac{z_J}{2}}
        		- 2 \alpha' \eta_I \cdot \eta_J \theta_I \theta_J \frac{1}{4} \frac{1}{\left({ \frac{z_I}{2} - \frac{z_J}{2} }\right)^2}
                }\right\}
             }\right]						\:.
        	\label{N_SA:z_Itoz_J}
\ea

\newpage


\end{document}